\documentclass[a4paper]{article}

\usepackage{geometry}
\usepackage{mathpartir}
\usepackage{import}
\usepackage{multirow}
\usepackage{amssymb}
\usepackage{stmaryrd}
\SetSymbolFont{stmry}{bold}{U}{stmry}{m}{n}
\usepackage{mathtools}
\usepackage[makeroom]{cancel}
\usepackage[table,xcdraw,dvipsnames]{xcolor}
\usepackage{verbatim}
\usepackage{bussproofs}
\usepackage{url}
\usepackage{graphicx}
\usepackage[inline]{enumitem}
\usepackage{lmodern}
\usepackage{cmll}
\usepackage{pdflscape}
\usepackage{float}
\usepackage{tikz}
\usepackage[framemethod=tikz]{mdframed}
\mdfsetup{%
    innerleftmargin=2pt,innerrightmargin=2pt,
    innertopmargin=2pt,innerbottommargin=2pt,
}
\usepackage[titletoc,title]{appendix}

\usepackage{xpatch}
\xpretocmd{\appendixpagename}{\sffamily}{}{}
\usepackage{hyperref}

\usepackage{array}
\usepackage{booktabs}
\usepackage{framed}
\usepackage{xspace}
\usepackage[appendix=inline]{apxproof}
\usepackage{cleveref}

\Crefname{theorem}{Theorem}{Theorems}
\Crefname{lemma}{Lemma}{Lemmata}
\Crefname{definition}{Definition}{Definitions}
\Crefname{section}{\S}{\S\S}

\Crefname{lemmaapx}{Lemma}{Lemmata}
\Crefname{lemmaapx}{Lemma}{Lemmata}
\Crefname{theoremapx}{Theorem}{Theorems}

\newtheorem{definition}{Definition} 
\newtheorem{theorem}[definition]{Theorem}
\newtheorem{lemma}[definition]{Lemma}
\newtheorem{example}[definition]{Example}

\newtheorem{corollary}[definition]{Corollary}

\newtheoremrep{theoremapx}[definition]{Theorem}
\newtheoremrep{propositionapx}[definition]{Proposition}
\newtheoremrep{lemmaapx}[definition]{Lemma}
\newtheoremrep{corollaryapx}[definition]{Corollary}

\Crefname{lemma}{Lemma}{Lemmata}
\Crefname{definition}{Definition}{Definitions}
\Crefname{theorem}{Theorem}{Theorems}

\Crefname{lemmaapx}{Lemma}{Lemmata}
\Crefname{theoremapx}{Theorem}{Theorems}

\definecolor{darkolivegreen}{rgb}{0.33, 0.42, 0.18}
\definecolor{MyGrey}{rgb}{115, 115, 115}

\allowdisplaybreaks
\usepackage{macros}
\newcommand{\mycode}[1]{\ensuremath{\textcolor{MidnightBlue}{\mathsf{#1}}\,}}

\hypersetup{
  colorlinks,
  filecolor=magenta,
  citecolor=Blue,
  linkcolor=CadetBlue,
  urlcolor=Blue}
  
\title{Contrasting Deadlock-Free Session Processes \\ (Extended Version)}
\author{Juan C. Jaramillo and Jorge A. P\'{e}rez \\ University of Groningen, The Netherlands}

\date{\today}

\usepackage{array} 
\usepackage{booktabs} 
\usepackage{graphicx} 
\newcolumntype{P}[1]{>{\centering\arraybackslash}p{#1}}

\begin{document}

\maketitle

\begin{abstract}

\emph{Deadlock freedom}  is a crucial property for  message-passing programs.
Over the years, several different type systems for concurrent processes that ensure deadlock freedom have been proposed; this diversity raises the question of how they compare.
We address this question, 
considering two type systems not covered in prior work: Kokke \etal's \HCP, a type system based on a linear logic with \emph{hypersequents}, 
and Padovani's \emph{priority-based} type system for asynchronous processes,  dubbed~\Padovani.
Their distinctive features make formal comparisons relevant and challenging.
Our findings are two-fold: (1)~the hypersequent setting does not drastically change the class of deadlock-free processes induced by linear logic, and (2)~we  relate the classes of deadlock-free processes induced by \HCP and~\Padovani.
We prove that our results hold under both synchronous and asynchronous communication.
Our results provide new insights into the essential mechanisms involved in statically avoiding deadlocks in concurrency.
\end{abstract}

   \section{Introduction}
   \label{sec: introduction}
    The \emph{deadlock freedom} property (DF, in the following) is a crucial guarantee for  message-passing programs. Informally, DF ensures  that processes ``never get stuck''---an essential requirement in  concurrent and distributed software systems, where even a single component that becomes permanently blocked while waiting for a message can compromise reliability.
     As illustration, consider a toy concurrent program in the  language of Gay and Vasconcelos~\cite{DBLP:journals/jfp/GayV10}:
     \begin{align*}
     \langle \mycode{let}(x_1,z)=\mycode{receive}x_1~\mycode{in~send}\,\mathtt{42}\,y_1 \rangle
     \parallel \langle \mycode{let}(y_2,k)=\mycode{receive}y_2~\mycode{in~send}\,`\mathtt{hello}\text{'}\,x_2 \rangle
         \end{align*}
We have two concurrent threads, each running an expression with two actions in sequence.
This program is evaluated under a context that binds together $x_1$ and $x_2$  and declares them as endpoints of the same linear channel (same for $y_1$ and $y_2$); we omit it for readability.  
Given a linear channel $x$, `$\mycode{send}\,\mathtt{v}\,x$' sends value $\mathtt{v}$ along $x$ and returns the continuation of $x$, whereas 
`$\mycode{receive}x$' returns a pair with the continuation of~$x$ and a received value. 
The threads are thus engaged in separate but intertwined protocols (one along $x_i$, the other along~$y_i$). Each thread should receive on a channel and then send a value on the other. However, the output matching the input in one thread is blocked in the other---the program is stuck. The key challenge in enforcing DF is thus performing the non-local analysis needed to avoid \emph{circular dependencies} between threads, which are often less apparent than in this program.
    
    An expressive model for concurrent processes, the $\pi$-calculus offers a convenient basis for developing analysis techniques for message-passing programs in different paradigms~\cite{ATtheoryofMobileProcesses}. Several type systems that enforce (forms of) DF have been proposed over the years (see, e.g.,~\cite{Kobayashi06,DBLP:conf/tgc/Dezani-CiancaglinidY07,DBLP:journals/corr/abs-1010-5566,padovani_linear_pi,DBLP:conf/fossacs/DardhaG18,DBLP:journals/toplas/ToninhoY18,DBLP:conf/esop/BalzerTP19}).
    They avoid circular dependencies in processes using different insights, which raises the  question of how they compare.
     In this paper, we rigorously compare some representative type systems for DF to better understand their underlying mechanisms.

A key motivation for studying type systems for concurrent processes is their suitability for developing advanced analysis  techniques at an appropriate level of abstraction---these typed frameworks have been directly incorporated into languages such as  Scala~\cite{scalas_et_al:LIPIcs.ECOOP.2017.24}, OCaml~\cite{DBLP:journals/jfp/Padovani17,imai_et_al:LIPIcs.ECOOP.2020.9}, and Rust~\cite{lagaillardie_et_al:LIPIcs.ECOOP.2022.4}.
Also, concurrent processes offer an adequate representation level for verifying liveness properties in  real Go programs~\cite{DBLP:conf/popl/LangeNTY17,DBLP:conf/icse/LangeNTY18}. Moreover, type systems for the $\pi$-calculus are the basis for static analyzers (such as Kobayashi's TyPiCal~\cite{TyPiCal}) and 
can be  ported to enforce DF for higher-order concurrent programs~\cite{DBLP:conf/forte/PadovaniN15}.
Type systems based on the Curry-Howard correspondence between session types and linear logic (\PaS~\cite{DBLP:conf/concur/CairesP10,Wadler14})  enforce DF; they entail topological invariants on processes, key to ensure deadlock and memory leak freedom in  concurrent programs~\cite{DBLP:journals/pacmpl/JacobsB23}. 
Further understanding the essential mechanisms that enforce DF in logic-based session types is a central topic in our work.

Prior work by Dardha and P\'{e}rez~\cite{ComparingDeadlock} contrasts type systems for DF by relating the classes of   processes they induce. 
They compare a \emph{priority-based} type system by Kobayashi~\cite{Kobayashi06} and a type system derived from \PaS.
Writing $\ClassK$ and $\ClassLDP$ to denote the respective classes of processes, a result   in~\cite{ComparingDeadlock} is that $\ClassLDP \subsetneq \ClassK$:
there exist deadlock-free processes  that cannot be typed by logic-based  type systems. 
The class $\ClassK \setminus \ClassLDP$ includes, for instance, process networks organized in circular topologies (e.g., the dinning philosophers). It also includes process $P_0$ below, in which two processes interact along independent protocols (\emph{sessions}):
\[P_0 \DefEq \news{x_1}{x_2}\news{y_1}{y_2}(
\InputSP{x_1}{z}\OutputSP{y_1}{\mathtt{42}}\Inact
\para
\OutputSP{x_2}{\mathtt{hello}}\InputSP{y_2}{k}\Inact)
\] 
Above, `$\news{x_1}{x_2}$' declares a new session with linear endpoints $x_1$ and $x_2$, 
and `$\InputSP{x}{y}Q$' (resp. `$\OutputSP{x}{v}Q$') denotes an input (resp. output of value $v$) along $x$ with continuation $Q$.
 $P_0$ is the deadlock-free variant of the program given above, now neatly specified as a process.
In  $\ClassLDP$,  processes can share at most one session in parallel, whereas processes in $P_0$ share two: the session `$x_1x_2$' for exchanging `$\mathtt{hello}$' and the session `$y_1y_2$' for exchanging `$\mathtt{42}$'.
In~\cite{ComparingDeadlock}, both compared classes rely on a reduction semantics; accordingly, DF concerns the absence of stable states with \emph{pending reduction steps} on linear names/channels.
  
  Here we extend the work in~\cite{ComparingDeadlock} by considering type systems not studied there: (i)~Kokke \etal's \HCP~\cite{betterlate} and  
  (ii)~Padovani's priority-based type system~\cite{padovani_linear_pi}, here dubbed~\Padovani. 
  Their distinctive features make formal comparisons relevant and challenging, 
  as we discuss next.

  Derived from \PaS, $\HCP$ is based on a presentation of classical linear logic ($\CLL$) with \emph{hypersequents}. 
  \HCP  has been studied in several works, which have deepened into its theory and applications~\cite{TakingLinearLogicApart,DBLP:conf/concur/0001KDLM21,client-server}.
  In $\HCP$,  the   semantics of processes is given by a Labeled Transition System (LTS), rather than by a reduction semantics. This   unlocks the  use of  well-established analysis techniques, notably observational equivalences, to reason about typed processes.  
  The (typed) LTS for \HCP includes the usual rules for observing actions and synchronizations, here dubbed \emph{regular transitions}, 
  but also rules that allow for \emph{delayed actions} (i.e., observable actions due to prefixes not at top-level) and \emph{self-synchronizations} (i.e., internal actions that do not require the interaction of two parallel processes). 
  Remarkably, the bisimilarity induced by the LTS coincides, in a fully-abstract way, with a \emph{denotational equivalence} over processes---this gives a strong justification for the design of the LTS.
  Typing ensures \emph{progress}: well-typed processes not equivalent to $\Inact$ always have a transition step.

An intriguing observation is that there are $\HCP$ processes that are stuck in a reduction-based setting but have synchronizations under $\HCP$'s LTS. 
We discuss two interesting examples, using `$\EmptyInputSP{x}Q$' and `$\EmptyOutputSP{x}Q$' to denote empty inputs and outputs, respectively: 
\begin{itemize}
	\item Process $P_1 \DefEq \news{x}{x'}\EmptyOutputSP{x'}\EmptyInputSP{x}\Inact$ is self-synchronizing: it does not have any reductions, whereas $P_1 \TransitionHCP{\tau} \Inact$, as in \HCP synchronizations do not need a parallel context to be enabled. 
\item Process
$P_2 \DefEq \news{x}{x'}\news{y}{y'}(
\EmptyOutputSP{y}\EmptyInputSP{x}\Inact
\para
\EmptyOutputSP{x'}\EmptyInputSP{y'}\Inact)
$ is arguably the paradigmatic example of a deadlock, with two independent sessions unable to reduce. 
In \HCP,   
thanks to delayed actions, 
we have both  $P_2 \TransitionHCP{\tau} 
\news{y}{y'}(
\EmptyOutputSP{y}\Inact
\para
\EmptyInputSP{y'}\Inact)
$
and
$P_2 \TransitionHCP{\tau} 
\news{x}{x'}(
\EmptyOutputSP{x}\Inact
\para
\EmptyInputSP{x'}\Inact)
$.
\end{itemize}

This observation shows that comparisons such as~\cite{ComparingDeadlock} should accommodate other semantics for processes, beyond reductions, which entail other definitions of DF and induce new deadlock-free processes (such as $P_1$ and $P_2$). This begs a first question: 

\begin{description}
	\item[(\textcolor{CadetBlue}{Q1})\label{lq1}]
What is the status of delayed actions and self-synchronizations, as typable in \HCP, with respect to DF? Do they induce a  distinct class of deadlock-free processes? 
\end{description}

Considering 
Padovani's \Padovani means addressing the case of processes with \emph{asynchronous communication}, widely used in real programs. 
In the $\pi$-calculus, asynchronous communication is obtained by defining outputs  as non-blocking actions; only inputs can block the execution of a process~\cite{Boudol92,HondaK:asycs}. Asynchrony and DF are then closely related: because an asynchronous setting involves the least amount of blocking operations, it offers the most elementary scenario for investigating  DF~\cite{DBLP:journals/corr/abs-2412-08232}. 
Now, the comparisons in~\cite{ComparingDeadlock} are limited to synchronous processes; we then observe that they should be extended to address asynchronous communication, too. 
Let us write $\ClassH$ and $\ClassLP$ to denote the class of processes induced by $\HCP$ and $\Padovani$, respectively, 
and $\ClassL$ to denote the class of processes induced by 
$\CP$ (a fragment of $\HCP$ without hypersequents and with a restricted form of parallel composition).
We consider the second question: 

\begin{description}
	\item[(\textcolor{CadetBlue}{Q2}) \label{lq2}] What is the status of $\ClassH$ and  $\ClassL$  with respect to asynchronous processes in $\ClassLP$?
\end{description}

Interestingly, 
our two questions are intertwined:
\HCP and \Padovani are related due to asynchrony. Our insight is that
delayed actions in $\HCP$ already implement an asynchronous semantics. 
To see this, consider the process
$P_3 \DefEq \news{x}{x'}\news{y}{y'}(
\EmptyOutputSP{y}\EmptyOutputSP{x}\Inact
\para
\EmptyInputSP{x'}\EmptyInputSP{y'}\Inact)
$, which
is not stuck: the output on $y$ is not really blocking, as delayed actions allow to anticipate the output (and synchronization) on~$x$. 
In other words, 
$P_3$ is morally the same as $\news{x}{x'}\news{y}{y'}(
\EmptyOutputSP{y}\Inact \para \EmptyOutputSP{x}\Inact
\para
\EmptyInputSP{x'}\EmptyInputSP{y'}\Inact)
$. 
We thus conclude that the (typed) operational semantics for \HCP admits many synchronous-looking processes (such as $P_3$) that actually behave in an asynchronous way; this naturally invites the question of how $\ClassH$ relates to $\ClassLP$.

   \subparagraph*{Contributions and Outline}
       This paper offers conceptual and technical contributions. From a conceptual view, we reflect upon the significance of formally comparing   different type systems for concurrency, while uniformly presenting various notions in the literature. On the technical side, we address  \hyperref[lq1]{\textbf{\textsf{(Q1)}}} and \hyperref[lq2]{\textbf{\textsf{(Q2)}}}, thus extending and complementing~\cite{ComparingDeadlock}.
       A key discovery is identifying  \emph{commuting conversions} and \emph{disentanglement} as key notions for characterizing deadlock-free processes in \HCP---as we will see, these notions define sound \emph{process optimizations} that   respect  causality and increase parallelism. As our results hold also in an asynchronous setting, not treated in prior work, our findings are most related to abstractions and mechanisms used in practical concurrent programming. More in detail:

            \begin{description}
            	\item[\Cref{sec: background}] We introduce $\SPExp$, a session $\pi$-calculus that  we use as baseline for comparisons, as well as $\CP$ and $\HCP$, the type systems resulting from the Curry-Howard correspondence with $\CLL$ and $\CLL$ with hypersequents, respectively. 

\item[\Cref{sec: results}] 
We  define~$\ClassL$ and $\ClassH$: the classes of deadlock-free  \SPExp processes induced by \CP and \HCP, respectively. We identify two relevant sub-classes of $\ClassH$:  
      \begin{description}
    \item[$\DeadlockFreedomHCPSet$:] Processes that use the $\mathsf{F}\text{ull}$ LTS.
     \item[$\DeadlockFreedomHCPRRSet$:] Processes that \emph{only}  use $\mathsf{R}$egular transitions (no delayed actions/self-synchronizations).
            \end{description}
 In our main result (\Cref{corollary: main corollary HCP}), we  prove that for every process $P \in \DeadlockFreedomHCPSet$, there exists an observationally equivalent process $P'\in \DeadlockFreedomHCPRRSet$. Enabled  by process optimizations induced via commuting conversions and disentanglement, this result shows that delayed actions and self-synchronizations are \emph{inessential} when it comes to DF, addressing \hyperref[lq1]{\textbf{\textsf{(Q1)}}}.

\item[\Cref{sec: charaterization in Pado}] 
To address \hyperref[lq2]{\textbf{\textsf{(Q2)}}}, we define $\ClassLP$: the class of deadlock free  \SPExp processes induced by \Padovani,
and its sub-class $\ClassMuP$ (`micro $\ClassLP$'), which is inspired by $\CP$. 
We prove that $\ClassL = \ClassMuP$ (\Cref{corollary: L equal to mupadovani}), thus casting Dardha and P\'{e}rez's key finding into the asynchronous setting. 
Our main results are: \Cref{lemma: comparing ClassH and Padovani}, which precisely  relates the classes $\ClassH$, $\DeadlockFreedomHCPSet$, $\DeadlockFreedomHCPRRSet$, and $\ClassLP$; and 
\Cref{corollary: main result mupado}, which strengthens \Cref{corollary: main corollary HCP} by considering DF in $\ClassLP$.
Moreover, we introduce $\SPExpA$, an asynchronous variant of $\SPExp$, and its corresponding classes of processes. We show how to transfer \Cref{lemma: comparing ClassH and Padovani,corollary: main result mupado} from $\SPExp$ to $\SPExpA$ (\Cref{c:transfer}).  
 \end{description}

\noindent 
\Cref{fig:resultsV1} offers a graphical description of the classes of processes and some of our main results (\Cref{corollary: main corollary HCP,corollary: L equal to mupadovani,lemma: comparing ClassH and Padovani}).
For simplicity, the figure does not depict several encodings needed to bridge the syntactic differences between the various classes of processes / type systems under consideration---\Cref{fig:summary} (Page~\pageref{fig:summary}) presents a detailed version. 
In  
\Cref{sec: concluding remarks} we collect some closing remarks. 

\begin{figure}[!t]
    \centering
\tikzset{every picture/.style={line width=0.75pt}} 

\begin{tikzpicture}[x=0.75pt,y=0.75pt,yscale=-1,xscale=1]

\draw   (20.6,74.95) .. controls (19.69,36.68) and (61.57,5.65) .. (114.15,5.65) .. controls (166.72,5.65) and (210.07,36.68) .. (210.98,74.95) .. controls (211.89,113.22) and (170.01,144.25) .. (117.44,144.25) .. controls (64.86,144.25) and (21.51,113.22) .. (20.6,74.95) -- cycle ;
\draw   (28.08,87.89) .. controls (27.34,56.76) and (65.81,31.52) .. (114,31.52) .. controls (162.2,31.52) and (201.87,56.76) .. (202.61,87.89) .. controls (203.34,119.01) and (164.87,144.25) .. (116.68,144.25) .. controls (68.48,144.25) and (28.81,119.01) .. (28.08,87.89) -- cycle ;
\draw   (38.42,100.75) .. controls (37.85,76.73) and (71.49,57.25) .. (113.55,57.25) .. controls (155.62,57.25) and (190.18,76.73) .. (190.75,100.75) .. controls (191.32,124.77) and (157.68,144.25) .. (115.62,144.25) .. controls (73.55,144.25) and (38.99,124.77) .. (38.42,100.75) -- cycle ;
\draw   (80.21,123.3) .. controls (79.93,111.72) and (95.82,102.34) .. (115.68,102.34) .. controls (135.55,102.34) and (151.88,111.72) .. (152.15,123.3) .. controls (152.43,134.87) and (136.55,144.25) .. (116.68,144.25) .. controls (96.81,144.25) and (80.48,134.87) .. (80.21,123.3) -- cycle ;
\draw [color={rgb, 255:red, 208; green, 2; blue, 27 }  ,draw opacity=1 ]   (114,49) -- (114,75) ;
\draw [shift={(114,77)}, rotate = 270.7] [color={rgb, 255:red, 208; green, 2; blue, 27 }  ,draw opacity=1 ][line width=0.75]    (10.93,-3.29) .. controls (6.95,-1.4) and (3.31,-0.3) .. (0,0) .. controls (3.31,0.3) and (6.95,1.4) .. (10.93,3.29)   ;
\draw [color=darkolivegreen  ,draw opacity=1 ]   (125,123) -- (317,123) ;
\draw [color=darkolivegreen ,draw opacity=1 ]   (125,125) -- (317,125) ;

\draw   (225.6,75.45) .. controls (224.69,37.18) and (266.57,6.15) .. (319.15,6.15) .. controls (371.72,6.15) and (415.07,37.18) .. (415.98,75.45) .. controls (416.89,113.72) and (375.01,144.75) .. (322.44,144.75) .. controls (269.86,144.75) and (226.51,113.72) .. (225.6,75.45) -- cycle ;
\draw   (285.96,123.8) .. controls (285.69,112.22) and (301.57,102.84) .. (321.44,102.84) .. controls (341.31,102.84) and (357.64,112.22) .. (357.91,123.8) .. controls (358.19,135.37) and (342.3,144.75) .. (322.44,144.75) .. controls (302.57,144.75) and (286.24,135.37) .. (285.96,123.8) -- cycle ;

\draw  [dash pattern={on 0.84pt off 2.51pt}]  (120,89) -- (320,89) ;
\draw  [draw opacity=0.6 ] (1.33,2) -- (433.33,2) -- (433.33,148.4) -- (1.33,148.4) -- cycle ;

\draw (320,118) node [anchor=north west][inner sep=0.75pt]   [align=left] {$\ClassMuP$};
\draw (320,10.12) node [anchor=north west][inner sep=0.75pt]   [align=left] {$\ClassLP$};
\draw (320,80) node [anchor=north west][inner sep=0.75pt]   [align=left] {$P''$};
\draw (406.8,7.67) node [anchor=north west][inner sep=0.75pt]   [align=left] {$\SPExp$};
\draw (48.28,51.06) node [anchor=north west][inner sep=0.75pt]   [align=left] {$\DeadlockFreedomHCPSet$};
\draw (51.53,103.33) node [anchor=north west][inner sep=0.75pt]   [align=left] {$\DeadlockFreedomHCPRRSet$};
\draw (106,10.12) node [anchor=north west][inner sep=0.75pt]  [xslant=-0.03]  {$\ClassH$};
\draw (106,35) node [anchor=north west][inner sep=0.75pt]   [align=left] {$P$};
\draw (106,118) node [anchor=north west][inner sep=0.75pt]   [align=left] {$\ClassL$};
\draw (106,80) node [anchor=north west][inner sep=0.75pt]    {$P'$};
\end{tikzpicture}
    \caption{Overview of main results (simplified version). 
   $\DeadlockFreedomHCPSet$, $\DeadlockFreedomHCPRRSet$, $\ClassLP$, $\ClassL$, and $\ClassMuP$ stand for classes of deadlock-free $\SPExp$ processes.
    The red arrow represents optimizations (commuting conversions and disentanglement) that relate $\DeadlockFreedomHCPSet$ and $\DeadlockFreedomHCPRRSet$.  
    The green lines denote the correspondence $\ClassL = \ClassMuP$. The dotted line   means that $P'$ and $P''$ are the representatives of   $P$ in $\DeadlockFreedomHCPRRSet$ and $\ClassLP$, up to an encoding.}
    \label{fig:resultsV1}
\end{figure}

\section{Session Processes and Their Type Systems}\label{sec: background}
We present \SPExp, our reference language of session processes.
As observed by Dardha and P\'{e}rez~\cite{ComparingDeadlock},
this is a convenient framework for formal comparisons,
because typing in \SPExp ensures communication safety but not DF.  
We also present \CP and \HCP, the two typed languages based on \PaS, and recall useful associated results. 

\subsection{Session Processes  \texorpdfstring{($\SPExp$)}{(SP)}}
\label{ss:sp}

$\SPExp$ is a core session-typed $\pi$-calculus,  based on the language defined by Vasconcelos~\cite{vasco_sess_typ}. 
We assume a base set of \emph{variables}, ranged over by $x, y, \ldots$, which denote \emph{channels} (or \emph{names}): they can be seen as \emph{endpoints} on which interaction takes place. Processes interact to exchange values $v, v', \ldots$;  for simplicity, values correspond to variables, i.e., $v  ::=   x$.  
 We write $\widetilde{x}$ to denote a finite sequence of variables.
The syntax of \emph{processes} is as follows:
\begin{eqnarray*}
  P       &::= \  \Inact \mybar  \OutputSP{x}{v} P   \mybar  \InputSP{x}{y}P  \mybar  P_1 \para P_2  \mybar  \news{x}{y} P  \mybar \EmptyOutputSP{x}P  
   \mybar  \EmptyInputSP{x}P    \end{eqnarray*}
\noindent
Process $\Inact$ denotes inaction.
Process $\OutputSP{x}{v} P$ sends  $v$ along  $x$ and continues as $P$.
Process $\InputSP{x}{y}P$ receives  $v$ along $x$ and continues as $P\subst{v}{y}$, i.e., the process resulting from the capture-avoiding substitution of  $y$ by $v$ in  $P$.
Process $P_1 \para P_2$ denotes the parallel execution of $P_1$ and $P_2$. 
Process $\news x y P$ declares  $x$ and $y$ as \emph{co-variables}, restricting their scope to $P$. That is, restriction declares co-variables as dual ends of a channel.
Finally, $\EmptyOutputSP{x}P$ and $\EmptyInputSP{x}P$ denote empty output and input (close and wait), which represent the closing  of session $x$.  

As usual, $ \InputSP{x}{y}P$ binds  $y$ in $P$ and $\news x y P$ binds  $x,y$ in $P$.
The sets of free and bound variables of  $P$, denoted $\mathtt{fv}(P)$ and $\mathtt{bv}(P)$, are defined accordingly.
The semantics of processes is given using a reduction relation, defined next. 
\begin{definition}[$\SPExp$: Reduction]\label{def: reduction semantics sp}
    \emph{Reduction} for $\SPExp$, denoted $\longrightarrow$,  is   defined by the rules   in \Cref{fig:sess_pi_semantics} (top).
     It relies on \emph{structural congruence}, denoted 
     $\equiv$, the smallest congruence   on processes that satisfies the axioms in \Cref{fig:sess_pi_semantics} (bottom).
\end{definition}

 We now define session types, which express  protocols associated to channels:
\begin{definition}[$\SPExp$: Session Types and Typing Environments]\label{def: SPExp}
The syntax of session types and typing environments is inductively defined as follows:
    \begin{eqnarray*}
T,S \  ::=   \ ?S.T \mybar ~!S.T \mybar \EndTypeI \mybar \EndTypeO \qquad \qquad  
    \Gamma\ ::=   \  \emptyset \mybar \Gamma, x : T  
\end{eqnarray*}
\end{definition}
\noindent

The type $!S.T$ (resp.  $?S.T$) is assigned to a channel able to send (resp. receive) a value of type $S$ and to continue as $T$. 
For convenience, we have two different types for  channels with terminated protocols, denoted $\EndTypeI$ and $\EndTypeO$.
Duality on types ensures that the  endpoints of a channel agree on the protocol they describe:

\begin{definition}[$\SPExp$: Duality]\label{def: dualSP}
The dual of $S$, denoted $\DualSPExp{S}$, is defined as follows:
\[\dual{\InputST{\empty}{S}{T}}\DefEq\, \OutputST{\empty}{S}{\dual{T}}  
\qquad \dual{\OutputST{\empty}{S}{T}}\DefEq\InputST{\empty}{S}{\dual{T}} 
\qquad \dual{\EndTypeI}\DefEq\EndTypeO
\qquad \dual{\EndTypeO}\DefEq\EndTypeI \]
\end{definition}

\begin{figure}[!t]  
    \footnotesize
    \begin{align*}
             \news x y(\OutputSP{x}{v} P \para  \InputSP{y}{z}Q\para S) &\ReductionSP \news x y(P \para Q\subst{v}{z} \para S) & \textsc{R-Com}
    \\
\news xy(\EmptyOutputSP{x}P\para \EmptyInputSP{y} Q)&\ReductionSP P \para Q & \textsc{R-EmptyCom}
    \end{align*}
    \begin{mathpar}
   \inferrule[\textsc{R-Par}]{P \longrightarrow P'}{ P \para Q \longrightarrow P' \para Q }
    \and
    \inferrule[\textsc{R-Res}]{P \longrightarrow P'}{ \news x y P\ \longrightarrow \news x yP'}
    \and 
   \inferrule[\textsc{R-Struct}]{ P \equiv P' \quad P' \longrightarrow Q'\quad Q' \equiv Q}{ P \longrightarrow Q}
    \end{mathpar}
            \vspace{-1mm}
        \rule[0.5ex]{13.25cm}{0.5pt}
        \vspace{-1mm}
    \begin{align*}
P \para Q\ &\equiv\ Q \para P                     &  \quad P \para \Inact\ &\equiv\ P\\
  P \para (Q \para R)\ &\equiv\ (P \para Q) \para R     &    \news x y P &\equiv \news y x P \\
  \news x yP \para Q\ &\equiv \ \news x y(P \para Q) \ \  \text{if $x, y \notin \mathtt{fv}(Q)$} & \news x y\news w z P\ &\equiv\ \news w z\news x yP
\end{align*}
   \vspace{-0.5cm}
    \caption{$\SPExp$: Reduction semantics (top) and structural congruence (bottom).}
    \label{fig:sess_pi_semantics}
  \end{figure}

\begin{figure}[t!]
  \footnotesize
  \begin{mathpar}
    \inferrule[\textsc{Inact}]{\phantom{\Gamma}}{\cdot\SPExpJud \Inact} 
    \and
    \inferrule[\textsc{EInput}]{\Gamma\SPExpJud P}{\Gamma, x: \EndTypeI\SPExpJud \EmptyInputSP{x}P}
    \and
    \inferrule[\textsc{EOutput}]{\Gamma\SPExpJud P}{\Gamma, x:\EndTypeO\SPExpJud \EmptyOutputSP{x}P}
    \and
\inferrule[\textsc{Par}]{\Gamma_1 \SPExpJud P \qquad \Gamma_2  \SPExpJud Q }{\Gamma_1,\Gamma_2 \SPExpJud P \para Q }
\\
\inferrule[\textsc{Res}] {\Gamma, x : T, y : \overline{T} \SPExpJud P  }{\Gamma\SPExpJud \news{x}{y} P }
\and
\inferrule[\textsc{Out}]{ \Gamma, x : U \SPExpJud P }{\Gamma, x: \LinOutputST{T}{U}, v:T \SPExpJud \OutputSP{x}{v} P  }
\and
    \inferrule[\textsc{In}] {\  \Gamma, y : T, x : U \SPExpJud P \quad  }{\Gamma, x: \LinInputST{T}{U}\SPExpJud \InputSP{x}{y} P  } 
\end{mathpar}
   \vspace*{-0.35cm}
  \caption{\SPExp: Typing Rules for Processes.}
  \label{fig:typ_rules}
\end{figure}

\begin{definition}[\SPExp: Typing Judgment and  Typing Rules]
  The judgment $\Gamma\SPExpJud P$ describes a well-typed process under the typing environment~$\Gamma$. The typing rules for $\SPExp$ are shown in \Cref{fig:typ_rules}.
We shall write  $P \in \SPExp$ if $\Gamma\SPExpJud P$, for some $\Gamma$.
\end{definition}

We now state the properties of well-typed $\SPExp$ processes, whose proofs follow~\cite{vasco_sess_typ}: 
\begin{theorem}[\SPExp: Preservation~\cite{vasco_sess_typ}]\label{theorem: preservation equivalence SP}
\label{theorem: preservation reduction SP}
Suppose $\Gamma\SPExpJud P$. 
    If $P\equiv Q$ then $\Gamma\SPExpJud Q$.
Also, if $P\ReductionSP Q$ then $\Gamma\SPExpJud Q$.
\end{theorem}

We will be interested in \emph{well-formed processes}, defined using two auxiliary notions.
First, we 
say that $P$ is \emph{prefixed at $x$}, written $\mathtt{Pr}(P)=x$, if $P$ has one of the following forms:  $\EmptyInputSP{x}P'$, $\EmptyOutputSP{x}P$, $\InputSP{x}{z}P'$, or $\OutputSP{x}{v}P'$.
Second, we say a process $\news{x}{y}(P\para Q)$ is a \emph{redex} if:
    \textup{(1)}~$P=\EmptyInputSP{x}P'$ and $Q=\EmptyOutputSP{y}Q'$, or 
    \textup{(2)}~$P=\InputSP{x}{z}P'$ and $Q=\OutputSP{y}{v}Q'$.
We may now define:

\begin{definition}[\SPExp: Well-formedness]
 A process is well-formed if for each of its structural congruent processes of the form 
 $\news{x_1}{y_1}\dots\news{x_n}{y_n}(P_1\para \cdots \para P_m)$ where  $\mathtt{Pr}(P_i)=x_k$ and $\mathtt{Pr}(P_j)=y_k$, then $\news{x_k}{y_k}(P_i\para P_j)$ is a redex.
\end{definition}
\begin{theorem}[Safety~\cite{vasco_sess_typ}]\label{theorem: Typable processes are not ill-formed}
    Typable processes are well-formed.
\end{theorem}

As a consequence of \Cref{theorem: preservation reduction SP,theorem: Typable processes are not ill-formed} we obtain the main guarantee of $\SPExp$: \emph{typable  processes only reduce to  well-formed processes}.
\begin{corollary}[\cite{vasco_sess_typ}]\label{corollary: wellformedness SP}
    If $\Gamma\SPExpJud P$ and $P\ReductionSP^* Q$, then $Q$ is well-formed.
\end{corollary}

Even though typability ensures well-formedness,
which excludes error processes such as
 $\news{x}{y}(\OutputSP{x}{v}P\para \EmptyInputSP{y}Q)$ and $\news{x}{y}(\EmptyInputSP{x}P\para \EmptyInputSP{y}Q)$,
 typability in \SPExp also admits well-typed process that exhibit unwanted behaviors, in particular deadlocks. 
Before illustrating this fact, we give the first definition of deadlock freedom that we will encounter, which follows~\cite{ComparingDeadlock}. 
We write $\new{\widetilde{xy}}Q$ to stand for $\news{x_1}{y_1}\cdots\news{x_n}{y_n}Q$, for some $n\geq 0$.

\begin{definition}[$\SPExp$: Deadlock Freedom]\label{def: deadlock freedom in SP}
    A process $P\in \SPExp$ is deadlock free, written $\DeadlockFreedomSP{P}$, if the following  condition holds: whenever $P\ReductionSP^* P'$ and one of the following holds:
        \textup{(1)}~$P'\equiv \new{\widetilde{xy}}(\OutputSP{x_i}{v}Q_1\para Q_2)$,
     \textup{(2)}~$P'\equiv \new{\widetilde{xy}}(\InputSP{x_i}{z}Q_1\para Q_2)$,
        \textup{(3)}~$P'\equiv \new{\widetilde{xy}}(\EmptyInputSP{x_i}Q_1\para Q_2)$, or
        \textup{(4)}~$P'\equiv \new{\widetilde{xy}}(\EmptyOutputSP{x_i}Q_1\para Q_2)$   
    (with $x_i \in \widetilde{x}$ in all cases), 
    then $P'\ReductionSP R$, for some $R$.
\end{definition}

Hence, $\DeadlockFreedomSP{P}$ is read as `$P$ is deadlock-free in \SPExp'.
Later on, we will encounter analogous notations for \HCP and \Padovani; they will be denoted $\DeadlockFreedomHCP{P}$ and $\DeadlockFreedomPado{P}$, respectively. 
\begin{example}[\SPExp: A Typable but Deadlocked Process]\label{example: type derivation} 
Consider the  process $P_{\ref{example: type derivation}}$ defined as \[P_{\ref{example: type derivation}} \DefEq\news{x_1}{y_1}\news{x_2}{y_2}\news{v_1}{k_1}\news{v_2}{k_2}(\InputSP{x_1}{z_1}\InputSP{x_2}{z_2}C_1\para  \OutputSP{y_2}{v_2}\OutputSP{y_1}{v_1}C_2)\]
where 
$C_1\DefEq\EmptyInputSP{x_1}\EmptyInputSP{x_2}\EmptyInputSP{z_1}\EmptyInputSP{z_2}\Inact$ and $C_2\DefEq\EmptyOutputSP{y_1}\EmptyOutputSP{y_2}\EmptyOutputSP{k_1}\EmptyOutputSP{k_2}\Inact$. 
Similar to   $P_2$ (\Cref{sec: introduction}), the structure of actions in process $P_{\ref{example: type derivation}}$ induces a circular dependency that prevents synchronizations: we have that $\neg\DeadlockFreedomSP{P_{\ref{example: type derivation}}}$
and yet $P_{\ref{example: type derivation}}$ is well-typed with the empty environment.
\end{example}

\subsection{\texorpdfstring{$\CP$}{CP} and \texorpdfstring{$\HCP$}{HCP}}\label{sec: CP and HCP}

The languages $\CP$ and $\HCP$ use the same syntax. Here we consider $\HCP$ as presented by Kokke \etal ~\cite{betterlate} (other variants have been studied in, e.g.,~\cite{separating}).
Assuming an infinite set of \emph{names} ($x, y, z,\dots$), the set of \emph{processes} ($P, Q,\dots$) in $\CP$ and $\HCP$ is defined as follows:
\begin{align*}
    P,Q::= \Inact& \mybar  (\nu xy)P \mybar  P\para Q \mybar  \forward{x}{y} \Sep  \boutHCP{y}{x}P \Sep \InputHCP{x}{y}P \Sep \EmptyOutputHCP{x}P\Sep \EmptyInputHCP{x}P    
\end{align*}
\noindent  
Most constructs are similar to those in 
$\SPExp$;  differences are 
the forwarder process $\forward{x}{y}$, which equates $x$ and $y$, and process  
$\boutHCP{y}{x}P$, 
which denotes the output of the private name $y$ along $x$, with continuation 
$P$.
Also, the empty output is now denoted $\EmptyOutputHCP{x}P$.
In $\boutHCP{y}{x}P$ 
and $\InputHCP{x}{y}P$, the name $y$ is bound in $P$. 
We write $P\subst{x}{y}$ to denote the capture-avoiding substitution of $y$ for $x$ in $P$.
$\FreeNamesHCP{P}$ denotes the free names of $P$. We use $\pi$ to range over  prefixes: $\boutHCPLabel{y}{x},\InputHCPLabel{x}{y},\EmptyInputHCPLabel{x}$, $\EmptyOutputHCPLabel{x}$.

The types assigned to names correspond to  formulas of classical linear logic~(\CLL):
\[A,B::= \unit \Sep  \bot \Sep  A\Tensor B \Sep  A\InputTypeHCP B\]
The assignment $x:A$ says that $x$ follows the input/output interactions described by~$A$.
Assignments $x: A\Tensor B$ and $x:A\InputTypeHCP B$ are read as sending and receiving an object of type $A$ along $x$, with continuation $B$, respectively.
There is a duality in the interpretation of the following pairs: $\Tensor$ and $\InputTypeHCP$; $\unit$ and $\bot$.  
 Formally, the dual type of $A$, denoted $\Dual{A}$, is defined as
\[\Dual{\unit}\DefEq  \bot
\qquad
\Dual{\bot} \DefEq \unit
\qquad
\Dual{(A\Tensor B)} \DefEq \Dual{A} \InputTypeHCP\Dual{B}
\qquad\Dual{(A\InputTypeHCP B)} \DefEq \Dual{A} \Tensor \Dual{B}\]
Let $\Gamma,\Delta$ range over \emph{environments}, unordered collections of assignments. 
Given an environment $\Delta=x_1:A_1,\dots,x_n:A_n$, its domain,  written $dom(\Delta)$, is the set $\{x_1,\dots,x_n\}$.
\begin{definition}[$\CP$]\label{def: CP}
Typing judgements for $\CP$  
are of the form $P\CPJud \Gamma$, with typing rules  as given in \Cref{fig:CP Rules}.
     We shall write  $P \in \CP$ if $P\CPJud \Gamma$, for some $\Gamma$.
\end{definition}

  \begin{figure}[t!]
    \centering
    \footnotesize{
    
\begin{mathpar}
    \inferrule[$\C\unit$]{ \empty}{\EmptyOutputCP{x}\Inact \CPJud x:\unit}
    \and
    \inferrule[$\C\bot$]{P\CPJud \Gamma}{\EmptyInputCP{x}P\CPJud \Gamma,x:\bot}
    \and
      \inferrule[$\C$\MixZero]{ \empty}{\Inact\CPJud \cdot}   
      \and 
    \inferrule[$\C\Tensor$]{P\CPJud\Gamma,y:A \and Q\CPJud \Delta,x:B}{\BoutCP{y}{x}(P\para Q)\CPJud \Gamma,\Delta,x:A\OutputTypeCP B}
    \and
    \inferrule[$\C\InputTypeCP$]{P\CPJud \Gamma ,y:A, x:B}{\InputCP{x}{y}P\CPJud \Gamma, x:A \InputTypeCP B}
    \and
    \inferrule[$\C$Cut]{P\CPJud \Gamma,x:A \and Q\CPJud \Delta,y:A^\bot}{\cut{xy}{P}{Q}\CPJud \Gamma,\Delta}
        \and
        \inferrule[$\C$Id]{ }{\forward{x}{y}\CPJud x:A, y:A^\bot} 
          \vspace{-0.5cm}
    \end{mathpar}
          }
  \vspace{-3mm}
    \caption{$\CP$: Typing rules for processes.}
    \label{fig:CP Rules}
\end{figure}

Observe how typing in \CP induces a specific shape for output processes: 
$\boutHCP{y}{x}(P\para Q)$.
This is not the case in $\HCP$, where processes are described by \emph{hyperenvironments}, unordered collections of environments:
$\HypH,\HypG::=\Gamma_1\SequentPara\cdots\SequentPara\Gamma_n$. 
\begin{definition}[$\HCP$]\label{def: type system HCP}
    Typing judgements for $\HCP$  
    are of the form $P\HCPJud \HypG$, with typing rules as given in \Cref{fig:type system HCP}. 
    We shall write  $P \in \HCP$ if $P\HCPJud \HypG$, for some $\HypG$.
\end{definition}

  \begin{figure}[!t]  
  \footnotesize
  \centering{
    {
    {
\begin{mathpar}
  \renewcommand{\and}{\hspace{12pt}}
    \inferrule[$\Hname$\MixTwo]{P\HCPJud \HypH \quad Q \HCPJud \HypG}{P\para Q \HCPJud\HypH \SequentPara \HypG}
    \and
    \inferrule[$\Hname$\MixZero]{\phantom{\forwardHCP{x}{y}}}{\Inact\HCPJud \cdot}
    \and
    \inferrule[$\Hname$id]{\phantom{\forwardHCP{x}{y}}}{\forwardHCP{x}{y}\HCPJud x:A , y: A^\bot}
      \and
    \inferrule[$\Hname\bot$]{P\HCPJud \HypH \SequentPara \Delta}{\EmptyInputHCP{x}P\HCPJud \HypH \SequentPara \Delta,x:\bot}
        \and
            \inferrule[$\Hname\unit$]{P\HCPJud \HypH}{\EmptyOutputHCP{x}P\HCPJud \HypH\SequentPara x:\unit}
      \and \\
    \inferrule[$\Hname\OutputTypeHCP$]{P\HCPJud \HypH\SequentPara \Delta, y:A \SequentPara \Gamma, x:B}
    {\boutHCP{y}{x}P \HCPJud \HypH \SequentPara \Delta,\Gamma ,x: A\OutputTypeHCP B}
    \and
    \inferrule[$\Hname\InputTypeHCP$]{P\HCPJud \HypG \SequentPara \Delta, y:B,x:A}
    {\InputHCP{x}{y}P\HCPJud \HypG \SequentPara \Delta, x:A\InputTypeHCP B}
    \and
    \inferrule[$\Hname$Cut]{P\HCPJud \HypH\SequentPara \Delta, x:A \SequentPara \Gamma, y:A^\bot}
    {\news{x}{y}P \HCPJud \HypH \SequentPara \Delta,\Gamma}
\end{mathpar}
    \vspace*{-0.35cm}
  }
    }
    }
    \caption{$\HCP$: Typing rules for processes.}
    \label{fig:type system HCP}
  \end{figure}
  
  \CP and \HCP differ in process composition, 
  which is key to exclude the circular dependencies that lead to deadlocks: it is realized by Rule~\C\textsc{Cut} in \CP and Rule~\Hname\textsc{Cut} in \HCP.  The former requires two premises describing   processes that share exactly one session; typing involves both composition and restriction and induces tree-like process topologies. The latter involves one premise only, which describes a process in which the separation between $x$ and $y$ is made explicit by `$\SequentPara$' at the level of hyperenvironments; typing involves restriction only.
  
Every $P$ typable in $\CP$ is also typable in $\HCP$, but the converse does not hold. We have:
\begin{lemma}[Relation Between $\CP$ and $\HCP$ \cite{betterlate}]\label{lemma: connection between CP and HCP}
    If $P\CPJud \Gamma$ then $P\HCPJud \Gamma$.
    \end{lemma}

 \begin{example}[The converse of \Cref{lemma: connection between CP and HCP} does not hold]\label{example: process in HCP not in CP}
 To illustrate a process typable in $\HCP$ but not in $\CP$, consider    
     $P_{\ref{example: process in HCP not in CP}} \DefEq    \news{x_1}{y_1}\news{x_2}{y_2}\news{v_1}{k_1}\news{v_2}{k_2}(P'_{\ref{example: process in HCP not in CP}}\para P''_{\ref{example: process in HCP not in CP}})$, where: 
\begin{align*}
P'_{\ref{example: process in HCP not in CP}}\DefEq&~\InputHCP{x_1}{z_1}\InputHCP{x_2}{z_2}\EmptyInputHCP{x_1}\EmptyInputHCP{x_2}\EmptyInputHCP{z_1}\EmptyInputHCP{z_2}\Inact \\
P''_{\ref{example: process in HCP not in CP}}\DefEq&~\boutHCP{v'_1}{y_1}(\forward{v'_1}{v_1}\para \boutHCP{v'_2}{y_2}(\forward{v'_2}{v_2}\para C))~~
&C\DefEq&~\EmptyOutputHCP{y_1}\Inact\para \EmptyOutputHCP{y_2}\Inact\para \EmptyOutputHCP{k_1}\Inact
           \para\EmptyOutputHCP{k_2}\Inact
\end{align*}

       \noindent
       We have 
       $P_{\ref{example: process in HCP not in CP}} \HCPJud \cdot ~$. 
       First,
$P'_{\ref{example: process in HCP not in CP}}\HCPJud x_1:\bot \InputTypeHCP\bot,x_2:\bot \InputTypeHCP\bot$ follows from several applications of Rules \Hname$\bot$ and \Hname$\InputTypeHCP$. 
To type $P''_{\ref{example: process in HCP not in CP}}$, we first 
obtain $C\HCPJud y_1:\unit \SequentPara y_2:\unit \SequentPara k_1:\unit \SequentPara k_2:\unit$, crucially using Rule \Hname\MixTwo several times to compose the four independent processes.
Because $\forward{v'_2}{v_2}\HCPJud v'_2:\unit,v_2:\bot$, by Rules \Hname\MixTwo and \Hname$\OutputTypeHCP$ we have: $\boutHCP{v'_2}{y_2}(\forward{v'_2}{v_2}\para C)\HCPJud y_1: \unit \SequentPara 
    y_2: \unit\OutputTypeHCP\unit, v_2:\bot 
    \SequentPara k_1:\unit\SequentPara k_2:\unit$. We proceed similarly to type the output on $y_1$ and complete the typing of $P''_{\ref{example: process in HCP not in CP}}$. 
        Finally, to compose  
        $P'_{\ref{example: process in HCP not in CP}}$
        and 
        $P''_{\ref{example: process in HCP not in CP}}$
        we apply \Hname\MixTwo and \Hname\textsc{Cut}.

Now, $C$ cannot be typed in $\CP$: there is no rule similar to Rule \Hname\MixTwo, which can compose independent processes.
Hence, $P_{\ref{example: process in HCP not in CP}}$ is not typable in \CP.

In general, the composition of two processes that share more than one session cannot be typed in $\CP$. 
Now consider $Q_{\ref{example: process in HCP not in CP}}$: 
\begin{align*}
Q_{\ref{example: process in HCP not in CP}}\DefEq\news{v_2}{k_2}(\news{v_1}{k_1}(\news{x_2}{y_2}(\news{x_1}{y_1}(& P'_{\ref{example: process in HCP not in CP}}  \para \BoutCP{v'_1}{y_1}(\forwardHCP{v'_1}{v_1}\para \EmptyOutputCP{y_1}\Inact)
\\
 \para & \BoutCP{v'_2}{y_2}(\forwardHCP{v'_2}{v_2}\para \EmptyOutputCP{y_2}\Inact))\para \EmptyOutputCP{k_1}\Inact)\para \EmptyOutputCP{k_2}\Inact)
\end{align*}
Observe how 
$Q_{\ref{example: process in HCP not in CP}}$ modifies 
$P_{\ref{example: process in HCP not in CP}}$ by
keeping 
$P'_{\ref{example: process in HCP not in CP}}$ unchanged
and
dividing the two outputs in $P''_{\ref{example: process in HCP not in CP}}$ into threads that can be composed separately using Rule~\C\textsc{Cut}.
We have 
       $Q_{\ref{example: process in HCP not in CP}} \CPJud \cdot ~$.
 \end{example}

Given a $P\HCPJud \HypG$, with $\HypG=\Gamma_1\SequentPara\cdots \SequentPara \Gamma_n$, we are interested in {partitioning} the free names in $P$ according to the environments $\Gamma_1, \ldots, \Gamma_n$.  
These \emph{name partitions} describe which names are used by each parallel component of $P$; they are defined considering all the different hyperenvironments under which $P$ is typable.

\begin{definition}[$\HCP$: Name Partitions]
\label{d:separation}
Given $\HypG=\Gamma_1\SequentPara\cdots\SequentPara\Gamma_n$, we define $\PartitionHCP{\HypG}=\{dom(\Gamma_1),\dots,dom(\Gamma_n)\}$. Given $P$, its \emph{name partitions} are $\NamePartitionHCP{P} = \{\PartitionHCP{\HypG}\Sep P\HCPJud \HypG\}$. 

$\Partition{x}{P}{y}$ holds
whenever there is a  $G\in \NamePartitionHCP{P}$ such that $x$ and $y$ belong to different
proper subsets of $G$.
Also,  $\NoPartition{x}{P}{y}$ holds whenever $x$ and $y$ belong to the same proper subset of $G$.
\end{definition}

\begin{example}\label{example: partitions}
Consider the process $P_{\ref{example: partitions}}\DefEq \EmptyInputHCP{x}(\EmptyOutputHCP{z}\Inact\para\EmptyOutputHCP{y}\Inact)$: it can be typed under the hyperenvironments
$\HypG\DefEq x:\bot,z:\unit\parallel y:\unit$ and $\HypH\DefEq z:\unit\parallel x:\bot,y:\unit$. 
Given this, $\NamePartitionHCP{P_{\ref{example: partitions}}}=\{  \{\{x,z\},\{y\}\} , \{\{z\},\{x,y\}\} \}$, and we have, e.g., $\Partition{y}{P_{\ref{example: partitions}}}{z}$ and $\NoPartition{x}{P_{\ref{example: partitions}}}{y}$.
\end{example}

The semantics of $\HCP$ is given in terms of a Labeled Transition System (LTS):

\begin{definition}[$\HCP$: LTS]\label{def:labelled Transition System HCP}
  The binary relation on $\HCP$ processes 
    $\TransitionHCP{l}$ is defined by the rules in \Cref{fig: labelled transitions HCP}, with action labels defined as follows:
 \[l,l'::= \forwardHCPLabel{x}{y}\Sep \boutHCPLabel{y}{x}\Sep \InputHCPLabel{x}{y}\Sep \EmptyInputHCPLabel{x}\Sep \EmptyOutputHCPLabel{x}\Sep l\para l  \Sep \tau\] 
In the following, $\FreeNamesHCP{l}$, $\BoundNamesHCP{l}$, and $\NamesHCP{l}$ represent the free, bound, and channel names of $l$, respectively, where 
$\FreeNamesHCP{\forwardHCPLabel{x}{y}}\DefEq\{x,y\}$, $\FreeNamesHCP{\tau}=\BoundNamesHCP{\tau}\DefEq\emptyset$, $\FreeNamesHCP{\boutHCPLabel{y}{x}}=\FreeNamesHCP{\InputHCPLabel{x}{y}}=\FreeNamesHCP{\EmptyInputHCPLabel{x}}=\FreeNamesHCP{\EmptyInputHCPLabel{x}}\DefEq x$, and  $\BoundNamesHCP{\boutHCPLabel{y}{x}}=\BoundNamesHCP{\InputHCPLabel{x}{y}}\DefEq\{y\}$. $\FreeNamesHCP{l_1\para l_2}\DefEq\FreeNamesHCP{l_1}\cup\FreeNamesHCP{l_2}$, and $\BoundNamesHCP{l_1\para l_2}\DefEq\BoundNamesHCP{l_1}\cup\BoundNamesHCP{l_2}$. Also, $\NamesHCP{l}\DefEq\FreeNamesHCP{l}\cup\BoundNamesHCP{l}$.
\end{definition}

The rules in \Cref{fig: labelled transitions HCP} are inspired by proof-theoretical transformations on $\HCP$'s proof derivations~\cite{betterlate}; they can be divided into two classes. The rules above the line are usual formulations for actions, structural rules, and communication. Actions can appear in parallel in a label, and compatible actions in parallel induce a $\tau$-transition. The five rules below the   line formalize \emph{delayed actions} and \emph{self-synchronizations}. Intuitively, the first four rules allow to observe an action $l$ in a delayed style; such an action must be independent from the prefix  at the top-level (on~$x$). The last rule, $\para\pi$, implements self-synchronization for a process $\pi.P$: the (top-level) prefix $\pi$ is considered in parallel with a (delayed) action $l$ emanating from $P$. 

\begin{figure}[!t]
\footnotesize
    \centering
\begin{mathpar}
    \inferrule*[Right=$\forwardHCPLabel{x}{y}$]{\empty}{\forwardHCP{x}{y}\TransitionHCP{\forwardHCP{x}{y}}\Inact}
    \and
     \inferrule*[Right=$\forwardHCPLabel{x}{y}$]{\empty}{\forwardHCP{x}{y}\TransitionHCP{\forwardHCP{y}{x}}\Inact} 
     \and 
     \inferrule*[Right=$\pi$]{\pi\in \{\boutHCPLabel{y}{x},\InputHCPLabel{x}{y}, \EmptyInputHCPLabel{x},\EmptyOutputHCPLabel{x}\}}{\pi.P\TransitionHCP{\pi}P}
     \\
\inferrule*[Right=$\text{par}_1$]{P\TransitionHCP{l}P'\quad \BoundNamesHCP{l}\cap \FreeNamesHCP{Q}=\emptyset}{P\para Q \TransitionHCP{l}P'\para Q}
\and
\inferrule*[Right=$\text{par}_2$]{Q\TransitionHCP{l}Q'\quad \BoundNamesHCP{l}\cap \FreeNamesHCP{P}=\emptyset}{P\para Q \TransitionHCP{l}P\para Q'}
\and
\inferrule*[Right=AxCut]{P\TransitionHCP{\forwardHCP{y}{z}}P'}{\news{x}{y}P\TransitionHCP{\tau}P'\subst{x}{z}}
\\
\inferrule*[Right=$\bot\unit$]{P\TransitionHCP{\EmptyOutputHCPLabel{x}\para \EmptyInputHCPLabel{y}}P'}{\news{x}{y}P\TransitionHCP{\tau}P'}
    \and
\inferrule*[Right=$\AlphaConversionHCP$]{P\AlphaConversionHCP Q\quad Q\TransitionHCP{l} R}{P\TransitionHCP{l}R}
\and
\inferrule*[Right=Res]{P\TransitionHCP{l}P'\quad x,y\notin \NamesHCP{l}\quad \Partition{x}{P'}{y}}{\news{x}{y}P\TransitionHCP{l}\news{x}{y}P'}
\\
 \inferrule*[Right=$\OutputTypeHCP\InputTypeHCP$]{P\TransitionHCP{\boutHCPLabel{x'}{x}\para\InputHCPLabel{y}{y'}}P'}{\news{x}{y}P\TransitionHCP{\tau}\news{x}{y}\news{x'}{y'}P'}
    \and\inferrule*[Right= syn]{P\TransitionHCP{l}P'\quad Q\TransitionHCP{l'}Q'\quad \BoundNamesHCP{l}\cap\BoundNamesHCP{l'}=\emptyset}{P\para Q\TransitionHCP{l\para l'}P'\para Q'}
\end{mathpar}
\rule[1ex]{12cm}{0.5pt}
\begin{mathpar}
    \inferrule*[Right=$\pi_1$]{P\TransitionHCP{l}P'\quad x\notin\NamesHCP{l}}{\EmptyOutputHCP{x}P\TransitionHCP{l}\EmptyOutputHCP{x}P'}
\and
\inferrule*[Right=$\pi_2$]{P\TransitionHCP{l}P'\quad \FreeNamesHCP{P'}\neq\emptyset}{\EmptyInputHCP{x}P\TransitionHCP{l}\EmptyInputHCP{x}P'}
\and
\inferrule*[Right=$\OutputTypeHCP$]{P\TransitionHCP{l}P'\quad x,x'\notin \NamesHCP{l}\quad \Partition{x}{P'}{x'}}{\boutHCP{x'}{x}P\TransitionHCP{l}\boutHCP{x'}{x}P'}
\and
\inferrule*[Right=$\InputTypeHCP$]{P\TransitionHCP{l}P'\quad x,x'\notin\NamesHCP{l}\quad \NoPartition{x}{P'}{x'}}{\InputHCP{x}{x'}P\TransitionHCP{l}\InputHCP{x}{x'}P'}
\and
\inferrule*[Right=$\para\pi$]{\pi.P\TransitionHCP{l}\pi.P'\quad \Partition{\FreeNamesHCP{\pi}}{\pi.P}{\FreeNamesHCP{l}}}{\pi.P\TransitionHCP{\pi\para l}P'}
\end{mathpar}
    \caption{The LTS for $\HCP$ (\Cref{def:labelled Transition System HCP}) includes rules for actions and communication (top) and for  delayed actions and self-synchronizations  (bottom).}
    \label{fig: labelled  transitions HCP}
\end{figure}

\begin{example} \label{example:intrigue} 
To  illustrate  the LTS of \HCP, recall the processes  $P_1$ and $P_2$ from \Cref{sec: introduction}.
In \HCP, they are denoted as:
$P_{\ref{example:intrigue}} \DefEq\news{x}{x'}(\EmptyOutputHCP{x}\EmptyInputHCP{x'}\Inact)$
and
$Q_{\ref{example:intrigue}}\DefEq\news{x}{x'}\news{y}{y'}(
\EmptyOutputHCP{y}\EmptyInputHCP{x}\Inact
\para
\EmptyOutputHCP{x'}\EmptyInputHCP{y'}\Inact)
$.
 Consider the following  type derivation, parametric on $w,z$:
      \begin{prooftree}
     \AxiomC{$\Inact\HCPJud$}
     \RightLabel{$\Hname\bot$}
   \UnaryInfC{$\EmptyInputHCP{w}\Inact\HCPJud w:\bot$}
       \LeftLabel{$\Pi_{w,z}=$}
       \RightLabel{$\Hname\unit$}
     \UnaryInfC{$\EmptyOutputHCP{z}\EmptyInputHCP{w}\Inact\HCPJud w:\bot\SequentPara z:\unit$}
      \end{prooftree}
\noindent
     Thus, $P_{\ref{example:intrigue}}$ is typed as $\Pi_{x',x}$ followed by an application of Rule~$\Hname$\textsc{Cut}. Also, $Q_{\ref{example:intrigue}}$ is typed by $\Pi_{x,y}$ and $\Pi_{y',x'}$ followed by an application of Rules~$\Hname$\MixTwo and $\Hname$\textsc{Cut}. Because Rule~$\C\unit$ only types empty outputs followed by $\Inact$, neither $P_{\ref{example:intrigue}}$ nor $Q_{\ref{example:intrigue}}$ are typable in $\CP$. 
     
     By Rules $\para\pi$ and $\bot\unit$ in \Cref{fig: labelled  transitions HCP}, $P_{\ref{example:intrigue}}\TransitionHCP{\tau}\Inact$. Similarly, $\EmptyOutputHCP{y}\EmptyInputHCP{x}\Inact
\para
\EmptyOutputHCP{x'}\EmptyInputHCP{y'}\Inact\TransitionHCP{\EmptyInputHCPLabel{x}\para \EmptyOutputHCPLabel{x'}}\EmptyOutputHCP{y}\Inact
\para
\EmptyInputHCP{y'}\Inact$
by Rules $\pi,\pi_1$ and \textsc{syn}.
Finally, by Rule $\bot\unit$, $Q_{\ref{example:intrigue}}\TransitionHCP{\tau}\news{y}{y'}(\EmptyOutputHCP{y}\Inact
\para
\EmptyInputHCP{y'}\Inact)$.
	\end{example}

We now recall some useful notions and results from~\cite{betterlate}.
\begin{definition}[$\HCP$: Weak LTS and Bisimilarity~\cite{betterlate}]\label{def: weak transition HCP}
\label{def: bisimilarity HCP}
    $\WeakTransitionHCP{l}$ is the smallest relation such that:
     \textup{(1)}~$P\WeakTransitionHCP{\tau} P$ for all $P$; 
    \textup{(2)}~and if $P\WeakTransitionHCP{\tau} P'$, $P'\TransitionHCP{l}Q'$, and $Q'\WeakTransitionHCP{\tau} Q$, then $P\WeakTransitionHCP{l}Q$.
\noindent
A symmetric relation $\mathcal{R}$   is a bisimulation if $P \mathrel{\mathcal{R}} Q$ implies that if $P\WeakTransitionHCP{l}P'$ then $Q\WeakTransitionHCP{l}Q'$ for some $Q'$ such that $P' \mathrel{\mathcal{R}} Q'$. Bisimilarity is the largest relation $\BisimilarityHCP$ that is a bisimulation.
   \end{definition}

Well-typed processes satisfy readiness, defined as follows:
\begin{theorem}[$\HCP$: Readiness~\cite{betterlate}]\label{theorem: readiness HCP}
    Let $P\HCPJud \Gamma_1\SequentPara \cdots \SequentPara \Gamma_n$. For every $i \in[1,n]$, there exist $x\in dom(\Gamma_i)$, $l_x$, and $P'$ such that $P\WeakTransitionHCP{l_x}P'$. 
\end{theorem} 

In $\HCP$, transitions associated to sessions in different partitions can be fired in parallel:

\begin{lemma}[\cite{betterlate}]\label{lemma: parallel transitions}
    If $P\HCPJud \HypG\parallel \Gamma,x:A\parallel \Delta,y:B$, $P\WeakTransitionHCP{l_x}P'$, and $P\WeakTransitionHCP{l_y}P''$, then there exists $Q$ s.t.  $P\WeakTransitionHCP{l_x\para l_y}Q$ (up to $\alpha$-renaming). 
\end{lemma}

\begin{definition}[\HCP: Structural Congruence]\label{ldef: zero equive}  
Let $\EquivHCP$ be the smallest congruence relation generated by the axioms: 
        \begin{align*}
P \para Q\ &\EquivHCP\ Q \para P                     &  \quad P \para \Inact\ &\EquivHCP\ P\\
  P \para (Q \para R)\ &\EquivHCP\ (P \para Q) \para R     &    \news x y P &\EquivHCP \news y x P \\
  \news x yP \para Q\ &\EquivHCP \ \news x y(P \para Q) \ \  \text{if $x, y \notin \FreeNamesHCP{Q}$} & \news x y\news w z P\ &\EquivHCP\ \news w z\news x yP
\end{align*}
\end{definition}

Because of Readiness (\Cref{theorem: readiness HCP}), $\HCP$ also satisfies \emph{progress}:
\begin{corollary}[$\HCP$: Progress~\cite{betterlate}]\label{corollary:progress}
    If $P \in \HCP$ and $P\not\EquivHCP\Inact$, then $P\TransitionHCP{l}P'$ for some $l$, $P'$.
\end{corollary}

In order to relate with reduction-based semantics (in \SPExp), we define the following:
\begin{definition}[\HCP: Reduction]
\label{hcp:reduction}
    We write $\TransitionHCPClo$ for the reflexive, transitive closure of $\TransitionHCP{\tau}$. 
\end{definition}

\begin{lemmaapxrep}\label{lemma: deadlock freedom HCP}
 Let $P\in$ $\HCP$   s.t. $P\HCPJud\cdot~$, then exists $P'$ s.t. $P\TransitionHCP{\empty}^*P'\EquivHCP\Inact$.
\end{lemmaapxrep}
\begin{proof}
	Note that the judgement $P
	\HCPJud\cdot$ can only be obtained via  Rules $\Hname$\MixZero and $\Hname$\textsc{Cut}. In the former case $P \EquivHCP \Inact$ and there is nothing to prove. 
	In the latter case, we split the proof in two parts: (i) we prove that there exists an $R$ such that $P\TransitionHCP{\tau}R$, with the transition involving a cut on session $xy$, therefore specializing \Cref{corollary:progress} (which ensures the existence of some transition, not necessarily a $\tau$-transition); (ii) we prove that there exist $k \geq 0$ and $Q$ s.t. $R\longrightarrow^k Q$ and there are no cuts on $xy$ in $Q$.
 
	    For part (i), we have $P\HCPJud\cdot~$ and $P\not\EquivHCP\Inact$; we proceed by induction on the number $n$ of parallel components of $P$.
    \begin{itemize}
        \item \emph{Base case ($n = 1$)}.        
    In this case, the last rule applied is $\Hname$\textsc{Cut}:
	\begin{prooftree}
		\AxiomC{$P'\HCPJud x:A\SequentPara y: \LinearDual{A}$}
		\UnaryInfC{$\news{x}{y}P'\HCPJud \cdot$}
	\end{prooftree}
	 By \Cref{theorem: readiness HCP},  there exist labels $l_x,l_y$ and processes $Q_1,Q_2$ s.t. $P'\WeakTransitionHCP{l_x}Q_1$ and $P'\WeakTransitionHCP{l_y}Q_2$.  \Cref{lemma: parallel transitions} ensures that there exists $Q$ s.t. $P'\WeakTransitionHCP{l_x\para l_y}Q$. Hence, by \Cref{def: bisimilarity HCP}, there exist processes $P''$ and $Q'$ s.t.
     $P'\WeakTransitionHCP{\tau}P''\TransitionHCP{l_x\para l_y}Q'\WeakTransitionHCP{\tau}Q$.
     To show the existence of $R$, we consider two possibilities.
     First, if the first sequence of $\tau$ transitions is not empty, then we must have
     $P' \TransitionHCP{\tau}R\TransitionHCPClo P''\TransitionHCP{l_x\para l_y}Q' \TransitionHCPClo Q$, and $P\TransitionHCP{\tau}R$. 
     Second, if 
  $P''=P'$ then we have  $P' \TransitionHCP{l_x\para l_y}Q' \TransitionHCPClo Q$, and $R$ results from a synchronization on $xy$, as follows.   
  By typability of $P'$, the actions $l_x$ and $l_y$ are dual, thus $\news{x}{y}P'$ has a $\tau$-transition due to  either Rule $\bot\unit$ or $\OutputTypeHCP\InputTypeHCP$, depending on the type $A$. Hence, $R = \news{x}{y}Q'$. So far we have proven that a well-typed closed process always can execute a $\tau$-transition. As a consequence of \emph{subject reduction} (cf. \cite[Theorem 3.9]{betterlate}), we also have $R \HCPJud\cdot~$.
  
  \item \emph{Inductive step ($n>1$)}. In this case, the last rule applied is $\Hname$\MixTwo. Hence,  $P=P_1\para P_2$:
  \begin{prooftree}
      \AxiomC{$P_1\HCPJud\HypG$}
      \AxiomC{$P_2\HCPJud\HypH$}
      \BinaryInfC{$P_1\para P_2\HCPJud\HypG\parallel\HypH$}
  \end{prooftree}
Clearly, $P_1$ and $P_2$ have less than $n$ parallel components, and the result follows by IH.
   \end{itemize}
We now proceed to prove part (ii); we require an auxiliary definition:
    \begin{definition}[Degree]
    The degree of a type $A$ is inductively defined by:
    \begin{itemize}
        \item $\DegreeType{\unit}\DefEq\DegreeType{\bot}\DefEq1$.
        \item  $\DegreeType{ A\OutputTypeHCP B}\DefEq\DegreeType{ A\InputTypeHCP B}\DefEq
        \max(\DegreeType{A},\DegreeType{B})+1$.
    \end{itemize}
        The degree of an $\Hname$\textsc{Cut} on the session $xy$ is defined to be the degree of the type which it eliminates (from the judgement). The degree of an $\Hname$\textsc{Cut} on a session $xy$ that does not occur in the process is 0. 
    \end{definition}
    This way, we prove:
given $R\HCPJud\cdot$ with a $\Hname$\textsc{Cut} of degree $d>0$ along the session $xy$, then there exist $k\geq 0$ and $Q$ s.t. $R\longrightarrow^k Q$, $Q\HCPJud\cdot$ and  the degree of the $\Hname$\textsc{Cut} along $xy$ is 0.
 We proceed by induction on the number $n$ of parallel components of $P$.
\begin{itemize}
    \item \emph{Base case $(n=1)$},
 The last rule applied must be $\Hname$\textsc{Cut}.
We proceed by induction on the degree of the type $A$ involved in the cut $xy$.
  \begin{itemize}
      \item Base case ($\DegreeType{A}= 1$). We must then have:
      \begin{prooftree}
		\AxiomC{$R'\HCPJud x:\unit\SequentPara y: \bot$}
		\UnaryInfC{$\news{x}{y}R'\HCPJud \cdot$}
	\end{prooftree}
     By the analysis in the first part,  there exist $R'',Q'$ s.t. 
 $R'\WeakTransitionHCP{\tau}R''\TransitionHCP{l_x\para l_y}Q'$. Since the first chain of $k'$ $\tau$-transitions does not occur along session $xy$, its degree remains unchanged. We have $\news{x}{y}R'\WeakTransitionHCP{\tau} \news{x}{y}R''\TransitionHCP{\tau}Q'$ and $Q'\HCPJud\cdot$, where the degree of $\Hname$\textsc{Cut} along $xy$ is 0, thus $Q=Q'$ and $k=k'+1$.

    \item Inductive step ($\DegreeType{A}>1$). We must then have:
    \begin{prooftree}
		\AxiomC{$R'\HCPJud x:A\OutputTypeHCP B\SequentPara y: \Dual{A}\InputTypeHCP \Dual{B}$}
		\UnaryInfC{$\news{x}{y}R'\HCPJud \cdot$}
	\end{prooftree}
	 Thus, the degree of $\Hname$\textsc{Cut} along $xy$ is $\DegreeType{ A\OutputTypeHCP B}=
        \max(\DegreeType{A},\DegreeType{B})+1>0$. By the analysis in the first part,   there exist $R'',Q'$ s.t. 
 $R'\WeakTransitionHCP{\tau}R''\TransitionHCP{l_x\para l_y}Q'$. Since the first chain of $k'$ $\tau$-transitions does not occur along session $xy$, its degree remains unchanged and we have:
 \begin{prooftree}
		\AxiomC{$R''\HCPJud x:A\OutputTypeHCP B\SequentPara y: \Dual{A}\InputTypeHCP \Dual{B}$}
		\UnaryInfC{$\news{x}{y}R''\HCPJud \cdot$}
	\end{prooftree}
    By Rule $\Hname\OutputTypeHCP\InputTypeHCP$ we have $\news{x}{y}R''\TransitionHCP{\tau}\news{x}{y}\news{x'}{y'}R^*$, typed as follows:
     \begin{prooftree}
		\AxiomC{$R^*\HCPJud x:A\SequentPara x':B\SequentPara y: \Dual{A},y': \Dual{B}$}
		\UnaryInfC{$\news{x'}{y'}R^*\HCPJud  x:A\SequentPara y: \Dual{A}$}
        \UnaryInfC{$\news{x}{y}\news{x'}{y'}R^*\HCPJud \cdot$}
	\end{prooftree}
    where the degree of $\Hname$\textsc{Cut} along $xy$ is $\DegreeType{A}<\DegreeType{ A\OutputTypeHCP B}$. By IH, there exist $k''$ and $Q$ s.t. $\news{x}{y}R''\longrightarrow^{k+m+1}Q$, whose degree of $\Hname$\textsc{Cut} on $xy$ is 0, thus $k=k'+k'+1$.
  \end{itemize}
 \item \emph{Inductive step ($n>1$)}. In this case, the last rule applied is $\Hname$\MixTwo. We have $P=P_1\para P_2$ and:
  \begin{prooftree}
      \AxiomC{$P_1\HCPJud\HypG$}
      \AxiomC{$P_2\HCPJud\HypH$}
      \BinaryInfC{$P_1\para P_2\HCPJud\HypG\parallel\HypH$}
  \end{prooftree}
Clearly, $P_1$ and $P_2$ have less than $k$ parallel components, thus the result follows by IH.
  \end{itemize}

  By the repeated application of part~(ii) for all the instances of Rule $\Hname$\textsc{Cut}, we finally find a process with no instances of Rule $\Hname$\textsc{Cut}. This concludes the proof, since a type derivation for a well-typed closed process $R$ with no instances of Rule $\Hname$\textsc{Cut} 
  must necessarily be $R\EquivHCP\Inact$.
\end{proof}

Inspired by \Cref{lemma: deadlock freedom HCP} we have the following definition of deadlock freedom  for \HCP.
 \begin{definition}[$\HCP$: Deadlock Freedom]\label{def: deadlock freedom delayed}
     Let $P\in\HCP$. We say that $P$ is deadlock free, written $\DeadlockFreedomHCP{P}$, iff there exists $P'$ s.t. $P\TransitionHCPast P'\EquivHCP \Inact$. 
 \end{definition}

$\HCP$ has a \emph{denotational semantics} that interprets well-typed processes as sets of possible interactions on  their free names. 
Given a process $P$, its set of denotations is written $\HCPDenotations{P}$. 

\begin{toappendix}
Below, we write $\FinitePowerSet{S}$ to denote the finite powerset of the set $S$.
\begin{definition}[$\HCP$: Denotations of Types~\cite{betterlate}] We define:
\label{d:denot}
\begin{align*}
    \HCPDenotations{\unit}&=\{\triangleleft\}\times \mathbb{N}\times \{*\} & 
     \HCPDenotations{A\OutputTypeHCP B}&=\{\triangleleft\}\times \mathbb{N}\times(\FinitePowerSet{ \mathbb{N}}\times \HCPDenotations{A}\times\FinitePowerSet{ \mathbb{N}}\times \HCPDenotations{B})
    \\
       \HCPDenotations{\bot}&=\{\triangleright\}\times \mathbb{N}\times \{*\}
 & \HCPDenotations{A\InputTypeHCP B}&=\{\triangleright\}\times \mathbb{N}\times (\HCPDenotations{A}\times \HCPDenotations{B})
\end{align*}
We write $a \DualDenotations b$ whenever $a\in\HCPDenotations{A}$ and $b\in\HCPDenotations{A^\bot}$.
\end{definition}

Natural numbers act as identifiers that describe dependencies with other channels. 
Note that `$\triangleleft$' and `$\triangleright$' are associated to output- and input-like behaviors, respectively; in particular, this entails that 
$\unit$ and $\bot$, which both capture the closing of a session, have different denotations.
The denotation of a process $P$ is written $\HCPDenotations{P}$; the definition  relies on a notion of \emph{antiderivatives} for sets of denotations. Intuitively, given an action label $l$ and a set of denotations $D$, $\int_{l}D$ returns a set of denotations $D'$ such that for some transformation $T_l$ on denotations, if $g\in D$, then $T_l(g)\in D'$. For instance:
$ \int_{\EmptyOutputHCPLabel{x}} D\DefEq\{g\cdot\SingleDenotation{\SingleDenotation{\MapstoDenotation{x}{(\triangleleft,n,*)}}}\Sep g\in D\}$.
\Cref{def: antiderivatives} gives 
a detailed definition.

\begin{definition}[$\HCP$: Antiderivates~\cite{betterlate}]\label{def: antiderivatives}
For $D$ a set of denotations, the antiderivate $\int_l D$ of $D$ wrt an action label $l$ is the set defined by: 
    \begin{align*}
        \int_{\EmptyOutputHCPLabel{x}} D&\DefEq\{g\cdot\SingleDenotation{\SingleDenotation{\MapstoDenotation{x}{(\triangleleft,n,*)}}}\Sep g\in D\}\\
        \int_{\EmptyInputHCPLabel{x}} D&\DefEq\{g\cdot\SingleDenotation{\gamma\cdot\SingleDenotation{\MapstoDenotation{x}{(\triangleright,n,*)}}}\Sep g\cdot\SingleDenotation{\gamma}\in D\}\\
         \int_{\boutHCPLabel{y}{x}} D&\DefEq
         \{g\cdot\SingleDenotation{\gamma\cdot\delta\cdot\SingleDenotation{\MapstoDenotation{x}{(\triangleleft,n,ids(\gamma),a,ids(\delta),b)}}}\Sep g\cdot\SingleDenotation{\gamma\cdot\SingleDenotation{\MapstoDenotation{x}{b}},\delta\cdot\SingleDenotation{\MapstoDenotation{y}{a}}}\in D\}\\
         \int_{\InputHCPLabel{x}{y}} D&\DefEq\{g\cdot\SingleDenotation{\gamma\cdot\SingleDenotation{\MapstoDenotation{x}{(\triangleright,n,a,b)}}} \Sep g\cdot\SingleDenotation{\gamma\cdot\SingleDenotation{\MapstoDenotation{x}{b},\MapstoDenotation{y}{a}}} \in D\\
         \int_{\forwardHCPLabel{x}{y}} D&\DefEq\{g\cdot\SingleDenotation{\SingleDenotation{\MapstoDenotation{x}{a},\MapstoDenotation{y}{a^\bot}}}\Sep g\in D\}\\
         \int_{\tau} D&\DefEq D
    \end{align*}
\end{definition}

\begin{definition}[$\HCP$: Denotational Equivalence~\cite{betterlate}]\label{def: denoations in HCP}
\label{def: denotatinal equivalence}
Let $\pi$ be a prefix. Denotations of processes are defined as follows:
\begin{align*}
\HCPDenotations{\Inact}&\DefEq\{\SingleDenotation{\empty}\}\qquad 
\HCPDenotations{\pi.P} \DefEq\int_\pi\HCPDenotations{P}\qquad 
\HCPDenotations{P\para Q} \DefEq\HCPDenotations{P}\times\HCPDenotations{Q}\\ 
\HCPDenotations{\forwardHCP{x}{y}} &\DefEq \{\SingleDenotation{\MapstoDenotation{x}{a},\MapstoDenotation{y}{a'}}\Sep a \DualDenotations a' \}\\  \HCPDenotations{\news{x}{y}P}&\DefEq \{g \cdot \SingleDenotation{\gamma\cdot \delta} \Sep g\cdot \SingleDenotation{\gamma\cdot \SingleDenotation{\MapstoDenotation{x}{a}},\delta\cdot\SingleDenotation{\MapstoDenotation{y}{a'}}}\in \HCPDenotations{P} \text{ and } a\DualDenotations a' \}
\end{align*}

Denotational equivalence is the relation on typed processes defined as: $P\ObservationalEq Q$ iff $\HCPDenotations{P}=\HCPDenotations{Q}$.
\end{definition}
\begin{example}[Denotations at work]\label{example: denotations}
Consider the processes $P_{\ref{example: denotations},1}\DefEq \EmptyOutputHCP{x}P$ and 
$P_{\ref{example: denotations},2}\DefEq \EmptyOutputHCP{x}\Inact \para P$, for some process
$P$ s.t. $P\HCPJud \HypH$. The following holds: $P_{\ref{example: denotations},1}\HCPJud \HypH\parallel x:\unit$ and  $P_{\ref{example: denotations},2}\HCPJud \HypH\parallel x:\unit$. Thus we have
\begin{align*}
    \HCPDenotations{P_{\ref{example: denotations},1}}= 
    \int_{\EmptyOutputHCPLabel{x}}\HCPDenotations{P}&=
    \{g\cdot\SingleDenotation{\SingleDenotation{\MapstoDenotation{x}{(\triangleleft,n,*)}}}\Sep g\in \HCPDenotations{P}\}\\
    &=
    \{\SingleDenotation{\SingleDenotation{\MapstoDenotation{x}{(\triangleleft,n,*)}}}\}\times\HCPDenotations{P} 
    =\HCPDenotations{\EmptyOutputHCP{x}\Inact \para P}
     =\HCPDenotations{P_{\ref{example: denotations},2}}
\end{align*}
Hence, $P_{\ref{example: denotations},1}\ObservationalEq P_{\ref{example: denotations},2}$.
\end{example}
    \end{toappendix}

We notice that  $\EquivHCP$ is sound wrt $\ObservationalEq$ and that  our results are reflected by bisimilarity, as ensured by the following result:

\begin{theorem}[$\HCP$: Full Abstraction \cite{betterlate}]\label{theorem: full abstraction}
    For well-typed processes: $\BisimilarityHCP\,\mathrel{=}\,\ObservationalEq$.
\end{theorem}

\section{The Role of Delayed Actions and Self-Synchronizations in DF}\label{sec: results}

We have seen that 
\SPExp defines a flexible typing discipline that admits  deadlocked and deadlock-free processes. Also, we saw the way in which \CP and \HCP enforce DF by typing, and that the latter's LTS allows for delayed actions and self-synchronizations. 
By defining two classes of \SPExp processes, $\ClassL$ and $\ClassH$, which capture typability   as in \CP and \HCP, respectively (\Cref{Def: ClassL and ClassH}), we study 
whether delayed actions and self-synchronizations influence DF. 

We will proceed as follows. 
As we have seen, $\DeadlockFreedomHCP{P}$ means that $P$ is deadlock free by considering all sources of behavior,  including delayed actions and self-synchronizations; we shall also define $\DeadlockFreedomHCPRR{P}$ to mean that $P$ is deadlock free by considering \emph{regular transitions only}. 
Our main result is \cref{corollary: main corollary HCP}, which ensures: 
    \textup{(1)}~Delayed actions and Self-synchronizations are \emph{inessential} when it comes to DF, and
    \textup{(2)}~given an encoding $\SPExptoHCP{\cdot}: \SPExp  \to \HCP$, 
    for a  process $P \in \SPExp$ such that $\DeadlockFreedomHCP{\SPExptoHCP{P}}$, there is a $P'$ that \emph{simulates} $P$, and 
    $\DeadlockFreedomHCPRR{\SPExptoHCP{P'}}$.

\subsection{From \texorpdfstring{$\SPExp$}{SP} to \texorpdfstring{$\HCP$}{HCP}}

We start by translating processes in $\SPExp$ into processes in $\HCP$:
\begin{definition}\label{def: encoding SPexp to HCP} Let $P\in \SPExp$. The encoding
$\SPExptoHCP{\cdot}: \SPExp \to \HCP$ is a homomorphism except for the case of free output, defined as
$\SPExptoHCP{\OutputSP{x}{v}P}\DefEq\boutHCP{y}{x}(\forwardHCP{y}{v}\para \SPExptoHCP{P})$. 
For types, we define:
$$\SPExptoHCP{\EndTypeI}\DefEq\bot
\qquad \SPExptoHCP{\EndTypeO}\DefEq\unit
\qquad \SPExptoHCP{\InputST{\empty}{T}{S}}\DefEq\SPExptoHCP{T}\InputTypeHCP \SPExptoHCP{S}
\qquad \SPExptoHCP{\OutputST{\empty}{T}{S}}\DefEq\Dual{\SPExptoHCP{T}}\OutputTypeHCP \SPExptoHCP{S}$$
For typing environments, we define: 
$\SPExptoHCP{\emptyset}\DefEq\emptyset$ and
$\SPExptoHCP{\Gamma,x:T}\DefEq\SPExptoHCP{\Gamma},x:\SPExptoHCP{T}$.
\end{definition}

 \emph{Regular Transitions}  do not depend on self-syn\-chro\-ni\-za\-tion or delayed actions:

\begin{definition}[$\HCP$: Regular Transitions]\label{def: regular reductions}
  A derivation $T$ for $P\TransitionHCP{l}Q$ is denoted $T:P\TransitionHCP{l}Q$. We define $\RegularReduction{T}$ iff $T$ is derived only using rules above the horizontal line in \Cref{fig: labelled  transitions HCP}.
\end{definition} 
\begin{example}
    Recall the processes 
    $P_{\ref{example:intrigue}}$
    and 
    $Q_{\ref{example:intrigue}}$ 
from \Cref{example:intrigue},
which use self-synchronizations and delayed actions, respectively:
we have that  $T_1:P_{\ref{example:intrigue}}\TransitionHCP{\tau}\Inact$, and  $T_2:Q_{\ref{example:intrigue}}\TransitionHCP{\tau}\news{y}{y'}(\EmptyOutputHCP{y}\Inact
\para
\EmptyInputHCP{y'}\Inact)$. Therefore, $\neg\RegularReduction{T_1}$ and $\neg\RegularReduction{T_2}$.
\end{example}

The encoding of free output induces transitions involving the forwarder process, not present in the source $\SPExp$ process. We handle the substitutions induced by forwarders in Rule~AxCut (\Cref{fig: labelled  transitions HCP}) as follows:

\begin{definition}[$\HCP$: Ready-for-Substitution]
    Let $P\in$ $\HCP$. We say that $P=\news{x}{y}P'$ is ready for substitution of $z$ for $x$, written $\ReadySubstitution{z}{x}{P}$, if there exists $z\in \FreeNamesHCP{P}$ and $P'$ s.t. $P\TransitionHCP{\forwardHCP{y}{z}}P'$, and $\news{x}{y}P\TransitionHCP{\tau}P'\subst{x}{z}$. Otherwise, we write $\NotReadySubstitution{P}$.
\end{definition}

We may now define transitions for processes that are ready for a substitution.

\begin{definition}[$\HCP$: Eager-to-Rename]\label{def: transition eager to rename}
The relation $\TransitionEagerHCP{l}$ on $\HCP$ processes is defined  as:
\begin{align*}
(P\TransitionHCP{l}Q)  \land \ReadySubstitution{x}{z}{Q}
& \Rightarrow P\TransitionEagerHCP{l}Q\subst{x}{z}
\qquad \text{and} \qquad
(P\TransitionHCP{l}Q)  \land \NotReadySubstitution{Q}  \Rightarrow P\TransitionEagerHCP{l}Q	
\end{align*}
\end{definition}

Intuitively, $\TransitionEagerHCP{l}$ executes outputs and forwarders in one single step.
When $\SPExptoHCP{P}$ only relies on regular transitions we have the following operational correspondence result.
\begin{toappendix}
	We start by proving a series of auxiliary results. First, we have that $\SPExptoHCP{\cdot}$ respects the sequential order of prefixes. We define contexts, with holes denoted `$\HoleSP$':
	\begin{definition}[Contexts in $\SPExp$ and $\HCP$] Contexts $\ContextSPC$ in $\SPExp$ and $\ContextHCPK$ in $\HCP$ are defined as: 
		\begin{align*}
			\ContextSPC::=&~ \HoleSP \Sep \ContextSPC\para P \Sep \news{x}{y}\ContextSPC \Sep \OutputSP{x}{v}\ContextSPC \Sep \InputSP{x}{y}\ContextSPC\\
			\ContextHCPK::=&~ \HoleHCP \Sep \ContextHCPK\para P \Sep \news{x}{y}\ContextHCPK \Sep \boutHCP{y}{x}(\forwardHCP{y}{z}\para \ContextHCPK) \Sep \InputHCP{x}{y}\ContextHCPK
		\end{align*}
	\end{definition}
	
	The encoding 
	$\SPExptoHCP{\cdot}$ (\Cref{def: encoding SPexp to HCP}) extends to contexts as expected, i.e., $\SPExptoHCP{\,\HoleSP\,} = \HoleHCP$.

	\begin{lemma}\label{lemma: translation preserves order of prefixes}
		Let $\alpha,\beta$ range over prefixes in \SPExp, and $\alpha'$ and $\beta'$ their corresponding in $\HCP$. 
		\[\alpha.\ContextSPIndex{C}{2}{\beta.Q}\Rightarrow \SPExptoHCP{\alpha.\ContextSPIndex{C}{2}{\beta.Q}}={\alpha'.\ContextHCPIndex{K}{2}{\beta'.Q'}}\]
	\end{lemma}
    \begin{proof}
        By case analysis in $\ContextSPC$ and \Cref{def: encoding SPexp to HCP}.
    \end{proof}
	
	Regular transitions concern synchronizations between actions that occur at top-level in different parallel components. 
	If $\SPExptoHCP{P}$ has a regular transition, then $P\in\SPExp$  has a reduction.

    \begin{lemma}\label{lemma: shape of processes in HCP with regular reductions}
		Suppose $P\in$ $\HCP$. If $T:P\TransitionHCP{\tau}Q$ s.t. $\RegularReduction{T}$, then either:
		\begin{enumerate}
			\item $P\EquivHCP \news{w}{z}(\EmptyInputHCP{w}Q\para\EmptyOutputHCP{z}R)\para 
            S$, or
			
			\item $P \EquivHCP\news{w}{z}(\InputHCP{w}{v}Q\para\boutHCP{v}{z}R)\para S$.
		\end{enumerate}
		
	\end{lemma}
	
	\begin{proof}
		By case analysis of $\tau$-transitions using the fact that $\RegularReduction{T}$.
	\end{proof}

	\begin{lemma}\label{lemma: substitution} 
		Let $P\in$ $\ClassH$. Then $\SPExptoHCP{P}\subst{y}{z}=\SPExptoHCP{P\subst{y}{z}}$.
	\end{lemma}
    \begin{proof}
        By induction on the structure of $P$ and \Cref{def: encoding SPexp to HCP}.
    \end{proof}
	
	\begin{lemma}\label{lemma: regular reductions and process}
		Let $P\in\ClassH$. If $T:\SPExptoHCP{P}\TransitionEagerHCP{\tau}P'$ and $\RegularReduction{T}$, then 
		
		\begin{enumerate}
			\item $P\equiv \news{x}{y}(\OutputSP{x}{v}Q\para \InputSP{y}{w}R)\para S$, or
			\item $P\equiv\news{x}{y}(\EmptyOutputSP{x}Q\para \EmptyInputSP{y}R)\para S$.
		\end{enumerate}    
	\end{lemma}
    \begin{proof}
        By a case analysis in $\TransitionEagerHCP{\tau}$ and \Cref{def: regular reductions,def: encoding SPexp to HCP}.
    \end{proof}
    \begin{lemma}\label{lemma: equivalence and encoding} Let $\SPExptoHCP{\cdot}$ be as in \Cref{def: encoding SPexp to HCP}. We have:
    \begin{enumerate}
        \item Let $P\in\ClassH$.  $P\equiv P'$ iff $\SPExptoHCP{P}\EquivHCP\SPExptoHCP{P'}$.
        \item Let $P\in\HCP$. If $P\EquivHCP P'$ and $P'\TransitionHCP{l}Q'$, then there exist $Q$ s.t. $P\TransitionHCP{l}Q$, and $Q\EquivHCP Q'$.
    \end{enumerate}
        
    \end{lemma}
    \begin{proof}
        Item~(1) follows by \Cref{def: encoding SPexp to HCP} and case analyses in $\equiv$ and $\EquivHCP$, respectively. Item~(2) follows by a case analysis in both $\EquivHCP$ and $\TransitionHCP{l}$.
    \end{proof}
	
	We may now   prove the desired result:
\end{toappendix}
 
\begin{theoremapxrep}[$\SPExptoHCP{\cdot}$: Operational Correspondence]\label{theorem: Operational correspondence SPExp-HCP}
Let $P\in\ClassH$.

\begin{enumerate}
\item If $T:\SPExptoHCP{P}\TransitionEagerHCP{\tau} R$ and $\RegularReduction{T}$, then $P\ReductionSP Q$ s.t. $\SPExptoHCP{Q}\EquivHCP R$.
    \item If $P\ReductionSP Q$, then $\SPExptoHCP{P}\TransitionEagerHCP{\tau}R$ s.t. $R \EquivHCP \SPExptoHCP{Q}$. 
\end{enumerate}
\end{theoremapxrep}
\begin{proof}
Part (1) follows by case analysis in $\TransitionEagerHCP{\tau}$.
There are two cases depending on the transition rule used: (i) $\OutputTypeHCP\InputTypeHCP$, or (ii) $\unit\bot$. We consider the case (i); the case (ii) follows similarly.

By assumption $\SPExptoHCP{P}\TransitionEagerHCP{\tau} R$ and $\RegularReduction{T}$, and
by \Cref{lemma: shape of processes in HCP with regular reductions}(2), $$\SPExptoHCP{P}\EquivHCP \news{x}{y}(\boutHCP{k}{x}(\forward{k}{v} \para\SPExptoHCP{Q_1})\para\InputHCP{y}{z}\SPExptoHCP{Q_2}\para \SPExptoHCP{S})\TransitionEagerHCP{\tau}\news{x}{y}( \SPExptoHCP{Q_1}\para\SPExptoHCP{Q_2\subst{v}{z}}\para \SPExptoHCP{S})$$ 

By \Cref{def: encoding SPexp to HCP} we have  $$P\equiv \news{x}{y}(\OutputSP{x}{v}Q_1\para\InputSP{y}{z}Q_2\para S)\ReductionSP \news{x}{y}(Q_1\para Q_2\subst{v}{z}\para S)$$
By \Cref{def: encoding SPexp to HCP} and \Cref{lemma: equivalence and encoding}(2),  $\SPExptoHCP{\news{x}{y}(Q_1\para Q_2\subst{v}{z}\para S)}= \news{x}{y}( \SPExptoHCP{Q_1}\para\SPExptoHCP{Q_2\subst{v}{z}}\para \SPExptoHCP{S})\EquivHCP R$, which concludes the proof.

Part (2) follows induction on the derivation of  $P\ReductionSP Q$. We detail the case  \textsc{R-Struct}. (The remaining cases follow immediately.)
\begin{prooftree}
\AxiomC{$P \equiv P' $}
    \AxiomC{$ P' \longrightarrow Q'$}
    \AxiomC{$Q' \equiv Q$}
    \RightLabel{\textsc{R-Struct}}
    \TrinaryInfC{$P \longrightarrow Q$}
\end{prooftree}
By IH $\SPExptoHCP{P'}\TransitionEagerHCP{\tau}R'$ s.t. $\SPExptoHCP{Q'}\EquivHCP R$. By \Cref{lemma: equivalence and encoding}(1) we have both $\SPExptoHCP{P}\EquivHCP \SPExptoHCP{P'}$ and $\SPExptoHCP{Q'}\EquivHCP \SPExptoHCP{Q}$. Thus, the proof concludes by  \Cref{lemma: equivalence and encoding}(2). 
\end{proof}

Following Dardha and P\'{e}rez~\cite{ComparingDeadlock}, 
 we now define $\ClassL$ and $\ClassH$ as the classes of \SPExp processes that are also typable in $\CP$ and \HCP, respectively: 
\begin{definition}[Classes $\ClassL$ and $\ClassH$] \label{Def: ClassL and ClassH}
    The classes  $\ClassL$ and $\ClassH$ of \SPExp processes are defined as:
    \begin{align*}
                 \ClassL&\DefEq\{P\in \text{$\SPExp$} \Sep \Gamma \SPExpJud P\,\wedge\,\SPExptoHCP{P}\CPJud\SPExptoHCP{\Gamma} \}\\
     \ClassH&\DefEq\{P \in \SPExp\Sep
     \Gamma\SPExpJud P\,\wedge\,\Gamma= \Gamma_1,\dots,\Gamma_n \,\wedge\,
     \SPExptoHCP{P}\HCPJud \SPExptoHCP{\Gamma_1}\SequentPara\cdots\SequentPara\SPExptoHCP{\Gamma_n} \}
    \end{align*}
\end{definition}
Hence, in writing `$P \in \ClassH$' we are not referring to an \HCP process directly, but rather to a \SPExp process $P$ (in the sense of \Cref{ss:sp}) that is related (up to $\SPExptoHCP{\cdot}$) to some (typable) \HCP process. This is how   \SPExp offers a  unifying framework for comparing different type systems.

As a direct consequence of \Cref{lemma: connection between CP and HCP} and \Cref{example: process in HCP not in CP} we have that
$\ClassL\subsetneq\ClassH$.

\subsection{A Refined Account of \texorpdfstring{$\ClassH$}{H}, Based on Deadlock Freedom}
As explained in the previous section, $\HCP$ enjoys liveness properties corresponding to the LTS. If we focus only on $\tau$-transitions (cf. \Cref{hcp:reduction}) we obtain the following results.

We extend \Cref{def: regular reductions} (regular transitions) to  sequences of transitions, as follows:
\begin{definition}[\HCP: Regular Sequence of Transitions]
Assume $P_1,\dots,P_n$  in $\HCP$ and transitions $T_i: P_i\TransitionHCP{l_i}P_{i+1}$. 
    We say that a transition sequence $\rho: P\TransitionHCP{l_1}\cdots\TransitionHCP{l_n}P_n$ transitions regularly iff $\RegularReduction{T_i}$ for all $i=1,\dots,n$. We write $\RegularReduction{\rho}$.
\end{definition}

We are interested in processes whose transitions are regular, thus we refine deadlock freedom in \HCP (\Cref{def: deadlock freedom delayed}) as follows: 
 \begin{definition}[$\HCP$: Deadlock Freedom based on Regular Transitions]\label{def: deadlock freedom no delayed}
    Let $P\in\HCP$. We say that $P$ is deadlock free without delayed actions and self-syn\-chro\-ni\-za\-tions, written $\DeadlockFreedomHCPRR{P}$, iff $\DeadlockFreedomHCP{P}$, $\rho: P\TransitionHCPClo P' \EquivHCP \Inact$, and $\RegularReduction{\rho}$.  
 \end{definition}

In turn, \Cref{def: deadlock freedom delayed,def: deadlock freedom no delayed} allow us to refine the definition of $\ClassH$ (\Cref{Def: ClassL and ClassH}):

\begin{definition}[$\HCP$: Sub-classes of $\ClassH$]\label{def: sub-classes of ClassH}
	We define the following classes of DF processes:
\begin{align*}
    \DeadlockFreedomHCPSet&\DefEq\{P\in\ClassH \Sep \DeadlockFreedomHCP{\SPExptoHCP{P}}\}
    \qquad \qquad 
    \DeadlockFreedomHCPRRSet \DefEq\{P\in\ClassH \Sep \DeadlockFreedomHCPRR{\SPExptoHCP{P}}\}
\end{align*}
\end{definition}

 By \Cref{Def: ClassL and ClassH,def: sub-classes of ClassH} we have $\DeadlockFreedomHCPRRSet\subsetneq\DeadlockFreedomHCPSet\subsetneq\ClassH$.

\begin{example}\label{example: Hrt and Hf}
Consider the \SPExp processes: 
\[
P_{\ref{example: Hrt and Hf}}\DefEq  ~\news{w}{z}(\EmptyOutputSP{w}\news{x}{y}(\EmptyOutputSP{x}\Inact \para \EmptyInputSP{y}\EmptyInputSP{z}\Inact))
\quad
 Q_{\ref{example: Hrt and Hf}}\DefEq  ~\news{w}{z}(\EmptyOutputSP{w}\Inact \para \news{x}{y}(\EmptyOutputSP{x}\Inact \para \EmptyInputSP{z}\EmptyInputSP{y}\Inact)) 
 \]

We have $P_{\ref{example: Hrt and Hf}} \in\DeadlockFreedomHCPSet \setminus   \DeadlockFreedomHCPRRSet$ and 
$Q_{\ref{example: Hrt and Hf}}  \in \DeadlockFreedomHCPRRSet$. 
Although $\neg \DeadlockFreedomSP{P_{\ref{example: Hrt and Hf}}}$, the $\HCP$ process $\SPExptoHCP{P_{\ref{example: Hrt and Hf}}}$ has a $\tau$-transition due to a delayed action, and therefore $\DeadlockFreedomHCP{\SPExptoHCP{P_{\ref{example: Hrt and Hf}}}}$. 
\end{example}

\begin{theorem}\label{theorem: deadlock freedom SPExp}
    If $P\in\DeadlockFreedomHCPRRSet$, then $\DeadlockFreedomSP{P}$.
\end{theorem}
\begin{proof}
     It follows from \Cref{theorem: Operational correspondence SPExp-HCP} and \Cref{def: sub-classes of ClassH,def: deadlock freedom in SP}.
\end{proof}

\subparagraph{From $\DeadlockFreedomHCPSet$ to  $\DeadlockFreedomHCPRRSet$ via process optimizations}
We now embark to show how to transform processes in $\DeadlockFreedomHCPSet$ (where self-synchronizations and delayed actions are used) into processes in $\DeadlockFreedomHCPRRSet$ (where they are not). 
Intuitively, we would like to transform, e.g.,  
\[\news{w}{z}(\EmptyOutputSP{w}\news{x}{y}(\EmptyOutputSP{x}\Inact \para \EmptyInputSP{y}\EmptyInputSP{z}\Inact))
~~~~\text{into}~~~~ 
    \news{w}{z}(\EmptyOutputSP{w}\Inact \para \news{x}{y}(\EmptyOutputSP{x}\Inact \para \EmptyInputSP{z}\EmptyInputSP{y}\Inact))\]
The transformation from $\DeadlockFreedomHCPSet$ to $\DeadlockFreedomHCPRRSet$  is formalized by \Cref{tehorem: deadlock free disentanglement HCP}. To achieve it, we leverage two existing  ingredients.
The \emph{first ingredient} is the notion of \emph{commuting conversions}: sound transformations on processes that involve actions from  independent sessions (cf.~\cite{DeYoungCPT12,LogicalRelations}). 
\begin{definition}[$\HCP$: Commuting Conversions]\label{def: commuting conversions}
We define $\CommutingConversion$  as the smallest congruence on \HCP processes induced by the   equalities in \Cref{fig: commuting conversions}. 
\end{definition}

\begin{figure}
\hspace{1.3cm}
\begin{minipage}{12cm}
\begin{align*}
    \EmptyInputHCP{x}\EmptyInputHCP{x'}P &\CommutingConversion \EmptyInputHCP{x'}\EmptyInputHCP{x}P
    &
        \EmptyInputHCP{x}\InputHCP{x'}{y'}P &\CommutingConversion \InputHCP{x'}{y'}\EmptyInputHCP{x}P
        \\
              \InputHCP{x}{y}\InputHCP{x'}{y'}P &\CommutingConversion \InputHCP{x'}{y'}\InputHCP{x}{y}P
          &
    \EmptyInputHCP{x}\boutHCP{y'}{x'}P&\CommutingConversion \boutHCP{y'}{x'}\EmptyInputHCP{x}P
    \\
      \boutHCP{y}{x}\InputHCP{x'}{y'}P&\CommutingConversion \InputHCP{x'}{y'}\boutHCP{y}{x}P 
     &
      \EmptyOutputHCP{x}\InputHCP{x'}{y'}P &\CommutingConversion \InputHCP{x'}{y'}\EmptyOutputHCP{x}P
      \\
    \boutHCP{y}{x}\boutHCP{y'}{x'}P &\CommutingConversion   \boutHCP{y'}{x'}\boutHCP{y}{x}P
    &
      \EmptyOutputHCP{x}\boutHCP{y'}{x'}P &\CommutingConversion \boutHCP{y'}{x'}\EmptyOutputHCP{x}P\\
\EmptyOutputHCP{x}\EmptyOutputHCP{x'}P &\CommutingConversion \EmptyOutputHCP{x'}\EmptyOutputHCP{x}P
        &
      \EmptyOutputHCP{x}\EmptyInputHCP{x'}P &\CommutingConversion \EmptyInputHCP{x'}\EmptyOutputHCP{x}P\\
      \news{x}{y}(\EmptyInputHCP{x'}P) &\CommutingConversion \EmptyInputHCP{x'}\news{x}{y}(P) 
      &
      \news{x}{y}(\EmptyOutputHCP{x'}P) &\CommutingConversion \EmptyOutputHCP{x'}\news{x}{y}(P) \\
      \news{x}{y}(\InputHCP{x'}{y'}P) &\CommutingConversion  \InputHCP{x'}{y'}\news{x}{y}(P)
      &
      \news{x}{y}(\boutHCP{y'}{x'}P) &\CommutingConversion \boutHCP{y'}{x'}\news{x}{y}(P)
    \end{align*}
    \end{minipage}
    \caption{\HCP: Commuting Conversions (\Cref{def: commuting conversions}).\label{fig: commuting conversions}}
    \vspace{-3mm}
    \end{figure}

The transformations induced by $\CommutingConversion$ have a proof-theoretical origin, and so they are type-preserving, which ensures that they do not alter the causality relations on processes.
We can show that that $\CommutingConversion$ is sound wrt $\BisimilarityHCP$, i.e., if $P\CommutingConversion Q$ then $P \BisimilarityHCP Q$.
\begin{example}\label{example: only coomuting conversions} 
Recall processes 
$P'_{\ref{example: process in HCP not in CP}}$,
$P''_{\ref{example: process in HCP not in CP}}$,
and 
$P_{\ref{example: process in HCP not in CP}}$
from \Cref{example: process in HCP not in CP}.
We illustrate the optimizations on processes induced by commuting conversions by combining 
 $P''_{\ref{example: process in HCP not in CP}}$ and the process
 $Q_{\ref{example: only coomuting conversions}}\DefEq  \InputHCP{x_2}{z_2}\InputHCP{x_1}{z_1}\EmptyInputHCP{x_1}\EmptyInputHCP{x_2}\EmptyInputHCP{z_1}\EmptyInputHCP{z_2}\Inact$,
 as follows:
  $$ 	 P_{\ref{example: only coomuting conversions}}\DefEq \news{x_1}{y_1}\news{x_2}{y_2}\news{v_1}{k_1}\news{v_2}{k_2}(Q_{\ref{example: only coomuting conversions}}\para P''_{\ref{example: process in HCP not in CP}})
$$
    Note that a $\tau$-transition along the sessions
$x_1y_1$ or $x_2y_2$ in $P_{\ref{example: only coomuting conversions}}$ would depend on a delayed action. 
Now, we can use a commuting conversion to swap the order of the inputs, effectively anticipating the input on $x_1$:  
$Q_{\ref{example: only coomuting conversions}}\CommutingConversion \InputHCP{x_1}{z_1}\InputHCP{x_2}{z_2}\EmptyInputHCP{x_1}\EmptyInputHCP{x_2}\EmptyInputHCP{z_1}\EmptyInputHCP{z_2}\Inact= P'_{\ref{example: process in HCP not in CP}}$;
therefore, $P_{\ref{example: only coomuting conversions}}\CommutingConversion P_{\ref{example: process in HCP not in CP}}$. Clearly,   there is a transition sequence $\rho: P_{\ref{example: process in HCP not in CP}}\TransitionHCPClo\EquivHCP\Inact$ s.t. $\RegularReduction{\rho}$.
\end{example}
The  \emph{second ingredient}  is  \emph{disentanglement}, which also induces optimizations on processes:
  \begin{toappendix}
  \begin{figure}[!t]
  \hspace{0.4cm}
        {\footnotesize
   \hspace*{-1.5cm}  
\begin{tabular}{l c r}
     \begin{minipage}{7cm}
         \begin{prooftree}
             \AxiomC{$P\HCPJud \HypG$}
             \RightLabel{$\Hname\unit$}
             \UnaryInfC{$\EmptyOutputHCP{x}P\HCPJud \HypG \SequentPara x:\unit$}
         \end{prooftree}
     \end{minipage}
     &$\Disentangle$&
     \begin{minipage}{7cm}
     \begin{prooftree}
         \AxiomC{$\Inact \HCPJud \cdot$}
         \RightLabel{\Hname$\unit$}
         \UnaryInfC{$\EmptyOutputHCP{x}\Inact\HCPJud x:\unit$}
         \RightLabel{$\Hname\unit$}
         \AxiomC{$P\HCPJud \HypG$}
         \RightLabel{$\Hname$\MixTwo}
         \BinaryInfC{$\EmptyOutputHCP{x}\Inact\para P\HCPJud \HypG\SequentPara x:\unit$}
          \end{prooftree}
     \end{minipage}
     \\[3em]
     \begin{minipage}{7cm}
         \begin{prooftree}
             \AxiomC{$P\HCPJud \HypG \SequentPara \Gamma$}
             \AxiomC{$Q\HCPJud \HypH $}
             \RightLabel{$\Hname$\MixTwo}
             \BinaryInfC{$P\para Q \HCPJud \HypG\SequentPara \HypH\SequentPara \Gamma $}
             \RightLabel{\Hname$\bot$}
             \UnaryInfC{$\EmptyInputHCP{x}(P\para Q )\HCPJud \HypG\SequentPara \HypH\SequentPara \Gamma, x:\bot $}
         \end{prooftree}
     \end{minipage}
     &$\Disentangle$&
     \begin{minipage}{7cm}
     \begin{prooftree}
     \AxiomC{$P\HCPJud \HypG \SequentPara \Gamma$}
     \RightLabel{\Hname$\bot$}
         \UnaryInfC{$\EmptyInputHCP{x}P\HCPJud \HypG \SequentPara \Gamma,x:\bot$}
         \AxiomC{$Q\HCPJud \HypH $}
         \RightLabel{$\Hname$\MixTwo}
         \BinaryInfC{$\EmptyInputHCP{x}P\para Q \HCPJud \HypG \SequentPara \HypH \SequentPara \Gamma,x:\bot$}
     \end{prooftree}
         
     \end{minipage}\\[3em]
    \begin{minipage}{7cm}
        \begin{prooftree}
            \AxiomC{$P\HCPJud \HypG \SequentPara\Gamma,x:A\SequentPara \Delta ,y:\Dual{A}$}
            \AxiomC{$Q\HCPJud \HypH$}
            \RightLabel{$\Hname$\MixTwo}
            \BinaryInfC{$P\para Q\HCPJud \HypG\SequentPara \HypH,\SequentPara\Gamma,x:A\SequentPara \Delta ,y:\Dual{A}$}
            \RightLabel{\Hname\textsc{Cut}}
            \UnaryInfC{$\cut{xy}{P}{Q}\HCPJud \HypG\SequentPara \HypH\SequentPara\Gamma\SequentPara \Delta$}
        \end{prooftree}
    \end{minipage} & $\Disentangle$& 
    
    \begin{minipage}{7cm}
              \begin{prooftree}
            \AxiomC{$P\HCPJud \HypG \SequentPara\Gamma,x:A\SequentPara \Delta ,y:\Dual{A}$}
            \RightLabel{\Hname\textsc{Cut}}
            \UnaryInfC{$\news{x}{y}  P\HCPJud \HypG\SequentPara \HypH,\SequentPara\Gamma\SequentPara \Delta$}
            \AxiomC{$Q\HCPJud \HypH$}
            \RightLabel{$\Hname$\MixTwo}
            \BinaryInfC{$\news{x}{y}P\para{Q}\HCPJud \HypG\SequentPara \HypH\SequentPara\Gamma\SequentPara \Delta$}
        \end{prooftree}
    \end{minipage}
    \\[3em]
     \begin{minipage}{7cm}
         \begin{prooftree}
             \AxiomC{$P\HCPJud \HypG \SequentPara \Gamma, y:A \SequentPara \Delta, x:B$}
             \AxiomC{$Q\HCPJud \HypH$}
             \RightLabel{$\Hname$\MixTwo}
             \BinaryInfC{$P\para Q \HCPJud \HypG \SequentPara \HypH \SequentPara \Gamma, y:A \SequentPara \Delta, x:B$}
             \RightLabel{\Hname$\OutputTypeHCP$}
             \UnaryInfC{$\boutHCP{y}{x}(P\para Q )\HCPJud \HypG \SequentPara \HypH \SequentPara \Gamma ,\Delta, x:A\OutputTypeHCP B$}
         \end{prooftree}
     \end{minipage}
     & $\Disentangle$ &
     \begin{minipage}{7cm}
        \begin{prooftree}
             \AxiomC{$P\HCPJud \HypG \SequentPara \Gamma, y:A \SequentPara \Delta, x:B$}
             \RightLabel{\Hname$\OutputTypeHCP$}
             \UnaryInfC{$\boutHCP{y}{x}P\HCPJud \HypG \SequentPara \Gamma ,\Delta, x:A\OutputTypeHCP B$}
             \AxiomC{$Q\HCPJud \HypH$}
             \RightLabel{$\Hname$\MixTwo}
             \BinaryInfC{$\boutHCP{y}{x}P\para Q \HCPJud \HypG \SequentPara \HypH \SequentPara \Gamma ,\Delta, x:A\OutputTypeHCP B$}
         \end{prooftree}
     \end{minipage}
    \\[3em]
           \begin{minipage}{7cm}
         \begin{prooftree}
             \AxiomC{$P\HCPJud \HypG \SequentPara \Gamma, y:A, x:B$}
             \AxiomC{$Q\HCPJud \HypH$}
             \RightLabel{$\Hname$\MixTwo}
             \BinaryInfC{$P\para Q \HCPJud \HypG \SequentPara \HypH \SequentPara \Gamma, y:A, x:B$}
             \RightLabel{\Hname$\InputTypeHCP$}
             \UnaryInfC{$\InputHCP{x}{y}(P\para Q )\HCPJud \HypG \SequentPara \HypH \SequentPara \Gamma ,\Delta, x:A\InputTypeHCP B$}
         \end{prooftree}
     \end{minipage}
     & $\Disentangle$ &
     \begin{minipage}{7cm}
        \begin{prooftree}
             \AxiomC{$P\HCPJud \HypG \SequentPara \Gamma, y:A \SequentPara \Delta, x:B$}
              \RightLabel{\Hname$\InputTypeHCP$}
             \UnaryInfC{$\InputHCP{x}{y}P\HCPJud \HypG \SequentPara \Gamma ,\Delta, x:A\OutputTypeHCP B$}
             \AxiomC{$Q\HCPJud \HypH$}
             \RightLabel{$\Hname$\MixTwo}
             \BinaryInfC{$\InputHCP{x}{y} P\para Q \HCPJud \HypG \SequentPara \HypH \SequentPara \Gamma ,\Delta, x:A\OutputTypeHCP B$}
         \end{prooftree}
     \end{minipage}
\end{tabular}
    }
     \caption{$\HCP$: Disentanglement (\Cref{def: disentanglement}).}
    \label{fig: full disnetanglement}
\end{figure}
  \end{toappendix}

\begin{definition}[$\HCP$: Disentanglement~\cite{TakingLinearLogicApart}]\label{def: disentanglement}
Disentanglement is described by the smallest relation $\Disentangle$ on (typed) processes closed under the rules in \Cref{fig: full disnetanglement}.
 We write $\DisentangleClo$ for the reflexive and transitive closure of $\Disentangle$.
\end{definition}

Disentanglement proceeds by the repeated application of the rules from \Cref{fig: full disnetanglement} to the derivation of $P\HCPJud \Gamma_1\SequentPara \cdots \SequentPara \Gamma_n$, which aims at moving downwards the instances of Rule \Hname\MixTwo (cf.~\Cref{fig:type system HCP}). If \Hname\MixTwo gets stuck in either $\Hname$\textsc{Cut} or $\Hname\OutputTypeHCP$, then it becomes an instance of $\C$\textsc{Cut} or $\C\OutputTypeHCP$, respectively (cf.~\Cref{fig:CP Rules}). 
 Disentanglement increases the parallelism in processes. This way, e.g., for  $P_{\ref{example: process in HCP not in CP}}$ and $Q_{\ref{example: process in HCP not in CP}}$  (\Cref{example: process in HCP not in CP}) we have: $P_{\ref{example: process in HCP not in CP}}\DisentangleClo Q_{\ref{example: process in HCP not in CP}}$.
Also, note that Rule $\Hname\unit$ types $\EmptyOutputHCP{x}P\HCPJud \Gamma \SequentPara x:\unit$, induces a discrepancy: at the process level, the rule places the empty output in sequential composition with $P$,  whereas at the level of typing  $x:\unit$ is placed in its own partition, enforcing independence from all names in $\FreeNamesHCP{P}$. Disentanglement solves this discrepancy. Consider, e.g., process $P_{\ref{example:intrigue}}=\news{x}{x'}(\EmptyOutputHCP{x}\EmptyInputHCP{x'}\Inact)$: we have  $P_{\ref{example:intrigue}}\DisentangleClo\,\news{x}{x'}(\EmptyOutputHCP{x}\Inact\para\EmptyInputHCP{x'}\Inact)$.

\begin{lemma}[Disentanglement~\cite{betterlate}]\label{lemma: disentanglement}
    If there exists a derivation $d$ of $P\HCPJud \Gamma_1\SequentPara \cdots \SequentPara \Gamma_n$, then there exist derivations $d_1,\dots,d_n$ of $P_1\CPJud \Gamma_1, \dots ,P_n\CPJud \Gamma_n$ in $\CP$, and 

    \begin{tabular}{ccc}
        $P\HCPJud \Gamma_1\SequentPara \cdots \SequentPara \Gamma_n$ & $\DisentangleClo$ & 
        \begin{minipage}{4cm}
        \begin{prooftree}
            \AxiomC{ $P_1\CPJud \Gamma_1\, \cdots\, P_n\CPJud \Gamma_n$ }
            \RightLabel{$\Hname$\MixTwo}
            \doubleLine
            \UnaryInfC{$P_1\para\cdots\para P_n\HCPJud \Gamma_1\SequentPara \cdots \SequentPara \Gamma_n$}
            \end{prooftree}
        \end{minipage}
    \end{tabular}
    \\
    where we use the double line to indicate multiple uses of the same rule.
\end{lemma}

The following result ensures that process observations are invariant under disentanglement. Moreover, since instances of $\Hname$\textsc{Cut} are turned into  instances of $\C$\textsc{Cut}, disentangled processes do not feature occurrences of restriction between sessions in sequential composition.
\begin{lemma}\label{lemma: denotations preserved in disentanglement}
     Suppose $P\in\HCP$. If  $P\Disentangle^* P'$ then $P\BisimilarityHCP P'$.
\end{lemma}
\begin{proof}

We prove that if  $P\Disentangle^* P'$ then $P\ObservationalEq P'$ by induction on the length $k$ of $\Disentangle^*$. For the base case ($k = 1$)  we proceed by a case analysis in the last rule applied in the type derivation of $P\HCPJud\HypG$. The fact that $P\BisimilarityHCP P'$ follows from \Cref{theorem: full abstraction}.
\end{proof}

Given a process $P$, the rules for disentanglement can be applied in any order; we therefore define a set of disentangled processes:
\begin{definition}[\HCP: Disentangled Processes]\label{def: disentangled process HCP} Given $P\in \HCP$, the set  $\Disentangled{P}$ is defined as follows:
\begin{mathpar}
 \Disentangled{P}     
    \DefEq\left\{P' \in \HCP ~\mathrel{\big |}~
    \begin{gathered}[l]
    P\DisentangleClo P',\\[-1.5em]
     P'=P_1\para\cdots\para P_n\HCPJud \Gamma_1\SequentPara\cdots\SequentPara \Gamma_n , ~~
      P_1\CPJud \Gamma_1, ~\cdots~ ,P_n\CPJud \Gamma_n
      \end{gathered}
\right\}
\end{mathpar}
\end{definition}
From \Cref{lemma: denotations preserved in disentanglement} we have that 
all disentangled processes are observationally equivalent:
\begin{lemma}
\label{lem:disentrans}
    Let $P\in\HCP$. If $Q,R\in\Disentangled{P}$, then $Q\BisimilarityHCP R$.
\end{lemma}
\begin{proof}
 From \Cref{lemma: denotations preserved in disentanglement},  $P\BisimilarityHCP P'$ for all $P'\in\Disentangled{P}$.  By \Cref{def: weak transition HCP}, $\BisimilarityHCP$  is transitive.
 Hence, for all $Q, R\in\Disentangled{P}$, $Q\BisimilarityHCP R$.  
\end{proof}

We use disentanglement to prove that if $\DeadlockFreedomHCP{P}$ then there is a  $P'$ s.t. $\DeadlockFreedomHCPRR{P'}$, with $P\BisimilarityHCP P'$. 

The following  main result establishes that the notions of $\DeadlockFreedomHCP{\cdot}$ 
(\Cref{def: deadlock freedom delayed})
and $\DeadlockFreedomHCPRR{\cdot}$ 
(\Cref{def: deadlock freedom no delayed})
are equivalent, up to commuting conversions and disentanglement. 

\begin{theoremapxrep}\label{tehorem: deadlock free disentanglement HCP}
For all processes  $P\in \HCP$ s.t. $\DeadlockFreedomHCP{P}$, then there exist $P',Q\in \HCP$ s.t.: 
\textup{(1)} $P'\in\Disentangled{P}$, 
\textup{(2)} $P'\CommutingConversion Q$,
\textup{(3)} $P\BisimilarityHCP Q$,
\textup{(4)} $\DeadlockFreedomHCPRR{Q}$.
\end{theoremapxrep}
\begin{proof}
    Part (1), on the existence of $P'$, is immediate from \Cref{def: disentangled process HCP}.  
    Part (2) follows from \Cref{def: commuting conversions} and the soundness of commuting conversions with respect to bisimilarity.
    Part (3) is immediate from \Cref{lemma: denotations preserved in disentanglement} and Part (2). 
    As for Part (4), because $P$ and $Q$ are bisimilar, the analysis of deadlock freedom reduces to the analysis of self-synchronizations and delayed actions in $Q$. We detail this analysis next, considering these two cases separately.

\subparagraph{Self-synchronizations}
    By assumption, $\DeadlockFreedomHCP{P}$. Therefore, $P\TransitionHCPClo\Inact$ relying on $\TransitionHCP{\tau}$. This means that self-synchronizations  must happen on dual names, inside a restriction. By the previous analysis, we know that for each self-synchronization there is a subprocess of $P$ $P''=\news{x}{y}\pi.R$ s.t.  $\FreeNamesHCP{\pi}=x, \FreeNamesHCP{l}=y$, and $\pi.R\TransitionHCP{\pi\para l}R'$, where $\Partition{x}{\pi.R}{y}$. 
    By the definition of $\Partition{x}{\pi.R}{y}$ (\Cref{d:separation}), we infer that 
    $\pi.R\HCPJud \HypG$, where $\HypG=\Gamma_1\parallel \cdots \parallel \Gamma_{n-1},x:A\parallel\Gamma_n,y:\Dual{A}$. 
    
Recall that disentanglement turns instances of Rule \Hname \textsc{Cut} into instances of Rule \C \textsc{Cut}:
\begin{prooftree}
    \AxiomC{$\pi.R\HCPJud\Gamma_1\parallel\cdots \parallel \Gamma_{n-1},x:A\parallel\Gamma_n,y:\Dual{A}$}
    \RightLabel{\Hname\textsc{Cut}}
    \UnaryInfC{$\news{x}{y}\pi.R\HCPJud\Gamma_1\parallel\cdots  \parallel \Gamma_{n-1},\Gamma_n$}
\end{prooftree}
    By \Cref{lemma: disentanglement}, there exist 
    $R_1, \cdots, R_n$  such that  $R_1\CPJud \Gamma_1, \cdots, R_{n-1}\CPJud \Gamma_{n-1}$ and also $x:A, R_{n}\CPJud$ and $\Gamma_{n},y:\Dual{A}$.  Thus we have:
    \begin{prooftree}
        \AxiomC{$R_1\para \cdots\para R_{n-2}\HCPJud\Gamma_1\parallel\cdots \parallel \Gamma_{n-2} $}
        \AxiomC{$ R_{n-1}\para R_n\HCPJud \Gamma_{n-1},x:A\parallel\Gamma_n,y:\Dual{A}$}
        \RightLabel{\C \textsc{Cut}}
        \UnaryInfC{$ \news{x}{y}(R_{n-1}\para R_n)\HCPJud \Gamma_{n-1},\Gamma_n$}
        \BinaryInfC{$R_1\para \cdots\para R_{n-2}\para\news{x}{y}(R_{n-1}\para R_n)\HCPJud\Gamma_1\parallel\cdots \parallel \Gamma_{n-2}\para \Gamma_{n-1},\Gamma_n$}
    \end{prooftree}

    By \Cref{lemma: denotations preserved in disentanglement},  
    $\news{x}{y}\pi.R\BisimilarityHCP R_1\para \cdots\para R_{n-2}\para\news{x}{y}(R_{n-1}\para R_n)$. Moreover, by typability ensures that $x$ and $y$ are free names only in $R_{n-1}$ and $R_n$, respectively. Therefore $R_{n-1}\para R_n\TransitionHCP{\pi\para l} R''$, for some $R''$ s.t. $R'\BisimilarityHCP R''$. Note that $R_{n-1}\para R_n\TransitionHCP{\pi\para l} R''$ is not an instance of Rule $|\pi$ (which induces self-synchronizations), because the actions $\pi$ and $l$ occur in different parallel components. Thus, if we take a process $P'\in\Disentangled{P}$ in which all the instances of Rule \Hname    {\textsc{Cut}} have been transformed into instances of  Rule \C{\textsc{Cut}}, then we obtain a process  that does not rely in self-synchronizations. 
   
 \subparagraph{Delayed actions} If $P'$ does not rely on delayed actions, then we take $Q=P'$. Otherwise, $P'$ has transitions that rely on Rules $\pi_1,\pi_2,\otimes$ or $\parr$. Two observations: 
  \begin{enumerate}
   \item Free names are invariant under commuting conversions; and
       \item We can always bring to the top-level those actions that are happening under some guard.
 \end{enumerate}
While Item (1) follows by straightforward inspection of the the rules in \Cref{def: commuting conversions}, Item (2) follows by induction in the number of prefixes/restrictions guarding the delayed action. The base case follows immediately by \Cref{def: commuting conversions}. The inductive step follows by a case analysis in the last rule applied and IH. Thus, by repeated application of commuting conversions, we can find a process $Q$ s.t. $Q\BisimilarityHCP P'$ (and by transitivity $Q\BisimilarityHCP P$) and $\DeadlockFreedomHCPRR{Q}$.
     \end{proof}

\Cref{tehorem: deadlock free disentanglement HCP} concerns $\HCP$ processes.  To lift this result to~$\SPExp$, we need some  results. 
\begin{lemma}\label{lemma: disentanglement in SP}
      Let $P\in \ClassH$. There exists $P'\in \ClassH$ s.t.:  
             \textup{(1)} $\SPExptoHCP{P}\Disentangle^*\SPExptoHCP{P'}$.
          \textup{(2)} $\SPExptoHCP{P}\CommutingConversion^*\SPExptoHCP{P'}$.
\end{lemma}
\begin{proof}
The items follow by induction on the length of $\Disentangle^*$ and $\CommutingConversion^*$, respectively. For 1. in the base cases we proceed by a case analysis in the last rule applied in the type derivation of $\SPExptoHCP{P}\HCPJud\SPExptoHCP{\Gamma}$; it shows the structure of $\SPExptoHCP{P}$ and by \Cref{def: encoding SPexp to HCP} we obtain the structure of $P$. The proof finishes by applying a single step of $\Disentangle$ to the type derivation of $\SPExptoHCP{P}\HCPJud\SPExptoHCP{\Gamma}$ and replicating it to the derivation of $\Gamma \SPExpJud P$. The proof of 2. proceeds similarly.
\end{proof}

\Cref{lemma: disentanglement in SP}  ensures that the following definition is sound:
\begin{definition}[Lifting Commuting Conversions and Disentanglement to $\SPExp$]\label{def: lifting disentanglement}
    Let $P, P' \in\ClassH$. We write $P\DisentangleClo P'$ and $P\CommutingConversion^*P'$, whenever $\SPExptoHCP{P}\DisentangleClo  \SPExptoHCP{P'}$ and $\SPExptoHCP{P}\CommutingConversion^* \SPExptoHCP{P'}$ hold, respectively. 
\end{definition}
Having disentanglement in $\SPExp$, we aim to prove that the resulting disentangled process \emph{simulates} the original process. To define simulation in $\SPExp$, we need the following notion:
\begin{definition}[$\SPExp$: Labelled Reduction Semantics]\label{def: labelled reduction semantics SPExp}
Let $P\in\SPExp$, then $P \LabelledReduction{x}{y} Q$   whenever:
\textup{(1)}~$P\equiv\news x y(\OutputSP{x}{v}P' \para \InputSP{y}{z}Q'\ \para\ R) \longrightarrow \news x y(P'\ \para\ Q'\subst{v}{z} \para R)\equiv Q $, or
    \textup{(2)}~$P\equiv\news xy(\EmptyInputSP{x}P'\para \EmptyOutputSP{y} Q'\para R)\ReductionSP \news xy (P' \para Q'\para R)\equiv Q$.
\end{definition}
\begin{definition}[$\SPExp$: Simulation]
\label{def:simulation}
    A binary relation $\mathbb{S}$ in $\SPExp$ is a (typed)  simulation if whenever $(P,Q)\in \mathbb{S}$ then 
    $\Gamma\SPExpJud P$, $\Gamma\SPExpJud Q$ and $P\LabelledReduction{x}{y}P'$ implies
    $Q\LabelledReduction{x}{y} Q'$ for some $Q'$ s.t. $(P',Q')\in \mathbb{S}$. We write $P \ReductionSimulation Q$ whenever $(P,Q)\in \mathbb{S}$ for a (typed) simulation $\mathbb{S}$.
\end{definition}
In the LTS semantics of $\HCP$, the processes $P$ and $Q$ from \Cref{tehorem: deadlock free disentanglement HCP} are bisimilar; however, in a reduction semantics, any transitions of $P$ that use delayed actions or self-synchronization would be considered as deadlocks, and then $Q$ would only simulate $P$. Formally:

\begin{theoremapxrep}\label{theorem: Rs in a reduction simulation}
The relation
$\mathbb{S}=\{(P,R) \,\mid\, P\in \ClassH, \text{and } P\Disentangle R\}$ is a    simulation.
\end{theoremapxrep}
\begin{proof}
We check the conditions of \Cref{def:simulation} by induction on the structure of $P$. 
We consider two interesting cases; the remaining cases proceed similarly.
\begin{itemize}
    \item Case $P=\EmptyOutputSP{x}P'$ . 
    By \Cref{def: disentanglement} we have that disentanglement is applied as follows:\\
    \begin{tabular}{ccc}
\hspace*{-1.3cm}  
 \begin{minipage}{7cm}
         \begin{prooftree}
             \AxiomC{$\SPExptoHCP{P'}\HCPJud \HypG$}
             \UnaryInfC{$\SPExptoHCP{\EmptyOutputSP{x}P'}\HCPJud \HypG \SequentPara x:\unit$}
         \end{prooftree}
     \end{minipage}
     &$\Disentangle$&
     \begin{minipage}{7cm}
     \begin{prooftree}
         \AxiomC{$\SPExptoHCP{\Inact} \HCPJud \cdot$}
         \UnaryInfC{$\SPExptoHCP{\EmptyOutputSP{x}\Inact}\HCPJud x:\unit$}
         \AxiomC{$\SPExptoHCP{P}\HCPJud \HypG$}
         \RightLabel{$\Hname$\MixTwo}
         \BinaryInfC{$\SPExptoHCP{\EmptyOutputSP{x}\Inact\para P}\HCPJud \HypG\SequentPara x:\unit$}
          \end{prooftree}
     \end{minipage}
\end{tabular}

 Thus $(\EmptyOutputSP{x}P', \EmptyOutputSP{x}\Inact \para P')\in\mathbb{S}$. By \Cref{def: reduction semantics sp}
is clear that   $\EmptyOutputSP{x}P'$ has no reduction, which concludes this case. Cases $P=\EmptyInputHCP{x}P',P=\InputHCP{x}{y}P'$ and $P=\boutHCP{v}{x}P'$ follow similarly.

\item Case $P=\news{x}{y}(Q\para R)$. 

By \Cref{def: disentanglement} we have that disentanglement is applied as follows:\\
\begin{tabular}{ccc}
\hspace*{-1.3cm}  
\begin{minipage}{7cm}
	\begin{prooftree}
		\AxiomC{$\SPExptoHCP{Q}\HCPJud \HypG \SequentPara\Gamma,x:A\SequentPara \Delta ,y:\Dual{A}$}
		\AxiomC{$\SPExptoHCP{R}\HCPJud \HypH$}
		\BinaryInfC{$\SPExptoHCP{Q}\para \SPExptoHCP{R}\HCPJud \HypG\SequentPara \HypH,\SequentPara\Gamma,x:A\SequentPara \Delta ,y:\Dual{A}$}
		\UnaryInfC{$\cut{xy}{\SPExptoHCP{Q}}{\SPExptoHCP{R}}\HCPJud \HypG\SequentPara \HypH\SequentPara\Gamma\SequentPara \Delta$}
	\end{prooftree}
    
\end{minipage}
	&
$\Disentangle$&
\begin{minipage}{7cm}
	\begin{prooftree}
		\AxiomC{$\SPExptoHCP{Q}\HCPJud \HypG \SequentPara\Gamma,x:A\SequentPara \Delta ,y:\Dual{A}$}
		
		\UnaryInfC{$\news{x}{y}  \SPExptoHCP{Q}\HCPJud \HypG\SequentPara \HypH,\SequentPara\Gamma\SequentPara \Delta$}
		\AxiomC{$\SPExptoHCP{R}\HCPJud \HypH$}
		\BinaryInfC{$\news{x}{y}(\SPExptoHCP{Q})\para{\SPExptoHCP{R}}\HCPJud \HypG\SequentPara \HypH\SequentPara\Gamma\SequentPara \Delta$} 
	\end{prooftree}
\end{minipage}
\end{tabular}
\\
Thus $(\news{x}{y}(Q\para R),\news{x}{y}(Q)\para R)\in\mathbb{S}$. If $\news{x}{y}(Q\para R)\ReductionSP S$, \Cref{def: reduction semantics sp} we have the following cases:
\begin{enumerate}

    \item $R\ReductionSP R'$ and $S=(\news{x}{y}(Q\para R')$,
    \item $Q\ReductionSP Q'$ in a session different to $xy$ and $S=(\news{x}{y}(Q'\para R)$, or
 \item $\news{x}{y}(Q\para R)\ReductionSP S$ on session $xy$, and we have either
           \begin{enumerate}
      
           \item \sloppy reduction due to \textsc{R-Com}, then $Q\equiv \InputSP{y}{z}Q_1\para \OutputSP{x}{v}Q_2\para Q_3$
           and  $S\equiv\news{x}{y}(Q_1\subst{v}{z}\para Q_2\para Q_3\para R)$, or 
           \item reduction due to \textsc{R-EmptyCom}, then $Q\equiv \EmptyInputSP{y}Q_1\para \EmptyOutputSP{x}Q_2\para Q_3$
           and $S\equiv Q_1\para Q_2\para Q_3\para R$.
           \end{enumerate}
\end{enumerate}
Since $\news{x}{y}(Q\para R)\equiv\news{x}{y}(Q)\para R$, it is clear that $\news{x}{y}(Q\para R)\ReductionSimulation\news{x}{y}(Q)\para R$. It remains to check that if $\news{x}{y}(Q\para R)\LabelledReduction{a}{b} S$ and $\news{x}{y}(Q)\para R \LabelledReduction{a}{b} S'$, then $(S,S')\in\mathbb{S}$. We proceed as follows:
In Case (1) we have:\\
\begin{tabular}{ccc}
\hspace*{-1.3cm}  
\begin{minipage}{7cm}
	\begin{prooftree}
		\AxiomC{$\SPExptoHCP{Q}\HCPJud \HypG \SequentPara\Gamma,x:A\SequentPara \Delta ,y:\Dual{A}$}
		\AxiomC{$\SPExptoHCP{R'}\HCPJud \HypH$}
		\BinaryInfC{$\SPExptoHCP{Q}\para \SPExptoHCP{R'}\HCPJud \HypG\SequentPara \HypH,\SequentPara\Gamma,x:A\SequentPara \Delta ,y:\Dual{A}$}
		\UnaryInfC{$\cut{xy}{\SPExptoHCP{Q}}{\SPExptoHCP{R'}}\HCPJud \HypG\SequentPara \HypH\SequentPara\Gamma\SequentPara \Delta$}
	\end{prooftree}
    
\end{minipage}
	&
$\Disentangle$&
\begin{minipage}{7cm}
	\begin{prooftree}
		\AxiomC{$\SPExptoHCP{Q}\HCPJud \HypG \SequentPara\Gamma,x:A\SequentPara \Delta ,y:\Dual{A}$}
		
		\UnaryInfC{$\news{x}{y}  \SPExptoHCP{Q}\HCPJud \HypG\SequentPara \HypH,\SequentPara\Gamma\SequentPara \Delta$}
		\AxiomC{$\SPExptoHCP{R'}\HCPJud \HypH$}
		\BinaryInfC{$\news{x}{y}(\SPExptoHCP{Q})\para{\SPExptoHCP{R'}}\HCPJud \HypG\SequentPara \HypH\SequentPara\Gamma\SequentPara \Delta$} 
	\end{prooftree}
\end{minipage}
\end{tabular}\\
and  $(\news{x}{y}(Q\para R'),\news{x}{y}(Q)\para R')\in\mathbb{S}$.

Case (2) is similar to case (1).

In Case (3a) we have:\\

\begin{tabular}{c}
\hspace*{-1.3cm}  
\begin{minipage}{7cm}
	\begin{prooftree}
		\AxiomC{$\SPExptoHCP{Q_1\subst{v}{z}}\para\SPExptoHCP{Q_2}\para \SPExptoHCP{Q_3}\HCPJud \HypG \SequentPara\Gamma,x:A\SequentPara \Delta ,y:\Dual{A}$}
		\AxiomC{$\SPExptoHCP{R}\HCPJud \HypH$}
		\BinaryInfC{$\SPExptoHCP{Q_1\subst{v}{z}}\para\SPExptoHCP{Q_2}\para \SPExptoHCP{Q_3}\para \SPExptoHCP{R}\HCPJud \HypG\SequentPara \HypH,\SequentPara\Gamma,x:A\SequentPara \Delta ,y:\Dual{A}$}
		\UnaryInfC{$\cut{xy}{\SPExptoHCP{Q_1\subst{v}{z}}\para\SPExptoHCP{Q_2}\para \SPExptoHCP{Q_3}}{\SPExptoHCP{R}}\HCPJud \HypG\SequentPara \HypH\SequentPara\Gamma\SequentPara \Delta$}
	\end{prooftree}
    
\end{minipage}\\
	
$\Disentangle$\\
\begin{minipage}{7cm}
	\begin{prooftree}
		\AxiomC{$\SPExptoHCP{Q_1\subst{v}{z}}\para\SPExptoHCP{Q_2}\para \SPExptoHCP{Q_3}\HCPJud \HypG \SequentPara\Gamma,x:A\SequentPara \Delta ,y:\Dual{A}$}
		
		\UnaryInfC{$\news{x}{y}  \SPExptoHCP{Q_1\subst{v}{z}}\para\SPExptoHCP{Q_2}\para \SPExptoHCP{Q_3}\HCPJud \HypG\SequentPara \HypH,\SequentPara\Gamma\SequentPara \Delta$}
		\AxiomC{$\SPExptoHCP{R}\HCPJud \HypH$}
		\BinaryInfC{$\news{x}{y}(\SPExptoHCP{Q_1\subst{v}{z}}\para\SPExptoHCP{Q_2}\para \SPExptoHCP{Q_3})\para{\SPExptoHCP{R}}\HCPJud \HypG\SequentPara \HypH\SequentPara\Gamma\SequentPara \Delta$} 
	\end{prooftree}
\end{minipage}
\end{tabular}\\

Thus, $(\news{x}{y}(Q_1\subst{v}{z}\para Q_2\para Q_3\para R),\news{x}{y}(Q_1\subst{v}{z}\para Q_2\para Q_3)\para R)\in\mathbb{S}$. The case (3b) follows similarly.
\end{itemize}
\end{proof}

Finally, \Cref{corollary: main corollary HCP}, given below, brings \Cref{tehorem: deadlock free disentanglement HCP} to $\SPExp$: it shows that even though in $\HCP$ both $\DeadlockFreedomHCP{\cdot}$ and $\DeadlockFreedomHCPRR{\cdot}$ are equivalent, in $\SPExp$ those notions are related via a simulation.
\begin{corollary}\label{corollary: main corollary HCP}
Given  $P\in\DeadlockFreedomHCPSet$, then there exist $Q\in\ClassH$ and $P'\in \DeadlockFreedomHCPRRSet$ s.t.:
    \textup{(1)} $\SPExptoHCP{Q}\in \Disentangled{\SPExptoHCP{P}}$,
    \textup{(2)} $\SPExptoHCP{Q}\CommutingConversion^*\SPExptoHCP{P'}$,
    \textup{(3)} $P\ReductionSimulation P'$,
    \textup{(4)} $\SPExptoHCP{P}\BisimilarityHCP\SPExptoHCP{P'}$. 
\end{corollary}
\begin{proof}
    It follows by \Cref{lemma: disentanglement in SP,tehorem: deadlock free disentanglement HCP,theorem: Rs in a reduction simulation}
\end{proof}

\begin{example}
[\Cref{corollary: main corollary HCP}, At Work]\label{example: section 3 at work}
Consider the  process $P_{\ref{example: section 3 at work}}$ and its encoding into $\HCP$:
\begin{align*}
     P_{\ref{example: section 3 at work}}&\DefEq\news{w}{z}(\EmptyOutputSP{w}\news{x}{y}(\EmptyOutputSP{x}\Inact \para\news{k}{u}(\EmptyInputSP{k}\Inact \para \EmptyInputSP{y}\EmptyInputSP{z}\EmptyOutputSP{u}\Inact)))\\
 \SPExptoHCP{P_{\ref{example: section 3 at work}}}&\DefEq\news{w}{z}(\EmptyOutputHCP{w}\news{x}{y}(\EmptyOutputHCP{x}\Inact \para\news{k}{u}(\EmptyInputHCP{k}\Inact \para \EmptyInputHCP{y}\EmptyInputHCP{z}\EmptyOutputHCP{u}\Inact)))
\end{align*}         

 Note that  $\neg\DeadlockFreedomSP{P_{\ref{example: section 3 at work}}}$ but $\SPExptoHCP{P_{\ref{example: section 3 at work}}}$ has two   $\tau$-transitions:
\begin{align}
    \SPExptoHCP{P_{\ref{example: section 3 at work}}}&\TransitionHCP{\tau}\news{x}{y}(\EmptyOutputHCP{x}\Inact \para\news{k}{u}(\EmptyInputHCP{k}\Inact \para \EmptyInputHCP{y}\EmptyOutputHCP{u}\Inact))\TransitionHCPClo\EquivHCP\Inact\label{first transition}\\
    \SPExptoHCP{P_{\ref{example: section 3 at work}}}&\TransitionHCP{\tau}\news{w}{z}(\EmptyOutputHCP{w}\news{k}{u}(\EmptyInputHCP{k}\Inact \para \EmptyInputHCP{z}\EmptyOutputHCP{u}\Inact))\TransitionHCPClo\EquivHCP\Inact \label{second transition}
\end{align}

    where the transition in (\ref{first transition}) is due to a delayed action on $z$ and a self-synchronization along the session $wz$, and  the transition in (\ref{second transition}) is due to a delayed $\tau$-transition along the session $xy$. Hence, $\DeadlockFreedomHCP{\SPExptoHCP{P_{\ref{example: section 3 at work}}}}$.
   Via disentanglement we obtain:
\[
    \SPExptoHCP{P_{\ref{example: section 3 at work}}}\Disentangle\news{w}{z}(\EmptyOutputHCP{w}\Inact \para\news{x}{y}(\EmptyOutputHCP{x}\Inact \para \news{k}{u}(\EmptyInputHCP{k}\Inact\para \EmptyInputHCP{y}\EmptyInputHCP{z}\EmptyOutputHCP{u}\Inact)))\DefEq R'
\]
In $R'$, neither a delayed action nor a self-synchronization are needed, thus $\DeadlockFreedomHCPRR{R'}$. We could also give priority to the $\tau$-transition along $wz$; however, because the input on $z$ is not at the top-level, a synchronization on $wz$ still requires a delayed action. This shows the need for commuting conversions. Indeed, by using $\CommutingConversion$  we obtain:
\[
     R' \CommutingConversion \news{w}{z}(\EmptyOutputHCP{w}\Inact \para \news{x}{y}(\EmptyOutputHCP{x}\Inact \para \news{k}{u}(\EmptyInputHCP{k}\Inact\para \EmptyInputHCP{z}\EmptyInputHCP{y}\EmptyOutputHCP{u}\Inact)))\DefEq R
\]
We have that $\DeadlockFreedomHCPRR{R}$. Moreover, 
there is a process $P'$ in $\SPExp$ s.t. $\SPExptoHCP{P'}=R$: 
\[
    P'\DefEq\news{w}{z}(\EmptyOutputSP{w}\Inact \para \news{x}{y}(\EmptyOutputSP{x}\Inact \para \news{k}{u}(\EmptyInputSP{k}\Inact\para \EmptyInputSP{z}\EmptyInputSP{y}\EmptyOutputSP{u}\Inact)))\]
    Notice that $P_{\ref{example: section 3 at work}}\ReductionSimulation P'$ follows trivially since  $\DeadlockFreedomSP{P'}$.
\end{example}

\section{The Case of Asynchronous Processes}\label{sec: charaterization in Pado}

\newcommand{\colorshade}{EFEFEF}
\begin{figure}[t!]
    \centering
	 \begin{tabular}{|P{1.6cm}|P{2cm}||P{1cm}|P{2cm}||P{2.8cm}|P{2cm}|}
	 \hline
	 $\SPExp$ & \Cref{def: SPExp}& $\ClassL$& \Cref{Def: ClassL and ClassH} &$\SPExptoHCP{\cdot}:\SPExp\rightarrow \HCPalt$ &\Cref{def: encoding SPexp to HCP}
    \\
	\hline
$\CP$ & \Cref{def: CP}& $\ClassH$& \Cref{Def: ClassL and ClassH}& 	   \multicolumn{2}{l|}{} 
         \\ 
	 \hline
	 $\HCP$ & \Cref{def: type system HCP} & $\DeadlockFreedomHCPSet, \DeadlockFreedomHCPRRSet$ & \cref{def: sub-classes of ClassH} & \multicolumn{2}{l|}{}
       \\
	 \hline \hline 
	 \rowcolor[HTML]{\colorshade}
$\Padovani$ & \Cref{def: type system LP}&  $\ClassLP$ & \Cref{def: class muP} &     $\SPExptoPado{f}{\cdot}:\SPExp\rightarrow \Padovani$  &\Cref{def: encoding SPExp to Padovani}
         \\
	 \hline  \rowcolor[HTML]{\colorshade}
	   $\MuPadovani$ &   \Cref{def: type system MuLP}& $\ClassMuP$ & \Cref{def: class muP} & 
	  $\SPAtoSP{\cdot}:\SPExpA \rightarrow \SPExp$ &   \Cref{def: encoding SPA to SP}
         \\
	 \hline \rowcolor[HTML]{\colorshade}
  $\SPExpA$ &   \Cref{def: type system ASP}&  $\ClassLA, \ClassHA$ &   \Cref{d:asyncclasses} & \multicolumn{2}{l|}{} 
\\
	 \hline
    \end{tabular}
          \caption{Inventory of type systems, classes of processes, and encodings considered in the paper. Shaded cells concern elements used in the case of asynchronous  processes (\Cref{sec: charaterization in Pado}).}
    \label{fig:diagram}
    \vspace{-3mm}
\end{figure}

Up to here we have addressed \hyperref[lq1]{\textbf{\textsf{(Q1)}}} by considering the relation between 
$\DeadlockFreedomHCPSet$ and $\DeadlockFreedomHCPRRSet$ in \HCP. We now turn our attention to \hyperref[lq2]{\textbf{\textsf{(Q2)}}} and consider the interplay of DF and \emph{asynchronous processes} as typable by Padovani's type system for  asynchronous processes, here dubbed~$\Padovani$~\cite{padovani_linear_pi}. 

We proceed as follows. 
In \Cref{sec: padovani} we recall Padovani's type system and  introduce $\ClassLP$, the class of $\SPExp$ processes induced by $\Padovani$, and 
its sub-class $\ClassMuP$, derived from the type system $\MuPadovani$.
In \Cref{sec: mupadovani}, we show that $\ClassMuP = \ClassL$ (\Cref{corollary: L equal to mupadovani}). 
In \Cref{sec: comparing H and MuLP}, 
we establish a key result: the precise  relations between $\ClassH$, $\DeadlockFreedomHCPSet$, $\DeadlockFreedomHCPRRSet$, and $\ClassLP$ (\Cref{lemma: comparing ClassH and Padovani}).
Also, we revisit  \Cref{corollary: main corollary HCP} from the standpoint of $\Padovani$: we prove that the process $P'\in \DeadlockFreedomHCPRRSet$  (obtained via commuting conversions and disentanglement) is also in $\ClassLP$ (\Cref{corollary: main result mupado}). In \Cref{sec: async} we present $\SPExpA$,  an asynchronous version of $\SPExp$, and show how to extend our results for $\SPExp$ to $\SPExpA$ (\Cref{c:transfer}). 

\Cref{fig:diagram} summarizes all technical ingredients, including those introduced in this section.

\subsection{The Type System \texorpdfstring{$\Padovani$ }{LP}}\label{sec: padovani}

We summarize the linear fragment of the type system $\Padovani$ by Padovani~\cite{padovani_linear_pi}, which  streamlines  Kobayashi's type system for deadlock freedom~\cite{Kobayashi06}. We assume sets of channels $a, b, \ldots$ and variables $x, y, \ldots$; then, names  $u, v, \ldots$ are either channels or variables. We find it convenient to use processes with polyadic communication: 
    \[P::= \OutputPado{u}{x_1,\dots,x_n}
     \Sep \InputPado{u}{x_1,\dots,x_n}P 
    \Sep P\para Q \Sep \new{a}P
    \Sep 
    \Inact
    \]

Two salient features are the output $\OutputPado{u}{x_1,\dots,x_n}$, which does not have a continuation, and `$\new{a}P$', the standard  restriction construct of the $\pi$-calculus.
Notions of free and bound names are as usual: in the process $\InputPado{u}{x_1,\dots,x_n}P $,   $x_1,\dots,x_n$ are bound in $P$; in $\new{a}P$,  $a$ is bound in $P$. We write $\FreeNamePado{P}$ and $\BoundNamePado{P}$ for the free and bound names of $P$, respectively.

The reduction semantics is closed under a structural congruence, denoted $\EquivPado$, which is standardly defined by the following axioms:
(i)~$P \para Q \EquivPado Q \para P$, 	
(ii)~$P \para \Inact \EquivPado P$,
(iii)~$P \para (Q \para R) \EquivPado (P \para Q) \para R$,
(iv)~$\new{x}P \para Q \EquivPado  \new{x}(P \para Q)$ (if $x \notin  \FreeNamePado{Q}$), and 
(v)~$\new{x}\new{y}  \EquivPado \new{y}\new{x}P$.

Reduction is then defined as
    \[\OutputPado{a}{v_1,\dots,v_n}\para \InputPado{a}{x_1,\dots,x_n}P\ReductionPado P\subst{v_1,\dots,v_n}{x_1,\dots,x_n}\]

\begin{example}\label{example:pado}

Differently from \SPExp,   channels in Padovani's framework  are used exactly once. Structured communications can of course be still expressed, using continuations.
Consider the process $P_{\ref{example:pado}}\DefEq\new{x}( Q_{\ref{example:pado}}\para R_{\ref{example:pado}})$ with 
\begin{align*}
  Q_{\ref{example:pado}}&\DefEq\new{a}\new{c}(\OutputPado{x}{a,c}\para\OutputPado{c}{a})& R_{\ref{example:pado}}&\DefEq \InputPado{x}{z_1,w}\InputPado{w}{z_2}(\new{r}\OutputPado{z_1}{r}\para\InputPado{z_2}{w}\Inact)  
\end{align*}
$Q_{\ref{example:pado}}$ sends the two endpoints of a channel: the first is sent along $x$ (together with a continuation channel $c$), and the other is sent along $c$. Channel $a$ occurs in both outputs, but these occurrences actually denote different endpoints; this difference will be justified by typing (cf. \Cref{exmaple: typing in LP}).
$R_{\ref{example:pado}}$ receives these endpoints to implement a local exchange of name $r$ along $a$.
\end{example} 

Padovani's type system ensures the \emph{absence of pending communications} in a process with respect to a channel.
To formalize this notion, we rely on reduction contexts, defined as $
\mathcal{C}  ::=   \HoleP  \Sep(\mathcal{C} \para P)  \Sep\new x \mathcal{C}$.
We may now define the following auxiliary predicates:
  \begin{align*} 
  \mathsf{out}(a, P)\ & \Leftrightarrow \ P \EquivPado \mathcal{C}[\OutputPado{a}{e}]\ \wedge\ a \notin \BoundNamePado{C}
  & 
    ~~  \mathsf{in}(a, P)\ & \Leftrightarrow   \ P \EquivPado \mathcal{C}[\InputPado{a}{x} Q]\ \wedge\ a \notin \BoundNamePado{C}
  \\
  \mathsf{sync}(a, P)\ & \Leftrightarrow  \ (\mathsf{in}(a, P) \wedge\ \mathsf{out}(a, P))
\end{align*}
Intuitively, 
	predicate $\mathsf{in}(a, P)$  holds if $a$ is free in $P$ and there is a subprocess of $P$ that can perform a linear input on $a$.
 Predicate $\mathsf{out}(a, P)$ holds if $a$ is free in $P$ and a subprocess of $P$ is waiting to send a value $v$. Predicate $\mathsf{sync}(a, P)$ denotes a pending input/output on $a$ for which a synchronization on $a$ is immediately available. 
 This way, to formalize a pending input/output for which a synchronization on $a$ is not immediately possible, we define:

\[  \mathsf{wait}(a, P)\ \Leftrightarrow  \ (\mathsf{in}(a, P)\ \vee\ \mathsf{out}(a, P))\ \wedge\ \neg \mathsf{sync}(a, P)
\] 
Thus, ensuring no pending communications in  $P$ means ensuring  
$\neg\mathsf{wait}(a, P)$  for every  $a$.  

\begin{example}[Predicates at work]
\label{exam:padopred}
	Consider the process $P_{\ref{exam:padopred}}\DefEq\InputPado{a}{x}\OutputPado{b}{v_1}\para \InputPado{b}{y}\OutputPado{a}{v_2} $, which has a circular dependency. 
	 We have: 
   \textup{(1)}~$ \mathsf{out}(a, P_{\ref{exam:padopred}})$ and $\mathsf{out}(b, P_{\ref{exam:padopred}})$. 
   \textup{(2)}~$\mathsf{in}(a, P_{\ref{exam:padopred}}) $ and $\mathsf{in}(b, P_{\ref{exam:padopred}})$. 
   \textup{(3)}~$\neg\mathsf{sync}(a,P_{\ref{exam:padopred}})$ and $\neg\mathsf{sync}(b,P_{\ref{exam:padopred}})$.
   \textup{(4)}~$\mathsf{wait}(a,P_{\ref{exam:padopred}}) $ and $\mathsf{wait}(b,P_{\ref{exam:padopred}})$.
\end{example}

\subparagraph{Type System}
To ensure absence of pending communications, $\Padovani$ relies on \emph{priorities}.
Let  $p,q,\ldots$ denote subsets of $\{?,!\}$ (\emph{polarities}).
Types for channels are defined  as follows: 
    \[t::= \TypePado{p}{t_1,\dots,t_m}{n}\]
    where $n\in\mathbb{Z}$ is the {priority} of a channel, which  indicates urgency for its use: the lower the number, the higher the urgency.
 We write \#  for  $\{?,!\}$. This way, \# and $\emptyset$ are even polarities.

\begin{example}[Priorities]\label{example: deadlock in pado}
To illustrate types and priorities, recall again the process
	$P_{\ref{exam:padopred}}\DefEq\InputPado{a}{x}\OutputPado{b}{v_1}\para \InputPado{b}{y}\OutputPado{a}{v_2}$.
	 As we have seen, we have $\mathsf{wait}(a, P_{\ref{exam:padopred}})$ and $\mathsf{wait}(b, P_{\ref{exam:padopred}})$.
	 
	     Let us assume that $x,y,v_1,v_2$ are all typed with some type $t$. By inspecting the left thread of $P_{\ref{exam:padopred}}$, we infer that $a$ and $b$ must be typed as $\TypePado{?}{t}{m}$ and 
    $\TypePado{!}{t}{n}$, respectively, for some $m,n$. 
    Then, by inspecting the right thread, we infer that   $a$ and $b$ must then be typed with $\TypePado{!}{t}{m}$ and $\TypePado{?}{t}{n}$, respectively. This way, their types in the composed process are $\TypePado{\#}{t}{m}$ (for $a$) and  $\TypePado{\#}{t}{n}$ (for $b$).
    Looking again at the left thread, as the input along $a$ is blocking the output along $b$, we must have $m < n$; in the right thread we have the exact opposite situation, thus $n < m$. These unsolvable inequalities reveal the circular dependency between $a$ and $b$, which will make $P_{\ref{exam:padopred}}$ not typable in the type system $\Padovani$.
\end{example}

 We now summarize how typing excludes pending communications.
 Type environments $\Gamma, \Gamma', \ldots$ are finite maps from channels to types.
    We write $dom(\Gamma)$ for the domain of $\Gamma$. We write $\Gamma,\Gamma'$ for the union of $\Gamma$ and $\Gamma'$ when $dom(\Gamma)\cap dom(\Gamma')=\emptyset$. We write $\PartialComposition$ to denote a partial composition operator on types defined as: $\TypePado{p}{s_1,\dots,s_m}{n}\PartialComposition \TypePado{q}{s_1,\dots,s_m}{n}=\TypePado{(p\cup q)}{s_1,\dots,s_m}{n}$ if $p\cap q =\emptyset$, and undefined otherwise. For type environments we have: 
         $\Gamma \PartialComposition \Gamma'\DefEq\Gamma,\Gamma'$ if $dom(\Gamma)\cap dom(\Gamma')=\emptyset$, and
         $(\Gamma,u:t)\PartialComposition (\Gamma',u:s)\DefEq (\Gamma\PartialComposition\Gamma'),u:t\PartialComposition s$. 
Let $\mathbb{Z}_{\top}=\mathbb{Z}\cup\{\top\}$, where $ n< \top$ for every $n\in \mathbb{Z}$. 
    The functions $\PriorityFunction{\cdot}$, from types to $\mathbb{Z}_{\top}$, and  $\Shift{h}{\empty}$ from types to types are defined as follows:
    
\begin{tabular}{ll}
  \begin{minipage}{7cm}
     \[ \PriorityFunction{t} \DefEq
  \begin{cases}
    n      & \quad \text{if } t=\TypePado{p}{s}{n} \text{and } p\neq \emptyset\\
    \top & \quad \text{otherwise} 
  \end{cases}\] 
  \end{minipage}
  &
 \hspace*{-1.4cm}
  \begin{minipage}{7cm}
      \[\Shift{h}{t} \DefEq
  \begin{cases}
    \TypePado{p}{s}{n+h}     & \quad \text{if } t=\TypePado{p}{s}{n} \text{and } p\neq \emptyset\\
    t & \quad \text{otherwise} 
  \end{cases}\]
  \end{minipage}
\smallskip
\end{tabular}\\
Intuitively, the  function $\Shift{h}{\empty}$ allows us to shift priorities. 
    We say that a type $t$ is \emph{unlimited}, written $\UnrestrictedPado{t}$, iff $\PriorityFunction{t}=\top$.
    We write $\UnrestrictedPado{\Gamma}$ iff all the types in the range of $\Gamma$ are unlimited.

We are now ready to give the typing rules, considering only processes with biadic communication, for simplicity. The rules for monadic communication are as expected.

\begin{definition}[$\Padovani$]\label{def: type system LP}
    The typing rules for processes 
    use judgements of the form $\Gamma \PadoJud P$  and $\Gamma \PadoJud \widetilde{e}:\widetilde{t}$ and  
    are presented in \Cref{fig:type system padovani}.  We shall write  $P \in \Padovani$ if $\Gamma \PadoJud P$, for some $\Gamma$.
\end{definition}
 Rule \textsc{T-In} is used to type inputs $\InputPado{u}{x_1,x_2}P$, where $u$ has type $\InputTypePado{t_1,t_2}{n}$. The condition $n<\PriorityFunction{\Gamma}$
 ensures that $u$ has been assigned the highest urgency; accordingly, the types of $x$ and $y$ are shifted by $n$.
 Rule \textsc{T-Out}, used to type outputs on channels of type $\OutputTypePado{t_1,t_2}{n}$, follows a similar principle: the types of $x$ and $y$ are shifted by $n$, since they must have lower urgency than $u$.
 Rule \textsc{T-New} is used for restricting new channels with full (\#) or empty ($\emptyset$) polarity.

\begin{example}\label{exmaple: typing in LP}
Recall processes $Q_{\ref{example:pado}}$ and $R_{\ref{example:pado}}$ from \Cref{example:pado}. 
Consider types $s'=\Shift{1}{\OutputTypePado{\TypePado{\emptyset}{\empty}{4}}{2}}$ and $s=\Shift{2}{\InputTypePado{\TypePado{\emptyset}{\empty}{4}}{1}}$. 
Note that $s'\PartialComposition s=\TypePado{\#}{\TypePado{\emptyset}{\empty}{4}}{3}$. 
Process $Q_{\ref{example:pado}}$ is typed as follows:
   \begin{prooftree}
       \AxiomC{$a:s',c:\Shift{1}{\InputTypePado{s}{1}}\PadoJud a:s',c:\Shift{1}{\InputTypePado{s}{1}}$}
       \RightLabel{\textsc{T-Out}}
       \UnaryInfC{$a:s',c:\Shift{1}{\InputTypePado{s}{1}},x:\OutputTypePado{s',\Shift{1}{\InputTypePado{s}{1}}}{1}\PadoJud \OutputPado{x}{a,c}$}
           \AxiomC{$a:s\PadoJud a:s$}
           \RightLabel{\textsc{T-Out}}
       \UnaryInfC{$a:s,c:\OutputTypePado{s}{2}\PadoJud \OutputPado{c}{a}$}
       \RightLabel{\textsc{T-Par}}
       \BinaryInfC{$a:\TypePado{\#}{\TypePado{\emptyset}{\empty}{4}}{3},x:\OutputTypePado{s',\Shift{1}{\InputTypePado{s}{1}}}{1},c:\TypePado{\#}{s}{2}\PadoJud\OutputPado{x}{a,c}\para\OutputPado{c}{a}$}
       \RightLabel{\textsc{T-New}}
       \UnaryInfC{$a:\TypePado{\#}{\TypePado{\emptyset}{\empty}{4}}{3},x:\OutputTypePado{s',\Shift{1}{\InputTypePado{s}{1}}}{1}\PadoJud\new{c}(\OutputPado{x}{a,c}\para\OutputPado{c}{a})$}
       \RightLabel{\textsc{T-New}}
       \UnaryInfC{$x:\OutputTypePado{s',\Shift{1}{\InputTypePado{s}{1}}}{1}\PadoJud\new{a}\new{c}(\OutputPado{x}{a,c}\para\OutputPado{c}{a})$}
   \end{prooftree} 
   Similarly we can check that $x:\InputTypePado{s',\Shift{1}{\InputTypePado{s}{1}}}{1}\PadoJud R_{\ref{example:pado}}$. Note that $x$ is typed with opposite polarities in$Q_{\ref{example:pado}}$ and $R_{\ref{example:pado}}$; thus by applying Rules \textsc{T-Par} and \textsc{T-New} we obtain  $\PadoJud P_{\ref{example:pado}}$.
\end{example}

  \begin{figure}[!t]  
  \centering{
    \footnotesize{
    {
\begin{mathpar}
    \inferrule[T-Idle]{\UnrestrictedPado{\Gamma}}{\Gamma\PadoJud \Inact}
     \and
    \inferrule[T-Par]{\Gamma_1\PadoJud P_1\quad \Gamma_2\PadoJud P_2}{\Gamma_1\PartialComposition \Gamma_2\PadoJud P_1\para P_2}
    \and 
    \inferrule[T-out]{\Gamma,e_1:\Shift{n}{t_1},e_2:\Shift{n}{t_2}\PadoJud e_1:\Shift{n}{t_1},e_2:\Shift{n}{t_2}\quad \UnrestrictedPado{\Gamma}}{\Gamma\PartialComposition u: \OutputTypePado{t_1,t_2}{n}\PadoJud \OutputPado{u}{e_1,e_2}}
       \\
         \inferrule[T-In]{\Gamma,x:\Shift{n}{t_1}, y:\Shift{n}{t_2}\PadoJud P\qquad n<|\Gamma|}
    {\Gamma\PartialComposition u:\InputTypePado{t_1,t_2}{n}\PadoJud \InputPado{u}{x,y}P}
    \and 
    \inferrule[T-New]{\Gamma, a:\TypePado{p}{t}{n}\PadoJud P\qquad \text{$p$ even}}{\Gamma\PadoJud \new{a}P}
\end{mathpar}
  }
    }
    }
  \vspace*{-0.5cm}
    \caption{$\Padovani$: Typing rules for asynchronous processes. 
    }
    \label{fig:type system padovani}
    \vspace*{-0.3cm}
  \end{figure}
  
We may now characterize  deadlock freedom.
Let us write $\ReductionPado^*$ to denote the reflexive, transitive closure of $\ReductionPado$. Also, write $P \longarrownot\ReductionPado$ if there is no $Q$ such that $P  \ReductionPado Q$. We have:

\begin{definition}[$\Padovani$: Deadlock Freedom \cite{padovani_linear_pi}]\label{def: deadlock freedompadovani body}
We say that $P\in \Padovani$ is deadlock free, written $\DeadlockFreedomPado{P}$, if for every $Q$ such that $P\ReductionPadoClo \new{\tilde{a}}Q\longarrownot\ReductionPado$ we have $\neg\mathsf{wait}(a, Q)$ for every $a$. 
\end{definition}

We have that  processes typed in the empty environment are deadlock free:
  \begin{theorem}[$\Padovani$: Deadlock Freedom~\cite{padovani_linear_pi}]\label{theorem: deadlock freedom padovani}
      If $~\PadoJud P$, then $\DeadlockFreedomPado{P}$.
  \end{theorem}

\subsection{Relating \texorpdfstring{$\ClassLP$}{P} and \texorpdfstring{$\ClassL$}{L}}
\label{sec: mupadovani}

We now
consider $\MuPadovani$ (`micro $\Padovani$'), which is a fragment of $\Padovani$ that is similar to $\CP$.
 We also define the encoding $\SPExptoPado{f}{\cdot}:\SPExp\rightarrow\Padovani$, and $\ClassLP$ and $\ClassMuP$ the classes of \SPExp processes that are captured by  $\Padovani$ and $\MuPadovani$, respectively. These are all ingredients for our results  in \Cref{sec: comparing H and MuLP}.
\begin{definition}[$\MuPadovani$]\label{def: type system MuLP}
    The typing rules for processes are presented in \Cref{fig:type system mu padovani}. A judgement $\Gamma \MuPadoJud P$ denotes that $P$ is well typed in $\Gamma$. We shall write  $P \in \MuPadovani$ if $\Gamma \MuPadoJud P$, for some $\Gamma$.  

\end{definition}
  \begin{figure}[!t]  

    \footnotesize{
    
\begin{mathpar}
    \inferrule[M-Idle]{\UnrestrictedPado{\Gamma}}{\Gamma\MuPadoJud\Inact}
    \and
     \inferrule[M-Out]{\Gamma,e_1:\Shift{n}{t_1},e_2:\Shift{n}{t_2}\MuPadoJud e_1:\Shift{n}{t_1},e_2:\Shift{n}{t_2}\quad \UnrestrictedPado{\Gamma}}{\Gamma\PartialComposition u: \OutputTypePado{t_1,t_2}{n}\MuPadoJud \OutputPado{u}{e_1,e_2}}
    \and
    \inferrule[M-In]{\Gamma,x:\Shift{n}{t_1}, y:\Shift{n}{t_2}\MuPadoJud P\qquad n<|\Gamma|}
    {\Gamma\PartialComposition u:\InputTypePado{t_1,t_2}{n}\MuPadoJud \InputPado{u}{x,y}P}
           \and 
    \inferrule[M-New-Par]{\Gamma_1,a:\TypePado{p}{t}{n}\MuPadoJud P\qquad 
    \Gamma_2, a:\TypePado{q}{t}{n}\MuPadoJud Q\qquad dom(\Gamma_1)\cap dom(\Gamma_2)=\emptyset,\,p\cup q \text{ even}}
    {\Gamma_1\PartialComposition \Gamma_2 \MuPadoJud \new{a}(P\para Q)}\\
     \vspace*{-0.7cm}
   \end{mathpar}
    }\vspace*{-0.5cm}
    \caption{$\MuPadovani$: Modified typing rules for asynchronous processes. 
    }
    \label{fig:type system mu padovani}
     \vspace*{-0.3cm}
  \end{figure}

   $\MuPadovani$ results from $\Padovani$ by replacing Rules \textsc{T-Par} and \textsc{T-New} in \Cref{fig:type system padovani} with Rule \textsc{M-New-Par},  which combines restriction and parallel  just as Rule $\C$\textsc{Cut} in \Cref{fig:CP Rules}.  
\begin{lemmaapxrep}\label{lemma: mupadovani in padovani}
   If $\Gamma \MuPadoJud P$ then  $\Gamma \PadoJud P$. 
\end{lemmaapxrep}
\begin{proof}
By induction on the type derivation, with a case analysis in the last rule applied.
We only consider two cases.
\begin{itemize}
	\item Rule~\textsc{M-Idle}: This case is immediate: if $\Gamma\MuPadoJud\Inact$ then $\Gamma\PadoJud\Inact$.
	\item Rule \textsc{M-New-Par}: This is the only interesting case. We have:
	\begin{prooftree}
	\AxiomC{$\Gamma_1,a:\TypePado{p}{t}{n}\MuPadoJud P$}
	\AxiomC{$\Gamma_2, a:\TypePado{q}{t}{n}\MuPadoJud Q$} 
		\AxiomC{ $\Gamma_1\cap \Gamma_2=\emptyset,\,p\cup q \text{ even}$}
		\TrinaryInfC{$ {\Gamma_1\PartialComposition \Gamma_2 \MuPadoJud \new{a}(P\para Q)}$}
	\end{prooftree}
	By IH, $\Gamma_1,a:\TypePado{p}{t}{n}\PadoJud P$ and $\Gamma_2, a:\TypePado{q}{t}{n}\PadoJud Q$. The proof concludes by applying Rules \textsc{T-Par} and \textsc{T-New} from \Cref{fig:type system padovani} in sequence. 
\end{itemize}
\end{proof}

\begin{corollary}
    If $\MuPadoJud P$, then $\DeadlockFreedomPado{P}$.
\end{corollary}
\begin{proof}
    It follows by \Cref{lemma: mupadovani in padovani,theorem: deadlock freedom padovani}.
\end{proof}

We now define an   encoding from $\SPExp$ to $\Padovani$, following a continuation-passing style (cf.~\cite{Dardha:2012:STR:2370776.2370794}). Let $f$  be a function from names to names. We use the following conventions: $f_x$ stands for $f(x)$, and $f,\{x\mapsto c\}$ stands for the function $f'$ s.t. $f'(x)=c$ and $f'(y)=f(y)$ if $x\neq y$.
\begin{definition}\label{def: encoding SPExp to Padovani}
The (partial) encoding $\SPExptoPado{f}{\cdot}: \SPExp \to \Padovani$  is parameterized by a function, ranging over $f,g$, from names to names, and is defined for processes as follows:
    \begin{align*}
    \SPExptoPado{f}{\OutputSP{x}{v}P}&\DefEq \new{m}(\OutputPado{f_x}{f_v,m}\para \InputPado{m}{k}\new{c}(\OutputPado{k}{c}\para \SPExptoPado{f,\{x\mapsto c\}}{P}))\\
    \SPExptoPado{f}{\InputSP{x}{z}P}&\DefEq \InputPado{f_x}{z,m}\new{a}(\OutputPado{m}{a}\para \InputPado{a}{c}\SPExptoPado{f,\{x\mapsto c\}}{P})
    \\
    \SPExptoPado{f}{\EmptyOutputSP{x}\Inact}&\DefEq \new{r}\OutputPado{f_x}{r}
    \\
    \SPExptoPado{f}{\EmptyInputSP{x}P}&\DefEq \InputPado{f_x}{r}\SPExptoPado{f}{P} \qquad \text{(with $r \notin \mathtt{fv}(P)\cup \mathtt{bv}(P)$)}
\\   
\SPExptoPado{f}{\news{x}{y}P}&\DefEq \new{d}\SPExptoPado{f,\{x,y\mapsto d\}}{P}
    \end{align*}
    and as a homomorphism for $P\para Q$ and $\Inact$.
  The encoding of session types is as follows:
    \begin{align*}
     \SPExptoPadoType{\EndTypeI}&\DefEq\TypePado{?}{\TypePado{\emptyset}{\empty}{\empty}}{n}
     &
          \SPExptoPadoType{\LinOutputST{S}{T}}&\DefEq \TypePado{?}{\SPExptoPadoType{T},\TypePado{!}{\TypePado{!}{\SPExptoPadoType{S}}{i}}{j}}{n}
\\
      \SPExptoPadoType{\EndTypeO}&\DefEq\TypePado{!}{\TypePado{\emptyset}{\empty}{\empty}}{n} & 
     \SPExptoPadoType{\LinInputST{S}{T}}&\DefEq \TypePado{!}{\SPExptoPadoType{T},\TypePado{!}{\TypePado{!}{\SPExptoPadoType{\DualSPExp{S}}}{i}}{j}}{n}
\end{align*}
Typing environments are encoded inductively as $\SPExptoPado{f}{\emptyset}\DefEq\emptyset$ and  $\SPExptoPado{f}{\Gamma,x:T}\DefEq\SPExptoPado{f}{\Gamma}, f_x:\SPExptoPadoType{T}$. The values of $i,j$ and $n$ are calculated when checking in $\MuPadovani$ the typablity of $\SPExptoPado{f}{\Gamma}\MuPadoJud\SPExptoPado{f}{P}$. 
\end{definition}

Intuitively, the encoding of processes in \Cref{def: encoding SPExp to Padovani} collapses the two endpoints of a session into a single name in $\Padovani$. The encoding of $\OutputSP{x}{v}P$ mimics both the sequential structure of a session and  synchronous communication.   $\SPExptoPado{f}{\OutputSP{x}{v}P}$ sends the (renamed) value $f_v$ and a new channel $m$ along which an input $\InputPado{m}{k}$ blocks the continuation $\SPExptoPado{f,\{x\mapsto c\}}{P}$, and receives the channel along which the the continuation $c$ of the session $x$ will be sent.
 Dually,  $\SPExptoPado{f}{\InputSP{x}{z}P}$ receives the intended value and the channel $m$, which sends the channel $a$, along which the continuation of the session $x$ will be received. 
 Finally, the encoding is partial because $\SPExptoPado{f}{\EmptyOutputSP{x}P}$ is only defined when $ P = \Inact$: this allows us to characterize the session processes captured in $\CP$,  where the empty output has no continuation (see \Cref{def: CP}).

\begin{example}\label{example: operational correspondence SP Pado}
Consider the process $P_{\ref{example: operational correspondence SP Pado}}\DefEq\news{x}{y}(\OutputSP{x}{v}R\para \InputSP{y}{z}Q)$ in $\SPExp$.
  We show $\SPExptoPado{f}{P_{\ref{example: operational correspondence SP Pado}}}$ and illustrate its reductions.
    Let us assume $P_{\ref{example: operational correspondence SP Pado}}$ is well-typed. Because of linearity of $x$ and $y$, they only appear in the left and right subprocess, respectively. Let $f=\{x,y\mapsto d\}$ and $g_1=\{x\mapsto c\}$,$g_2=\{y\mapsto c'\}$, then we have:
\begin{align*}
    \SPExptoPado{\emptyset}{P_{\ref{example: operational correspondence SP Pado}}} & = \new{d}(\SPExptoPado{f}{\OutputSP{x}{v}R}\para \SPExptoPado{f}{\InputSP{y}{z}Q})\\
    & =\new{d}( \new{m}(\OutputPado{d}{v,m}\para \InputPado{m}{k}\new{c}(\OutputPado{k}{c}\para \SPExptoPado{g_1}{R}))
    \\
    &\quad \phantom{=\new{d}}\para  \InputPado{d}{z,m'}\new{a}(\OutputPado{m'}{a}\para \InputPado{a}{c'}\SPExptoPado{g_2}{Q}))
    \\
    & \ReductionPado \new{m}(\InputPado{m}{k}\new{c}(\OutputPado{k}{c}\para \SPExptoPado{g_1}{R}) \para  \new{a}(\OutputPado{m}{a}\para \InputPado{a}{c'}\SPExptoPado{g_2}{Q}\subst{v}{z}))
    \\
    &  \ReductionPado \new{a}( \new{c}(\OutputPado{a}{c}\para \SPExptoPado{g_1}{R})\para \InputPado{a}{c'}\SPExptoPado{g_2}{Q}\subst{v}{z})\\
     & \ReductionPado \new{c}(\SPExptoPado{g_1}{R}\para \SPExptoPado{g_2}{Q}\BiSubstitution{v}{z}{c}{c'})
\end{align*}
\end{example}

Our main result concerns the classes of processes associated to $\Padovani$ and $\MuPadovani$, which we define by mirroring the definition of $\ClassL$ and $\ClassH$ (\Cref{Def: ClassL and ClassH}):
\begin{definition}[$\ClassLP$ and  $\ClassMuP$]\label{def: class muP} 
	The classes of processes $\ClassLP$ and $\ClassMuP$ are defined as follows:
	\begin{align*}
		\ClassLP&\DefEq\{P\in \text{$\SPExp$} \Sep \Gamma \SPExpJud P\,\wedge\,\SPExptoPado{f}{\Gamma}\PadoJud\SPExptoPado{f}{P} \} \qquad 
		\ClassMuP \DefEq\{P\in \text{$\SPExp$} \Sep \Gamma \SPExpJud P\,\wedge\,\SPExptoPado{f}{\Gamma}\MuPadoJud\SPExptoPado{f}{P} \} \label{class muLP}
	\end{align*}
\end{definition}

The following result establishes an operational correspondence between a process and its encoding, what will allows us to capture deadlock free session processes via the encoding.
\begin{theorem}[Operational Correspondence for $\SPExptoPado{f}{\cdot}$]\label{theorem: operational correspondence SP and pado}
    Let $P\in \ClassMuP$. We have:

    \begin{enumerate}
        \item 
        \begin{enumerate}
            \item If $P\equiv \new{\widetilde{xy}}(\EmptyOutputSP{x}Q_1\para\EmptyInputSP{y}Q_2\para S )$, then $\exists Q\in \Padovani$ such that $\SPExptoPado{f}{P}\ReductionPado Q$ and 
        $Q\EquivPado \SPExptoPado{f}{\new{\widetilde{xy}}(Q_1\para Q_2\para S )}$.
        \item If $P\equiv \new{\widetilde{xy}}(\OutputSP{x}{v}Q_1\para \InputSP{y}{z}Q_2\para S)$, then $\exists Q\in \Padovani$ such that $\SPExptoPado{f}{P}\ReductionPado^3 Q$ and 
        $Q\EquivPado \SPExptoPado{f}{\new{\widetilde{xy}}(Q_1\para Q_2\subst{v}{z}\para S)}$. 
        \end{enumerate}
        \item If $\SPExptoPado{f}{P}\ReductionPado R$, then either: 
        \begin{enumerate}
        \item  $\exists P'\in\SPExp$ such that $P\ReductionSP P'$ and $R\EquivPado \SPExptoPado{f}{P'}$, or
            \item   $\exists Q\in\Padovani$, and $P'\in\SPExp$ such that $R\ReductionPado^2 Q$, $P\ReductionSP P'$ and $Q\EquivPado \SPExptoPado{f}{P'}$.
        \end{enumerate}
    
    \end{enumerate}

\end{theorem}
\begin{proof}
 By assumption, $P\in\ClassMuP$ and so we have that $\Gamma\SPExpJud P$, for some $\Gamma$, and $\SPExptoPado{f}{\Gamma}\MuPadoJud\SPExptoPado{f}{P}$ is defined. Moreover, by \Cref{corollary: wellformedness SP} $P$ is well-formed. 
 We consider each item separately. 
 Item~1(b) follows by \Cref{def: encoding SPExp to Padovani} and the rest of the analysis proceeds as in \Cref{example: operational correspondence SP Pado}; Item 1(a) is similar.
 Item~(2): If $\SPExptoPado{f}{P}\ReductionPado R$, then $\SPExptoPado{f}{P}\EquivPado\new{\tilde{a}}(\OutputPado{d}{\tilde{v}}\para \InputPado{d}{\tilde{x}}K\para S)$. By typability of $P$ and \Cref{def: encoding SPExp to Padovani}  one of the two possibilities apply:

 \begin{enumerate}
 	\item[i] $P\equiv \new{\widetilde{xy}}(\EmptyOutputSP{x}\Inact\para 
\EmptyInputSP{y}K'\para S')$, $\SPExptoPado{f}{S'}=S$, $\SPExptoPado{f}{K'}=K$ and $f=f',\{x,y\mapsto d\}$,
\item[ii] $P\equiv \new{\widetilde{xy}}(\OutputSP{x}{v}S_1\para \InputSP{y}{z}K'\para S_2)$ and $f=f',\{x,y\mapsto d\}$
 \end{enumerate}
  \sloppy In case (i)  $\SPExptoPado{f}{P}\EquivPado\new{\tilde{a}}(\new{r}\OutputPado{d}{r}\para \InputPado{d}{r'}K\para S)\ReductionPado\new{\tilde{a}}(K\para S)$  
  and $P\equiv \new{\widetilde{xy}}(\EmptyOutputSP{x}\Inact\para 
\EmptyInputSP{y}K'\para S')\ReductionSP \new{\widetilde{xy}'}( 
K'\para S'') $, which verifies case (2a). In case (ii) we proceed as in \Cref{example: operational correspondence SP Pado}, which verifies case (2b).
\end{proof}

As a consequence of \Cref{theorem: operational correspondence SP and pado} we have the following result, which shows the relation between deadlock free processes in $\ClassMuP$.

\begin{corollary}\label{corollary: deadlock freedom in pado and SP}
     Let $P\in \ClassMuP$. $\DeadlockFreedomSP{P}$ iff $\DeadlockFreedomPado{\SPExptoPado{f}{P}}$.
\end{corollary}
\begin{proof}
For each direction we prove its contrapositive. 
We discuss only the right-to-left direction; the analysis for the left-to-right direction is similar.
By \Cref{def: deadlock freedompadovani body}, there exist $R$ and $a$ s.t. $\SPExptoPado{f}{P}\ReductionPadoClo \new{\tilde{a}}R\longarrownot\ReductionPado$ and $\mathsf{wait}(a,R)$. If $\mathsf{wait}(a,R)$ holds because $\mathsf{in}(a,R)$ then 
$R\EquivPado \new{\tilde{a}}(\InputPado{a}{x}Q\para S)$.
 By \Cref{theorem: operational correspondence SP and pado,def: encoding SPExp to Padovani}  there exists $P'\equiv \new{\widetilde{xy}}(\InputSP{z}{w}Q'\para S')$ s.t. $P\ReductionSP^*P'$ and $R\EquivPado \SPExptoPado{f}{P'}$. Note that $P'\longarrownot\ReductionSP$, since otherwise by \Cref{theorem: operational correspondence SP and pado}(1), $\new{\tilde{a}}R\ReductionPado R''$ for some process $R''$, contradicting our assumption, thus $\neg\DeadlockFreedomSP{P}$. The case when $\mathsf{wait}(a,R)$ holds because $\mathsf{out}(a,R)$ proceeds similarly.
\end{proof}

\Cref{corollary: deadlock freedom in pado and SP} allows us to detect DF session processes via the encoding.

\subsection{Comparing \texorpdfstring{$\ClassH$}{H} and \texorpdfstring{$\ClassLP$}{P}}
\label{sec: comparing H and MuLP}

From \Cref{sec: results} we infer that  a process $P\in\ClassH$, typed as $\SPExptoHCP{P}\HCPJud \Gamma_1\SequentPara \cdots\SequentPara\Gamma_n$, can be decomposed into processes $P_1\CPJud \Gamma_1$, $\ldots$, $P_n\CPJud \Gamma_n$. Moreover, by \Cref{lemma: disentanglement in SP} (disentanglement in $\SPExp$) we infer that processes $P_1, \ldots, P_n$ have a pre-image in $\SPExp$, therefore $\forall i\in[1,n]$ $P_i\in\ClassL\subsetneq\ClassH$.

As an important preliminary step, we first prove that $\ClassL = \ClassMuP$ (\Cref{corollary: L equal to mupadovani}). 
We split the proof in two separate inclusions. 
The following lemma is useful to prove that $\ClassL\subseteq\ClassMuP$.

\begin{lemma}\label{lemma: big lemma}
    Let $\Gamma$ and $P$ be an environment and a process in $\Padovani$, resp. The following hold:
    \begin{enumerate}
    \item   If $\Gamma \PadoJud P$, then $\Shift{n}{\Gamma}\PadoJud P$.
     \item   If $\Gamma \MuPadoJud P$, then $\Shift{n}{\Gamma}\MuPadoJud P$.
    \item  Let $P\in \Padovani$. If $\Gamma\MuPadoJud P$ and $\UnrestrictedPado{T}$, then $\Gamma,x:T \MuPadoJud P$, with $x$ fresh wrt $P$.

        \item   Let $\Gamma_1,\dots, \Gamma_n$ be typing environments in $\SPExp$, then 
    $\SPExptoPado{f}{\Gamma_1,\cdots,\Gamma_n}=\SPExptoPado{f}{\Gamma_1}\PartialComposition\cdots\PartialComposition\SPExptoPado{f}{\Gamma_n}$, 
    for some renaming function $f$.
    \item Let $T$ be a session type. Then $\SPExptoPado{f}{T}\PartialComposition\Shift{m}{\SPExptoPado{f}{\DualSPExp{T}}}=\TypePado{p}{T'}{n}$,  for some $m\geq 0$, $n\in\mathbb{Z}$, $T'$ a type in $\Padovani$, and $p$ even.
\end{enumerate}
\end{lemma}
\begin{proof}
    \textup{(1)}~Proven in \cite{padovani_linear_pi}. \textup{(2)}~ Follows from \Cref{lemma: mupadovani in padovani} and item 1.
    \textup{(3)}~Follows by induction on the type derivation. (\Cref{fig:type system mu padovani}).
    \textup{(4)}~ Follows immediately by linearity of names in $\Gamma_i$, for $i\in[1,n]$.
    \textup{(5)}~Follows by induction on the structure of $T$.
\end{proof}

\begin{theoremapxrep}\label{theorem: L equal to mupadovani}
    $\ClassL\subseteq\ClassMuP$.
\end{theoremapxrep}

\begin{proof}
    By induction on the type derivation $\Gamma\SPExpJud P$ with a case analysis in the last rule applied.
    \begin{itemize}
    \item \textsc{Inact}. It is immediate.
        \item \textsc{Res}. 
    By assumption, \Cref{def: encoding SPexp to HCP,Def: ClassL and ClassH},
 we have that $\SPExptoHCP{\news{x}{y}P'}\CPJud \SPExptoHCP{\Gamma}$, and there exist $P_1$, $P_2$, $\Gamma_1$, and $\Gamma_2$ s.t. $\Gamma=\Gamma_1,\Gamma_2$,  $dom(\Gamma_1)\cap dom(\Gamma_2)=\emptyset$, and  $P'=P_1\para P_2$ s.t:
    \begin{prooftree}
        \AxiomC{$\SPExptoHCP{P_1}\CPJud \SPExptoHCP{\Gamma_1},x:\SPExptoHCP{T}$}
        \AxiomC{$\SPExptoHCP{P_2}\CPJud \SPExptoHCP{\Gamma_1},y:\SPExptoHCP{\DualSPExp{T}}$}
        \RightLabel{\C\textsc{Cut}}
        \BinaryInfC{$\news{x}{y}(\SPExptoHCP{P_1}\para\SPExptoHCP{P_2})\CPJud\SPExptoHCP{\Gamma}$}
    \end{prooftree}
   and we want to prove that $\SPExptoPado{f}{\Gamma_1,\Gamma_2}\MuPadoJud\SPExptoPado{f}{\news{x}{y}P'}$.
    By IH we have that 
   \[
        \SPExptoPado{f}{\Gamma_1},d:\SPExptoPadoType{T}\MuPadoJud\SPExptoPado{f}{P_1}\quad  
        \text{and}
   \quad \SPExptoPado{f}{\Gamma_2},d:\SPExptoPadoType{\DualSPExp{T}}\MuPadoJud\SPExptoPado{f}{P_2} 
    \]
    where $f=f',\{x,y\mapsto d\}$.
    By \Cref{lemma: big lemma}(5), there exist $m$ s.t. $\SPExptoPado{f}{T}\PartialComposition\Shift{m}{\SPExptoPado{f}{\DualSPExp{T}}}=\TypePado{p}{T'}{n}$, for  an even polarity $p$. 
    By \Cref{lemma: big lemma}(2), we have   $\Shift{m}{\SPExptoPado{f}{\Gamma_2}},d:\Shift{m}{\SPExptoPadoType{\DualSPExp{T}}}\MuPadoJud\SPExptoPado{f}{P_2}$. Then, applying Rule \textsc{M-New-Par}, we have $\SPExptoPado{f}{\Gamma_1}\PartialComposition\Shift{m}{\SPExptoPado{f}{\Gamma_2}}\MuPadoJud\new{d}(\SPExptoPado{f}{P_1}\para \SPExptoPado{f}{P_2}$) holds. By \Cref{def: encoding SPExp to Padovani} the priorities of types are to be assigned; thus we can assume that $\SPExptoPado{f}{\Gamma_1}\PartialComposition{\SPExptoPado{f}{\Gamma_2}}\MuPadoJud\new{d}(\SPExptoPado{f}{P_1}\para \SPExptoPado{f}{P_2}$) and by \Cref{lemma: big lemma}(4) and \Cref{def: encoding SPExp to Padovani} we obtain $\SPExptoPado{f}{\Gamma_1,\Gamma_2}\MuPadoJud\SPExptoPado{f}{\news{x}{y}P'}$,  which concludes this case.

\item \textsc{EInput}.   $\Gamma, x:\EndTypeI\SPExpJud \EmptyInputSP{x}P'$ is derived as
\begin{prooftree}
    \AxiomC{$\Gamma\SPExpJud P'$}
    \UnaryInfC{$\Gamma, x:\EndTypeI\SPExpJud \EmptyInputSP{x}P'$}
\end{prooftree}
By assumption and \Cref{def: encoding SPexp to HCP,Def: ClassL and ClassH},
 we have that \[\SPExptoHCP{\EmptyInputSP{x}P'}\CPJud \SPExptoHCP{\Gamma,x:\EndTypeI}\]

 we want to prove
 \[\SPExptoPado{f}{\Gamma,x:\EndTypeI}\MuPadoJud\SPExptoPado{f}{\EmptyInputSP{x}P'}\]
 By IH $\SPExptoPado{f}{\Gamma}\MuPadoJud\SPExptoPado{f}{P'}$. By \Cref{lemma: big lemma}(3), we have that 
 $\SPExptoPado{f}{\Gamma},r:\Shift{n}{\TypePado{\emptyset}{\empty}{\empty}}\MuPadoJud\SPExptoPado{f}{P'}$, for some $n$ s.t.
 $n<\PriorityFunction{\SPExptoPado{f}{\Gamma}}$. Thus, applying Rule~\textsc{M-Input}, we obtain:
 \begin{prooftree}
     \AxiomC{ $\SPExptoPado{f}{\Gamma},r:\Shift{n}{\TypePado{\emptyset}{\empty}{\empty}}\MuPadoJud\SPExptoPado{f}{P'}\quad n<\PriorityFunction{\SPExptoPado{f}{\Gamma}}$}
     \RightLabel{\textsc{M-Input}}
     \UnaryInfC{$\SPExptoPado{f}{\Gamma},f_x:\TypePado{?}{\Shift{n}{\TypePado{\emptyset}{\empty}{\empty}}}{n}\MuPadoJud \InputPado{f_x}{r}\SPExptoPado{f}{P'}$}
 \end{prooftree}

 Finally, by \Cref{def: encoding SPExp to Padovani} we obtain $\SPExptoPado{f}{\Gamma,x:\EndTypeI}\MuPadoJud\SPExptoPado{f}{\EmptyInputSP{x}P'}$.

\item \textsc{EOutput}, \textsc{Out} and \textsc{In} follow similarly, using \Cref{lemma: big lemma}(3)  when $x$ has no continuation, since $c$ is typed with the empty polarity.

    \end{itemize}
    \end{proof}

To prove the other direction we need the following result.

\begin{lemma}\label{lemma: 63}
\label{lemma: duality session types and HCP}
    Let $S$ and $T$ be session types.  We have:
    	\textup{(1)}~$\SPExptoPado{f}{S}\PartialComposition\SPExptoPado{f}{T}$ is defined iff $T=\DualSPExp{S}$.
\textup{(2)}~$\SPExptoHCP{\DualSPExp{S}}=\LinearDual{\SPExptoHCP{S}}$.
\end{lemma}
\begin{proof}
 Item (1) follows by \Cref{def: encoding SPExp to Padovani} and def. of $\PartialComposition$.
Item (2) follows immediately by induction in the structure of $S$ and \Cref{def: encoding SPexp to HCP}.
\end{proof}

\begin{theoremapxrep}\label{theorem: mupadovani in L}
    $\ClassMuP\subseteq \ClassL$.
\end{theoremapxrep}

\begin{proof}
By induction on the type derivation  of $\Gamma\SPExpJud P$ with a case analysis in the last rule applied. The interesting case is  \textsc{Res}, i.e, $\Gamma\SPExpJud \news{x}{y}P$. By the assumption $P\in\ClassMuP$, we have that $\SPExptoPado{f}{\Gamma}\MuPadoJud\SPExptoPado{f}{\news{x}{y}P}$, therefore there exist $\Gamma_1,\Gamma_2$, $P_1,P_2$ s.t.  $\Gamma=\Gamma_1,\Gamma_2$,  $dom(\Gamma_1)\cap dom(\Gamma_2)=\emptyset$,  $P'=P_1\para P_2$, $\SPExptoPado{f}{\Gamma_1,x:T}\MuPadoJud \SPExptoPado{f}{P_1}$ and $\SPExptoPado{f}{\Gamma_1,y:\DualSPExp{T}}\MuPadoJud \SPExptoPado{f}{P_2}$. Thus $\Gamma\SPExpJud \news{x}{y}P$ can be split as  $\Gamma_1,x:T\SPExpJud P_1$ and $\Gamma_2,y:\DualSPExp{T}\SPExpJud P_2$. \Cref{lemma: 63} ensures that $\SPExptoHCP{\DualSPExp{T}}=\LinearDual{\SPExptoHCP{T}}$, thus the proof concludes by IH and applying Rule $\C$\textsc{Cut}. The rest of the cases follow by IH. 
\end{proof}

 \begin{corollary}\label{corollary: L equal to mupadovani}
     $\ClassL=\ClassMuP$.
 \end{corollary}
 \begin{proof}
     It follows by \Cref{theorem: L equal to mupadovani,theorem: mupadovani in L}.
 \end{proof}

\subparagraph{Main Results}
We may now return to considering $\ClassH$. Note that  $\ClassH\setminus \ClassLP\neq \emptyset$: to see this, consider, e.g., the self-synchronizing process $P_1\in\DeadlockFreedomHCPSet$ from \Cref{example:intrigue}. 
Using \Cref{corollary: L equal to mupadovani} and disentanglement (\Cref{def: disentanglement}) we show that given a process $P\in\DeadlockFreedomHCPSet$, we can find  a deadlock-free process $P'\in\ClassLP$. Formally, we have:

    \begin{theorem}\label{theorem: Hrt and Pado}
          Given $P\in\DeadlockFreedomHCPSet$,   $\exists P'\in\DeadlockFreedomHCPRRSet$ s.t. $P'\in\ClassLP$, $\SPExptoHCP{P}\BisimilarityHCP \SPExptoHCP{P'}$,   and $\DeadlockFreedomPado{\SPExptoPado{f}{P'}}$.
    \end{theorem}
    \begin{proof}
        By \Cref{corollary: main corollary HCP}, $\SPExptoHCP{P}\DisentangleClo\CommutingConversion^*\SPExptoHCP{P'}=P_1\para\cdots\para P_n$, and by \Cref{lemma: disentanglement in SP} $\forall i\in[1,n]$, there exists $P'_i\in \ClassL$ s.t. $\SPExptoHCP{P'_i}=P_i$. From \Cref{corollary: L equal to mupadovani,lemma: mupadovani in padovani} we infer $\SPExptoPado{f}{P'_i}\in\ClassLP$, $\forall i\in[1,n]$. Moreover, by repeatedly applying  Rule~\textsc{T-Par} we have that $P'\in\ClassLP$. Finally, $\DeadlockFreedomPado{\SPExptoPado{f}{P'}}$ follows from \Cref{theorem: deadlock freedom SPExp,corollary: deadlock freedom in pado and SP}.
    \end{proof}

The following result relates $\ClassH$, its subclasses $\DeadlockFreedomHCPSet$ and $\DeadlockFreedomHCPRRSet$, and $\ClassLP$.

\begin{theoremapxrep}\label{lemma: comparing ClassH and Padovani}
 The following hold:
 \textup{(1)}~$\ClassLP\setminus\ClassH\neq\emptyset$.
     \textup{(2)}~$\DeadlockFreedomHCPSet\setminus \ClassLP\neq \emptyset$.
      \textup{(3)}~$\ClassLP\cap\DeadlockFreedomHCPRRSet\neq\emptyset$.
\end{theoremapxrep}
    \begin{proof}
     We give witnesses for each statement:
    \begin{enumerate}
        \item Consider the $\SPExp$ process $P_{\ref{lemma: comparing ClassH and Padovani}}\DefEq \news{x}{y}(R_1\para R_2)$, with $R_1$ and $R_2$ defined as:
        \begin{align*}
            R_1\DefEq&\news{v_1}{v_2}(\OutputSP{x}{v_1}\OutputSP{x}{v_2}\EmptyOutputSP{x}\Inact)
            \qquad
            R_2\DefEq \InputSP{y}{z_1}\InputSP{y}{z_2}\EmptyInputSP{y}(\EmptyOutputSP{z_1}\Inact\para\EmptyInputSP{z_2}\Inact)
        \end{align*}
To check that $P_{\ref{lemma: comparing ClassH and Padovani}}\in\ClassLP$, we need to verify that $\PadoJud\SPExptoPado{f}{P_{\ref{lemma: comparing ClassH and Padovani}}}$. Its  type derivation is similar to the derivation for $ Q_{\ref{exmaple: typing in LP}}$ (\Cref{exmaple: typing in LP}); however $P_{\ref{lemma: comparing ClassH and Padovani}}\notin\ClassH$, since $R_1\notin\ClassH$:
\[\SPExptoHCP{R_1}=\news{v_1}{v_2}(\boutHCP{v'_1}{x}(\forward{v'_1}{v_1}\para \boutHCP{v'_2}{x}(\forward{v'_2}{v_2}\para \EmptyOutputHCP{x}\Inact)))\]
 we have $\boutHCP{v'_1}{x}(\forward{v'_1}{v_1}\para\boutHCP{v'_2}{x}(\forward{v'_2}{v_2}\para \EmptyOutputHCP{x}\Inact))
    \HCPJud x:\unit\OutputTypeHCP(\bot\OutputTypeHCP\unit),v_1:\bot , v_2:\unit$.
However the restriction $\news{v_1}{v_2}$ cannot be typed since $v_1$ and $v_2$ are in the same partition; see Rule~$\Hname$\textsc{Cut} in \Cref{fig:type system HCP}. 

\item Consider a self-synchronizing process  in $\DeadlockFreedomHCPSet$ such as $P_1$ (\Cref{example:intrigue}) or $P_{\ref{example: section 3 at work}}$ (\Cref{example: section 3 at work}).

\item      Consider the  $\SPExp$ process $P_3 \DefEq    \news{x_1}{y_1}\news{x_2}{y_2}\news{v_1}{k_1}\news{v_2}{k_2}(Q_1\para Q_2)$ with
\begin{align*}
Q_1&\DefEq\InputSP{x_1}{z_1}\InputSP{x_2}{z_2}\EmptyInputSP{x_1}\EmptyInputSP{x_2}\EmptyInputSP{z_1}\EmptyInputSP{z_2}\Inact \\
Q_2&\DefEq\OutputSP{y_1}{v_1}\OutputSP{y_2}{v_2}(\EmptyOutputSP{y_1}\Inact\para \EmptyOutputSP{y_2}\Inact\para \EmptyOutputSP{k_1}\Inact\para\EmptyOutputSP{k_2}\Inact)
\end{align*}
       It is easy to check that $P_3\in\ClassLP$ since $\DeadlockFreedomSP{P_3}$. Note that $\SPExptoHCP{P_3}=P_{\ref{example: process in HCP not in CP}}$, thus $P\in\ClassH$. It is straightforward to check that $P\in\DeadlockFreedomHCPRRSet$.
     \end{enumerate}
    \end{proof}

Finally, from \Cref{theorem: Hrt and Pado} we have the following result, which complements  \Cref{corollary: main corollary HCP}: given $P\in\DeadlockFreedomHCPSet$, there exists a bisimilar process $P'\in\DeadlockFreedomHCPRRSet\cap \ClassLP$ via disentanglement. Note that this result applies in particular  to processes in $\DeadlockFreedomHCPSet\setminus \ClassLP$.

\begin{corollary}\label{corollary: main result mupado}
Given $P\in \DeadlockFreedomHCPSet$,   there exists $P'\in \DeadlockFreedomHCPRRSet$ s.t.
\textup{(1)}~$P\ReductionSimulation P'$,
    \textup{(2)}~$\SPExptoHCP{P}\BisimilarityHCP\SPExptoHCP{P'}$, and  
    \textup{(2)}~$\DeadlockFreedomPado{\SPExptoPado{f}{P'}}$.
\end{corollary}

\subsection{Asynchronous Session Processes}\label{sec: async}

\begin{figure}[t]
  \footnotesize
  \begin{mathpar}
  \inferrule[\textsc{A-Out}]{\phantom{\cdot\ASPJud \mathbf{0}}}{v:T, c:S, x:\LinOutputST{T}{S}\ASPJud \AOutputSP{x}{v}{c}   }
\and
 \inferrule[\textsc{A-In}] {\Gamma, y : T, w:S \ASPJud P \quad  }{\Gamma, x: \LinInputST{T}{S} \ASPJud \AInputSP{x}{y}{w} P  } 
 \and
    \inferrule[\textsc{A-Inact}]{\phantom{\cdot\ASPJud \mathbf{0}}}{\cdot\ASPJud \mathbf{0}} \\
    \and
     \inferrule[\textsc{A-EInput}]{\Gamma\ASPJud P}{\Gamma, x: \EndTypeI\ASPJud \EmptyInputSP{x}P}
    \and
    \inferrule[\textsc{A-EOutput}]{\phantom{\cdot\ASPJud \mathbf{0}}}{ x:\EndTypeO\ASPJud \EmptyOutputSPA{x}}
\and
\inferrule[\textsc{A-Res}] {\Gamma, x : T, y : \overline{T} \ASPJud P  }{\Gamma\ASPJud \news{x}{y} P }
    \and
\inferrule[\textsc{A-Par}]{\Gamma_1 \ASPJud P \qquad \Gamma_2  \ASPJud Q }{\Gamma_1,\Gamma_2 \ASPJud P \para Q }
\end{mathpar}
   \vspace*{-0.25cm}

  \caption{\SPExpA: Typing Rules for Processes 
  }
  \label{fig:typ_rules asynchronous SP}
\end{figure}

Up to here, we have considered asynchronous communication somewhat indirectly: we have examined the type system for asynchronous processes $\Padovani$, defined its associated classes $\ClassLP$ and $\ClassMuP$, and related them to $\ClassL$ and $\DeadlockFreedomHCPSet$.
To treat asynchronous communication directly, here we consider $\SPExpA$, an asynchronous variant of $\SPExp$, and 
 discuss how to transfer our results from \Cref{sec: results,sec: comparing H and MuLP} from $\SPExp$ to $\SPExpA$ by leveraging an encoding $\SPAtoSP{\cdot}: \SPExpA \to \SPExp$.

\subparagraph{\texorpdfstring{$\SPExpA$}{ASP}: Asynchronous $\SPExp$}\label{sec: ASP}
We define $\SPExpA$ as a variant of $\SPExp$  with asynchronous communication.
Considerations about variables, endpoints, and values are as for $\SPExp$ (cf. \Cref{ss:sp}).
The syntax of {processes} is largely as before; 
only constructs for output and input are modified:

\begin{eqnarray*}
  P       &::= \  \Inact \mybar \EmptyInputSP{x}P \mybar\EmptyOutputSPA{x} \mybar \AOutputSP{x}{v}{c}    \mybar  \AInputSP{x}{y}{w}P  \mybar  P_1 \para P_2  \mybar  \news x y P 
  \end{eqnarray*}
\noindent
 
Process $\AOutputSP{x}{v}{c} $ sends names $v$ and $c$ along  $x$: the former is a message and the latter is a continuation. Because it has no continuation, this process may be seen as an ``output particle'' that carries a message.
Accordingly, process $\AInputSP{x}{y}{w}P$ receives two values $v$ and $c$ along $x$ and continues as $P\BiSubstitution{v}{y}{c}{w}$, i.e., the process resulting from the capture-avoiding substitution of  $y$ by $v$, and $w$ by $c$ in  $P$.
The construct for empty output is also modified. 
 In both $\AInputSP{z}{x}{y}P$ and $\news x y P$   $x,y$ are bound  in $P$.
\begin{definition}[$\SPExpA$]\label{def: type system ASP}
The typing rules for processes are presented in \Cref{fig:typ_rules asynchronous SP}. A judgement $\Gamma\ASPJud P$ denotes that $P$ is well typed in $\Gamma$. We shall write  $P \in \SPExpA$ if $\Gamma \ASPJud P$, for some $\Gamma$.  
\end{definition}

Most typing rules are self-explanatory. Rule \textsc{A-Eoutput} types an empty output without continuation. Rule  \textsc{A-Out} types the output particle $\AOutputSP{x}{v}{c}$ where $x:\LinOutputST{T}{S}$. Since the output has no continuation, the session name $c$ typed as $S$ serves a a continuation name for $x$. Dually, Rule \textsc{A-In} types $x$ as $\LinInputST{T}{S}$ provided that $y$ and $w$ are typed as $T$ and $S$ in $P$, respectively, thus, $w$ serves as a continuation for $x$ in $P$.

The reduction semantics for $\SPExpA$ is defined as follows:
\begin{align*}
     \news x y(\AOutputSP{x}{v}{c} \para  \AInputSP{y}{z}{w}Q\ \para\ S) &\ReductionSP ( Q\BiSubstitution{v}{z}{c}{w} \para S) &\textsc{A-Com}\\
     \news xy(\EmptyOutputSPA{x}\para \EmptyInputSP{y} Q\para S)&\ReductionSP   Q\para S &\textsc{A-EmptyCom}
\end{align*}

 Structural rules and the congruence  $\equiv$ are defined as in \Cref{fig:sess_pi_semantics}. 
  A
  labeled reduction semantics 
  $P \LabelledReduction{x}{y} Q$
  for  $\SPExpA$ 
  follows easily from \Cref{def: labelled reduction semantics SPExp}.

  \begin{example}\label{example: reductions asynchronous process}
Consider the process $P_{\ref{example: type derivation}}$ from \cref{example: type derivation}, which is typable and deadlocked in $\SPExp$.
 The following process can be seen as the analogue of  $P_{\ref{example: type derivation}}$ but in the asynchronous setting of $\SPExpA$. 
 We define  $P_{\ref{example: reductions asynchronous process}}$ as $\news{x_1}{y_1}\news{x_2}{y_2}\news{v_1}{k_1}\news{v_2}{k_2}P'$, where
\begin{align*}
    P'\DefEq&\AInputSP{x_1}{z_1}{w_1}\AInputSP{x_2}{z_2}{w_2}\EmptyInputSP{w_1}\EmptyInputSP{w_2}\EmptyInputSP{z_1}\EmptyInputSP{z_2}\Inact\\
    &\para \news{c_2}{c'_2}(\AOutputSP{y_2}{v_2}{c_2}\para \EmptyOutputSPA{c'_2})\para \news{c_1}{c'_1}(\AOutputSP{y_1}{v_1}{c_1}\para \EmptyOutputSPA{c'_1})\para \EmptyOutputSPA{k_1}\para\EmptyOutputSPA{k_2}
\end{align*}

Note how $c_1$ and $c_2$ serve as a continuation for $y_1$ and $y_2$, respectively.
There exist processes $Q_1, Q_2$ such that 
$P_{\ref{example: reductions asynchronous process}}~ \LabelledReduction{x_1}{y_1}~ Q_1~  \LabelledReduction{x_2}{y_2}~ Q_2 \ReductionSPA^*\equiv~  \Inact$.
\end{example}

We define DF for $\SPExpA$ by  adapting \Cref{def: deadlock freedom in SP}:

\begin{definition}[$\SPExpA$: Deadlock Freedom]\label{def: deadlock freedom in ASP}
    A process $P\in \SPExpA$ is deadlock free, written $\DeadlockFreedomSPA{P}$, if the following   holds: whenever $P\ReductionSP^* P'$ and one of the following holds:\\
        \textup{(1)}~$P'\equiv \new{\tilde{xy}}(\AOutputSP{x_i}{v}{c}\para Q_2)$, 
        \textup{(2)}~$P'\equiv \new{\tilde{xy}}(\InputSP{x_i}{y}Q_1\para Q_2)$,
        \textup{(3)}~$P'\equiv \new{\tilde{xy}}(\EmptyInputSP{x_i}Q_1\para Q_2)$, or
       \textup{(4)}~$P'\equiv \new{\tilde{xy}}(\EmptyOutputSPA{x_i}\para Q_2)$
      (with $x_i \in \widetilde{x}$ in all cases) 
    then there exists $R$ such that $P'\ReductionSP R$.
\end{definition}

\begin{example}[Comparing DF in $\SPExp$ and $\SPExpA$]
\label{ex:compare}
As just illustrated by means of processes $P_{\ref{example: type derivation}}$  (from \Cref{example: type derivation}) and $P_{\ref{example: reductions asynchronous process}}$, moving to an asynchronous setting may solve certain deadlocks: we have $\neg\DeadlockFreedomSP{P_{\ref{example: type derivation}}}$ and $\DeadlockFreedomSPA{P_{\ref{example: reductions asynchronous process}}}$.

On the other hand, moving to an asynchronous setting alone does not resolve the problem.
    Consider the process 
   $Q_{\ref{ex:compare}} \DefEq\news{x_2}{y_2}(\EmptyInputSP{y_2}\Inact\para\news{x_1}{y_1}(\EmptyInputSP{y_1}(\EmptyOutputSPA{x_2}\para \EmptyOutputSPA{x_1})))$. 
    Clearly,  
    $Q_{\ref{ex:compare}}$ is typable in \SPExpA.
    Also, 
        let 
    $Q^*_{\ref{ex:compare}}$
    be the $\SPExp$ variant of $Q_{\ref{ex:compare}}$ with $\Inact$ as continuation for outputs, i.e., 
    $Q^*_{\ref{ex:compare}} \DefEq\news{x_2}{y_2}(\EmptyInputSP{y_2}\Inact\para\news{x_1}{y_1}(\EmptyInputSP{y_1}(\EmptyOutputSP{x_2}{\Inact}\para \EmptyOutputSP{x_1}{\Inact})))$.
    We have that $Q^*_{\ref{ex:compare}}$ is typable in \SPExp and that $\neg\DeadlockFreedomSP{Q^*_{\ref{ex:compare}}}$ and $\neg\DeadlockFreedomSPA{Q_{\ref{ex:compare}}}$. 
\end{example}

Our goal is to transfer our technical results for $\SPExp$ to  the asynchronous setting of $\SPExpA$. To this end, we define the following encoding:
\begin{definition}\label{def: encoding SPA to SP}
    We define the encoding $\SPAtoSP{\cdot}$ from $\SPExpA$ to $\SPExp$ as follows:
 \begin{align*}
       \SPAtoSP{\AOutputSP{x}{v}{c}}&\DefEq\OutputSP{x}{v}\OutputSP{x}{c}\EmptyOutputSP{x}\Inact
    & \SPAtoSP{\EndTypeI}&\DefEq\EndTypeI
    \\
   \SPAtoSP{\AInputSP{x}{y}{w}P} &\DefEq \InputSP{x}{y}\InputSP{x}{w}\EmptyInputSP{x}\SPAtoSP{P} 
    & \SPAtoSP{\EndTypeO}&\DefEq\EndTypeO
    \\
      \SPAtoSP{\EmptyInputSP{x}P}&\DefEq\EmptyInputSP{x}\SPAtoSP{P} 
    & \SPAtoSP{\LinOutputST{T}{S}}&\DefEq\LinOutputST{T}{\LinOutputST{S}{\EndTypeO}}
    \\
    \SPAtoSP{\EmptyOutputSPA{x}}&\DefEq\EmptyOutputSP{x}\Inact 
    & \SPAtoSP{\LinInputST{T}{S}}&\DefEq\LinInputST{T}{\LinInputST{S}{\EndTypeI}}
\end{align*}
and as an homomorphism for $\Inact$, $P\para Q$, and $\news{x}{y}P$.
\end{definition}

\begin{example}[$\SPAtoSP{\cdot}$, At Work]\label{Example: SPAtoSP does not add deadlocks}
Consider again the process $P_{\ref{example: reductions asynchronous process}}$. We have seen that $\DeadlockFreedomSPA{P_{\ref{example: reductions asynchronous process}}}$. We now check that 
$\SPAtoSP{\cdot}$ does not add deadlocks by verifying that 
    $\DeadlockFreedomSP{\SPAtoSP{P_{\ref{example: reductions asynchronous process}}}}$:
\begin{align*}
    \SPAtoSP{P_{\ref{example: reductions asynchronous process}}}\DefEq&\news{x_1}{y_1}\news{x_2}{y_2}\news{v_1}{k_1}\news{v_2}{k_2}\SPAtoSP{P'}
\\
    \SPAtoSP{P'}\DefEq&\InputSP{x_1}{z_1}\InputSP{x_1}{w_1}\EmptyInputSP{x_1}\InputSP{x_2}{z_2}\InputSP{x_2}{w_2}\EmptyInputSP{x_2}\EmptyInputSP{w_1}\EmptyInputSP{w_2}\EmptyInputSP{z_1}\EmptyInputSP{z_2}\Inact\\
    &\para \news{c_2}{c'_2}(\OutputSP{y_2}{v_2}\OutputSP{y_2}{c_2}\EmptyOutputSP{y_2}\Inact\para \EmptyOutputSP{c'_2}\Inact)\\
    &\para \news{c_1}{c'_1}(\OutputSP{y_1}{v_1}\OutputSP{y_1}{c_1}\EmptyOutputSP{y_1}\Inact\para \EmptyOutputSP{c'_1}\Inact) \para \EmptyOutputSP{k_1}\Inact\para\EmptyOutputSP{k_2}\Inact
\end{align*}
One can show that there exist $Q_1, Q_2$ such that
$\SPAtoSP{P_{\ref{example: reductions asynchronous process}}}~  
    \LabelledReduction{x_1}{y_1}^3 \equiv Q_1 \LabelledReduction{x_2}{y_2}^3\equiv Q_2 \ReductionSPA^*\equiv\Inact$.

\end{example}

By a similar analysis to the one for \Cref{theorem: operational correspondence SP and pado}, we have the following result:
\begin{theorem}[$\SPAtoSP{\cdot}$: Operational Correspondence]\label{theorem: operational correspondence ASP and SP}
Let $P\in \SPExpA$. The following holds:

\begin{enumerate}
\item if $P\equiv \news{x}{y}(\AInputSP{x}{w}{z}P\para \AOutputSP{y}{v_1}{v_2}\para R)$, then   $\exists Q\in\SPExp$ s.t. $\SPAtoSP{P}\ReductionSP^3 Q$ and $Q=\SPAtoSP{P\BiSubstitution{v_1}{w}{v_2}{z}\para R}$.

\item If $P\equiv \news{x}{y}(\EmptyInputSP{x}P\para \EmptyOutputSP{y}\Inact\para R)$, then   $\exists Q\in\SPExp$ s.t. $\SPAtoSP{P}\ReductionSP Q$ and $Q=\SPAtoSP{P\para R}$.

\item If $\SPAtoSP{P}\ReductionSPA Q'$, then either:
(i)~there exists $Q\in\SPExp$ s.t. $Q'\ReductionSP^2 Q$, $P\ReductionSPA P'$ and $\SPAtoSP{P'}=Q$, or 
	(ii)~$P \ReductionSPA P'$ and $\SPAtoSP{P'}=Q'$.
\end{enumerate}
\end{theorem}

Using  \Cref{theorem: operational correspondence ASP and SP,def: deadlock freedom in SP,def: deadlock freedom in ASP} we formalize our claim from \Cref{Example: SPAtoSP does not add deadlocks}: $\SPAtoSP{\cdot}$ does not add deadlocks.
\begin{theoremapxrep}[Correspondence DF]\label{theorem: correspondence deadlock freedom ASP and SP}
For all $P\in\SPExpA$, $\DeadlockFreedomSPA{P}$ iff $\DeadlockFreedomSP{\SPAtoSP{P}}$.
\end{theoremapxrep}
\begin{proof}
	For each direction we prove its contrapositive and proceed by contradiction.
We discuss only the left-to-right direction; the analysis for the right-to-left direction proceeds similarly.
 If $\neg \DeadlockFreedomSPA{P}$, then there exists a $P'\in\SPExpA$, such that $P\ReductionSP^* P'$, $P'\longarrownot\ReductionSPA$
	, and $P'$ is equivalent to one of the four forms in \Cref{def: deadlock freedom in ASP}, thus $P'\not\equiv \Inact$. 
	 By (repeated applications of) \Cref{theorem: operational correspondence ASP and SP}(3), there exists $R\not\equiv\Inact$ s.t. $\SPAtoSP{P}\ReductionSPA^* R$ and $\SPAtoSP{P'}=R$.  If $\DeadlockFreedomSP{\SPAtoSP{P}}$, then there exists $Q$ s.t $\SPAtoSP{P'}\ReductionSPA Q$, and by \Cref{theorem: operational correspondence ASP and SP}(3), then $P'\ReductionSP Q'$ for some $Q'$, which is a contradiction. The right-to-left direction proceeds similarly using \Cref{def: deadlock freedom in SP}.
\end{proof}

Using the encoding $\SPAtoSP{\cdot}$, and  inspired by \Cref{theorem: operational correspondence ASP and SP,theorem: correspondence deadlock freedom ASP and SP}, we can define the following classes of asynchronous processes by relying on their synchronous counterparts:

     \begin{definition}
     \label{d:asyncclasses}
    The classes of asynchronous processes are defined as follows:\\
    \begin{minipage}{12cm}
            \begin{align*}
    \ClassLA&\DefEq\{P \in \SPExpA\Sep \SPAtoSP{P}\in \ClassL\} &  \ClassLPA&\DefEq\{P \in \SPExpA\Sep \SPAtoSP{P}\in \ClassLP\}  & 
       \ClassMuPA&\DefEq\{P \in \SPExpA\Sep \SPAtoSP{P}\in \ClassMuP\}
    \\
     \ClassHA&\DefEq\{P \in \SPExpA\Sep \SPAtoSP{P}\in \ClassH\}
 & 
    \DeadlockFreedomHCPSetA&\DefEq\{P \in\SPExpA\Sep \SPAtoSP{P}\in\DeadlockFreedomHCPSet\}& \DeadlockFreedomHCPRRSetA&\DefEq\{P \in\SPExpA\Sep \SPAtoSP{P}\in\DeadlockFreedomHCPRRSet\} 
    \end{align*}
    \end{minipage}

\end{definition}

Each asynchronous class is contained (up to $\SPAtoSP{\cdot}$) in its corresponding synchronous class:

\begin{lemma}\label{lemma: classes of processes ASP}
  Let $P\in\SPExpA$.  The following hold:
        \textup{(1)}~If $P\in\ClassLA$, then $\SPAtoSP{P}\in\ClassL$. 
        \textup{(2)}~If $P\in\ClassHA$, then $\SPAtoSP{P}\in\ClassH$.
        \textup{(3)}~If $P\in\ClassLPA$, then $\SPAtoSP{P}\in\ClassLP$.
        \textup{(4)}~If $P\in\ClassMuPA$, then $\SPAtoSP{P}\in\ClassMuP$.
        \textup{(5)}~If $P\in\DeadlockFreedomHCPSetA$, then $\SPAtoSP{P}\in\DeadlockFreedomHCPSet$.
        \textup{(6)}~If $P\in\DeadlockFreedomHCPRRSetA$, then $\SPAtoSP{P}\in\DeadlockFreedomHCPRRSet$.
\end{lemma}
\begin{proof}
	It follows straightforwardly from \Cref{d:asyncclasses}.
\end{proof}
Thus, we can transfer our main results (\Cref{corollary: main corollary HCP,corollary: main result mupado,corollary: L equal to mupadovani}) to $\SPExpA$ via $\SPAtoSP{\cdot}$.
For instance,  \Cref{corollary: main corollary HCP} is transferred to $\SPExpA$ as follows, using \Cref{lemma: classes of processes ASP}, and \Cref{theorem: operational correspondence ASP and SP}:
\begin{corollary}
\label{c:transfer}
Given  $P\in\DeadlockFreedomHCPSetA$, there exist $Q\in\ClassH$ and $P'\in\DeadlockFreedomHCPRRSetA$ s.t.:
    \textup{(1)}~$\SPExptoHCP{Q}\in \Disentangled{\SPExptoHCP{\SPAtoSP{P}}}$,
    \textup{(2)}~$\SPExptoHCP{Q}\CommutingConversion^*\SPExptoHCP{\SPAtoSP{P'}}$,
    \textup{(3)}~$\SPAtoSP{P}\ReductionSimulation \SPAtoSP{P'}$,
    \textup{(4)}~$\SPExptoHCP{\SPAtoSP{P}}\BisimilarityHCP\SPExptoHCP{\SPAtoSP{P'}}$. 
\end{corollary}

\section{Closing Remarks}\label{sec: concluding remarks}

\begin{figure}[t!]
    \centering
\tikzset{every picture/.style={line width=0.75pt}} 

\begin{tikzpicture}[x=0.75pt,y=0.75pt,yscale=-1,xscale=1]

\draw   (20.6,74.95) .. controls (19.69,36.68) and (61.57,5.65) .. (114.15,5.65) .. controls (166.72,5.65) and (210.07,36.68) .. (210.98,74.95) .. controls (211.89,113.22) and (170.01,144.25) .. (117.44,144.25) .. controls (64.86,144.25) and (21.51,113.22) .. (20.6,74.95) -- cycle ;
\draw   (28.08,87.89) .. controls (27.34,56.76) and (65.81,31.52) .. (114,31.52) .. controls (162.2,31.52) and (201.87,56.76) .. (202.61,87.89) .. controls (203.34,119.01) and (164.87,144.25) .. (116.68,144.25) .. controls (68.48,144.25) and (28.81,119.01) .. (28.08,87.89) -- cycle ;
\draw   (38.42,100.75) .. controls (37.85,76.73) and (71.49,57.25) .. (113.55,57.25) .. controls (155.62,57.25) and (190.18,76.73) .. (190.75,100.75) .. controls (191.32,124.77) and (157.68,144.25) .. (115.62,144.25) .. controls (73.55,144.25) and (38.99,124.77) .. (38.42,100.75) -- cycle ;
\draw   (80.21,123.3) .. controls (79.93,111.72) and (95.82,102.34) .. (115.68,102.34) .. controls (135.55,102.34) and (151.88,111.72) .. (152.15,123.3) .. controls (152.43,134.87) and (136.55,144.25) .. (116.68,144.25) .. controls (96.81,144.25) and (80.48,134.87) .. (80.21,123.3) -- cycle ;
\draw [color={rgb, 255:red, 208; green, 2; blue, 27 }  ,draw opacity=1 ]   (114,49) -- (114,75) ;
\draw [shift={(114,77)}, rotate = 270.7] [color={rgb, 255:red, 208; green, 2; blue, 27 }  ,draw opacity=1 ][line width=0.75]    (10.93,-3.29) .. controls (6.95,-1.4) and (3.31,-0.3) .. (0,0) .. controls (3.31,0.3) and (6.95,1.4) .. (10.93,3.29)   ;
\draw [color=darkolivegreen  ,draw opacity=1 ]   (125,123) -- (317,123) ;
\draw [color=darkolivegreen  ,draw opacity=1 ]   (125,125) -- (317,125) ;

\draw   (225.6,75.45) .. controls (224.69,37.18) and (266.57,6.15) .. (319.15,6.15) .. controls (371.72,6.15) and (415.07,37.18) .. (415.98,75.45) .. controls (416.89,113.72) and (375.01,144.75) .. (322.44,144.75) .. controls (269.86,144.75) and (226.51,113.72) .. (225.6,75.45) -- cycle ;
\draw   (285.96,123.8) .. controls (285.69,112.22) and (301.57,102.84) .. (321.44,102.84) .. controls (341.31,102.84) and (357.64,112.22) .. (357.91,123.8) .. controls (358.19,135.37) and (342.3,144.75) .. (322.44,144.75) .. controls (302.57,144.75) and (286.24,135.37) .. (285.96,123.8) -- cycle ;

\draw  [dash pattern={on 0.84pt off 2.51pt}]  (177,90) -- (311,90) ;
\draw  [draw opacity=0.6 ] (1.33,2) -- (433.33,2) -- (433.33,148.4) -- (1.33,148.4) -- cycle ;

\draw (320,118) node [anchor=north west][inner sep=0.75pt]   [align=left] {$\ClassMuP$};
\draw (320,10.12) node [anchor=north west][inner sep=0.75pt]   [align=left] {$\ClassLP$};
\draw (310.5,80.75) node [anchor=north west][inner sep=0.75pt]   [align=left] {$\SPExptoPado{f}{P'}$};
\draw (406.8,7.67) node [anchor=north west][inner sep=0.75pt]   [align=left] {$\SPExp$};
\draw (48.28,51.06) node [anchor=north west][inner sep=0.75pt]   [align=left] {$\DeadlockFreedomHCPSet$};
\draw (51.53,103.33) node [anchor=north west][inner sep=0.75pt]   [align=left] {$\DeadlockFreedomHCPRRSet$};
\draw (104.94,10.12) node [anchor=north west][inner sep=0.75pt]  [xslant=-0.03]  {$\ClassH$};
\draw (102,32.94) node [anchor=north west][inner sep=0.75pt]   [align=left] {$\SPExptoHCP{P}$};
\draw (110.53,118) node [anchor=north west][inner sep=0.75pt]   [align=left] {$\ClassL$};
\draw (57,80) node [anchor=north west][inner sep=0.75pt]    {$P_1\para\cdots\para P_n=\SPExptoHCP{P'}$};
\end{tikzpicture}
     \caption{Overview of main results (detailed version). The inclusions between the classes of   processes follow from \Cref{lemma: comparing ClassH and Padovani}. The red arrow stands for the relation $\DisentangleClo\CommutingConversion^*$ (disentanglement and commuting conversions) from \Cref{corollary: main corollary HCP}, where  $P_1,\dots,P_n\in\ClassL$. This  arrow,  together with $\SPExptoPado{f}{P'}\in\ClassLP$, represent \Cref{corollary: main result mupado}. The green lines represent \Cref{corollary: L equal to mupadovani} ($\ClassL = \ClassMuP$). }
    \label{fig:summary}
    \vspace{-3mm}
\end{figure}

Enforcing DF is a relevant issue for message-passing programs, and type systems for concurrent processes offer a convenient setting to develop analysis techniques that detect insidious circular dependencies in programs. 
Because there exist different type systems that enforce DF, it is natural to wonder how they compare. 
We have studied this question in the general and expressive setting of the $\pi$-calculus. 
Our approach aims at   establishing the key strengths and limitations of three representative type systems: \SPExp, the reference  language for session-typed processes; \HCP, the session-typed language based on \PaS and hypersequents; and \Padovani, the priority-based type system for asynchronous processes. 

    \Cref{fig:summary} gives an overview of our main technical results, refining \Cref{fig:resultsV1}. We briefly discuss their broader significance. Our work can be seen as covering three   dimensions of comparison: type systems (\SPExp, \HCP, \Padovani), process semantics  (reduction semantics and LTS), and several definitions of DF (cf. \Cref{def: deadlock freedom in SP,def: deadlock freedom delayed,def: deadlock freedom no delayed,def: deadlock freedompadovani body,def: deadlock freedom in ASP}). Clearly, each type system connects these dimensions in its own way via meta-theoretical results (e.g., type preservation). From this perspective, considering a type system such as \HCP, which comes with an LTS for typed processes with very distinctive features (delayed actions and self-synchronizations), brings significant value to our results. In contrast, the work of Dardha and P\'{e}rez~\cite{ComparingDeadlock} (the most closely related work) considers different type systems for synchronous processes but only under a reduction semantics, which  influences the definition(s) of DF to be considered. 
     
  Our key discovery is identifying the   role that commuting conversions and disentanglement have in the class of deadlock-free processes induced by \HCP (\Cref{corollary: main corollary HCP}). 
  Although studying the practical ramifications of our results goes beyond the scope of this paper,  two observations are relevant. 
  First, our results pave the way for developing extensions of the static analyzer TyPiCal~\cite{TyPiCal} that support the mechanized analysis of (session) processes with asynchronous communication.
  Second, commuting conversions are already important in practical developments, such as execution strategies in \emph{abstract machines} for session-typed programs~\cite{DBLP:conf/esop/CairesT24}
     and definitions of \emph{session subtyping}~\cite{DBLP:conf/ppdp/ChenDY14,10.1145/3434297} that enable flexible programming interfaces.
  An in-depth exploration of these latent connections between our results and tools for the analysis of message-passing programs is left for future work.  
     
     It is worth stressing that \HCP's LTS enjoys  strong logical and denotational justifications; by identifying disentanglement and commuting conversions as key notions for DF, we provide new insights into \HCP and its LTS. This way, our work sheds new light into the foundations of \HCP,  its positioning within the line of work on \PaS, and its expressive power. For instance, it is surprising that processes that share more than one session are typable in \HCP, in the sense of \Cref{example: process in HCP not in CP}. While Kokke \etal~\cite{betterlate,TakingLinearLogicApart} have studied disentanglement as a tool to relate \HCP and \CP, here we uncover its connection with DF. Also, their work on $\HCP^-$ (\HCP with a reduction semantics but without delayed actions and self-synchronizations) is complementary to our findings. In this sense, our work also offers good motivations for the use of LTSs with regular transitions as the semantics for logic-based session processes (which usually rely on reduction semantics).

    It is remarkable that \HCP and \Padovani are not just two  type systems relevant for a formal comparison, but that they are actually connected thanks to asynchrony (an important aspect not covered in Dardha and P\'{e}rez's work); the connection is made precise by   \Cref{lemma: comparing ClassH and Padovani} and \Cref{corollary: main result mupado}. Here again delayed actions in \HCP result to be insightful for an enhanced understanding of logic-based session types. In this line, we observe that De Young \etal~\cite{DeYoungCPT12} study asynchrony under \PaS, and in particular explain the role of commuting conversions, which in an asynchronous setting act as as structural congruences (cf. \Cref{def: commuting conversions}). In the synchronous case, they correspond to behavioral equivalences~\cite{LogicalRelations}. 
     
An interesting  point concerns the exact relation between our class $\ClassL$ (\Cref{Def: ClassL and ClassH}) and the  class $\ClassLDP$ in Dardha and P\'{e}rez's work. There is a key difference between the two: while they consider \CP with a typing rule for independent parallel composition (cf. Rule~$\Hname$\MixTwo in \Cref{fig:type system HCP}), we only have Rule \C\MixZero in \Cref{fig:CP Rules}. The reason is technical: it is known that having both \MixZero and \MixTwo induces the conflation of the units $\unit$ and $\bot$, which conflicts with the distinction made by \HCP's denotational semantics (\Cref{d:denot}). (In \cite{ComparingDeadlock}, the conflated type is denoted `$\bullet$', and there are no process constructs for session closing.)
 For the sake of consistency with the denotational semantics, we opted to not include this rule. 
Therefore, there are processes that are not in our class $\ClassL$ but are in Dardha and P\'{e}rez's $\ClassLDP$.

Having discovered the relation between DF and commuting conversions and disentanglement, an interesting item for future work is revisiting the process transformations proposed by Dardha and P\'{e}rez~\cite{ComparingDeadlock} using disentanglement. To focus on the fundamental issues of DF enforcement, we have not covered processes with recursion/sharing; we also plan to consider typed languages with constructs for shared servers and clients.

\bibliographystyle{plainurl}
\bibliography{references}

\end{document}